\documentclass[preprint,showpacs,preprintnumbers,amsmath,amssymb,aps,floatfix]{revtex4}

\usepackage{dcolumn}
\usepackage{graphicx}
\usepackage[mathscr]{euscript}
\usepackage[pdftex]{color}
\usepackage[sort&compress]{natbib}
\usepackage[colorlinks=true,linkcolor=blue,filecolor=blue,urlcolor=blue,citecolor=blue,pdftex=true,plainpages=false]{hyperref}
\usepackage{ulem}

\def\LB{\left\{}	
\def\RB{\right\}}	
\def\PAR#1#2{ {\frac{\partial #1}{\partial #2}} }

\newcommand{\BE}{\begin{displaymath}}
\newcommand{\EE}{\end{displaymath}}
\newcommand{\BNE}{\begin{equation}}
\newcommand{\ENE}{\end{equation}}
\newcommand{\BEA}{\begin{eqnarray}}
\newcommand{\EEA}{\nonumber\end{eqnarray}}
\newcommand{\EL}{\nonumber\\}

\newcommand{\ie}{{\em i.e.\ }}
\newcommand{\eg}{{\em e.\,g.\ }}

\def\chpt{\raise0.4ex\hbox{$\chi$}PT}
\def\schpt{S\raise0.4ex\hbox{$\chi$}PT}
\def\rschpt{rS\raise0.4ex\hbox{$\chi$}PT}
\def\figref#1{Fig.~\ref{fig:#1}}
\def\Figref#1{Figure~\ref{fig:#1}}
\def\figrefs#1#2{Figs.~\ref{fig:#1} and \ref{fig:#2}}

\def\figrefthree#1#2#3{Figs.~\ref{fig:#1}, \ref{fig:#2}, and \ref{fig:#3}}
\def\Figrefthree#1#2#3{Figures~\ref{fig:#1}, \ref{fig:#2}, and \ref{fig:#3}}

\def\figrefto#1#2{Figs.~\ref{fig:#1} -- \ref{fig:#2}}

\def\secref#1{Sec.~\ref{sec:#1}}
\def\secrefs#1#2{Secs.~\ref{sec:#1} and \ref{sec:#2}}

\def\Secref#1{Section~\ref{sec:#1}}
\def\tabref#1{Table~\ref{tab:#1}}

\def\gtwid{{\,\raise.3ex\hbox{$>$\kern-.75em\lower1ex\hbox{$\sim$}}\,}}
\def\ltwid{{\,\raise.3ex\hbox{$<$\kern-.75em\lower1ex\hbox{$\sim$}}\,}}

\def\ie{{\it i.e.},\ }
\def\eg{{\it e.g.},\ }
\def\et{{\it et al.}}

\def\cM{{\cal M}}

\def\cO{{\cal O}}

\def\rcite#1{Ref.~\cite{#1}}

\def\rcites#1{Refs.~\cite{#1}}

\def\eqn#1{\label{eq:#1}}
\def\Equation#1{Equation~(\ref{eq:#1})}

\def\eq#1{Eq.~(\ref{eq:#1})}
\def\eqsthru#1#2{Eqs.~(\ref{eq:#1}) through (\ref{eq:#2})}
\def\eqs#1#2{Eqs.~(\ref{eq:#1}) and (\ref{eq:#2})}

\def\aschpt{HMrAS\raise0.4ex\hbox{$\chi$}PT}

\begin{document}

\title{Charmed and light pseudoscalar meson decay constants from four-flavor lattice QCD with physical light quarks}

\author{A.~Bazavov}
\altaffiliation[Present address:~]{Department of Physics and Astronomy, University of Iowa, Iowa City, IA, USA}
\affiliation{Physics Department, Brookhaven National Laboratory, Upton, NY, USA}

\author{C.~Bernard}
\email[]{cb@lump.wustl.edu}
\affiliation{Department of Physics, Washington University, St.~Louis, MO, USA}

\author{C.M.~Bouchard}
\affiliation{Department of Physics, The Ohio State University, Columbus, OH, USA}

\author{C.~DeTar}
\affiliation{Department of Physics and Astronomy, University of Utah, Salt Lake City, UT, USA}

\author{D.~Du}
\affiliation{Department of Physics, Syracuse University, Syracuse, NY, USA}

\author{A.X.~El-Khadra}
\affiliation{Physics Department, University of Illinois, Urbana, IL, USA}

\author{J.~Foley}
\affiliation{Department of Physics and Astronomy, University of Utah, Salt Lake City, UT, USA}

\author{E.D.~Freeland}
\affiliation{Liberal Arts Department, School of the Art Institute of Chicago, Chicago, IL, USA}

\author{E.~G\'amiz}
\affiliation{CAFPE and Departamento de Fisica Te\'orica y del Cosmos, Universidad de Granada, Granada, Spain}

\author{Steven~Gottlieb}
\affiliation{Department of Physics, Indiana University, Bloomington, IN, USA}

\author{U.M.~Heller}
\affiliation{American Physical Society, Ridge, NY, USA}

\author{J.~Kim}
\altaffiliation[Present address:~]{Department of Physics and Astronomy, Seoul National University, Seoul, Korea}
\affiliation{Physics Department, University of Arizona, Tucson, AZ, USA}

\author{J.~Komijani}
\email[]{jkomijani@physics.wustl.edu}
\affiliation{Department of Physics, Washington University, St.~Louis, MO, USA}

\author{A.S.~Kronfeld}
\affiliation{Fermi National Accelerator Laboratory, Batavia, IL, USA}

\author{J.~Laiho}
\affiliation{Department of Physics, Syracuse University, Syracuse, NY, USA}

\author{L.~Levkova}
\affiliation{Department of Physics and Astronomy, University of Utah, Salt Lake City, UT, USA}

\author{P.B.~Mackenzie}
\affiliation{Fermi National Accelerator Laboratory, Batavia, IL, USA}

\author{E.T.~Neil}
\affiliation{Department of Physics, University of Colorado, Boulder, CO, USA}
\affiliation{RIKEN-BNL Research Center, Brookhaven National Laboratory, Upton, NY, USA}

\author{J.N.~Simone}
\affiliation{Fermi National Accelerator Laboratory, Batavia, IL, USA}

\author{R.~Sugar}
\affiliation{Department of Physics, University of California, Santa Barbara, CA, USA}

\author{D.~Toussaint}
\email[]{doug@physics.arizona.edu}
\affiliation{Physics Department, University of Arizona, Tucson, AZ, USA}

\author{R.S.~Van~de~Water}
\affiliation{Fermi National Accelerator Laboratory, Batavia, IL, USA}

\author{R.~Zhou}
\affiliation{Fermi National Accelerator Laboratory, Batavia, IL, USA}

\author{[Fermilab Lattice and MILC Collaborations]}
\noaffiliation

\date{\today}

\begin{abstract}
We compute the leptonic decay constants  $f_{D^+}$, $f_{D_s}$, and $f_{K^+}$, and the quark-mass ratios
$m_c/m_s$ and $m_s/m_l$ in unquenched lattice QCD using the experimentally determined value of
$f_{\pi^+}$ for normalization.  We use the MILC highly improved staggered quark (HISQ) ensembles with
four dynamical quark flavors---up, down, strange, and charm---and with both physical and unphysical
values of the light sea-quark masses.  The use of physical pions removes the need for a chiral
extrapolation, thereby eliminating a significant source of uncertainty in previous calculations.  Four
different lattice spacings ranging from $a\approx 0.06$ fm to $0.15$ fm are included in the analysis to
control the extrapolation to the continuum limit.  Our primary results are
$f_{D^+} = 212.6(0.4)({}^{+1.0}_{-1.2})\ \mathrm{MeV}$,
$f_{D_s} = 249.0(0.3)({}^{+1.1}_{-1.5})\ \mathrm{MeV}$, and
$f_{D_s}/f_{D^+} = 1.1712(10)({}^{+29}_{-32})$,
where the errors are statistical and total systematic, respectively.  The errors on our results for the charm decay constants 
and their ratio are approximately two to four times smaller than those of the most precise previous lattice calculations.
We also obtain $f_{K^+}/f_{\pi^+} = 1.1956(10)({}^{+26}_{-18})$, updating our previous result, 
and determine the quark-mass ratios 
$m_s/m_l = 27.35(5)({}^{+10}_{-7})$ and
$m_c/m_s = 11.747(19)({}^{+59}_{-43})$.
When combined with experimental measurements of the decay rates, our results lead to precise determinations of the CKM matrix elements $|V_{us}| = 0.22487(51) (29)(20)(5)$, $|V_{cd}|=0.217(1) (5)(1)$ and $|V_{cs}|= 1.010(5)(18)(6)$, 
where the errors are from this calculation of the decay constants, the uncertainty
in the experimental decay rates, structure-dependent electromagnetic corrections, and, in the case of $|V_{us}|$, the uncertainty in $|V_{ud}|$, respectively.

\end{abstract}

\maketitle

\section{Introduction and motivation}

The leptonic decays  of pseudoscalar mesons enable precise determinations of Cabibbo-Kobayashi-Maskawa (CKM) quark-mixing matrix elements within the Standard Model.
In  particular, experimental rates for the decays $D^+\to\mu^+\nu$, $D_s\to\mu^+\nu$ and
$D_s\to\tau^+\nu$, when combined with lattice calculations of the charm-meson decay constants  $f_{D^+}$  and $f_{D_s}$,
allow one to obtain
$|V_{cd}|$ and $|V_{cs}|$.  Indeed,
this approach results in the most precise current determination of $|V_{cd}|$. Similarly, the light-meson decay-constant
ratio $f_{K^+}/f_{\pi^+}$ can be used to extract $|V_{us}|/|V_{ud}|$ from the  experimental ratio of kaon and pion 
leptonic decay widths \cite{Marciano:2004uf,FPI04}.  Here we calculate the charm decay constants for the first time using physical values for the light sea-quark mass. We obtain $f_{D^+}$  and $f_{D_s}$ to about 0.5\% precision and their ratio $f_{D_s}/f_{D^+}$ to about  0.3\% precision; we also update our earlier calculation of $f_{K^+}/f_{\pi^+}$ \cite{FKPRL} to almost 0.2\% precision.  This is the most precise lattice calculation of the charm decay constants to date, and improves upon previous results by a factor of two to four.  We also compute the quark-mass ratios $m_c/m_s$ and $m_s/m_l$, which are fundamental parameters of the Standard Model.

We use the lattice ensembles generated by the MILC Collaboration with four flavors ($n_f=2+1+1$) of dynamical quarks using
the highly improved staggered quark (HISQ) action, and a one-loop tadpole improved Symanzik improved gauge
action~\cite{HPQCD_HISQ,milc_hisq,scaling09,HISQ_CONFIGS}.  
The generation algorithm uses the fourth-root procedure to remove the unwanted taste degrees of freedom
\cite{Marinari:1981qf,Prelovsek:2005rf,Bernard:2006zw,Bernard:2006ee,Bernard:2007qf,Shamir:2004zc,Shamir:2006nj,Follana:2004sz,Durr:2004as,Durr:2004ta,Wong:2004nk,Durr:2006ze,Donald:2011if}.
Our data set includes ensembles with four values of the lattice spacing ranging from approximately 0.15~fm to 0.06~fm, enabling good control over the continuum extrapolation.
The data set includes both ensembles with the light (up-down), strange, and charm sea-masses close to their physical values
(``physical-mass ensembles'') and ensembles where either the light sea-mass is heavier than in nature, or the
strange sea-mass is lighter than in nature, or both.  

The physical-mass ensembles enable us to perform first a straightforward analysis that does not require chiral fits.  This
analysis, which we refer to as the ``physical-mass analysis''  below, gives our results for $f_{K^+}/f_{\pi^+}$, as well as ratios of physical quark masses.  
The quark-mass ratios are then used as input to a more sophisticated analysis of the charm decay constants that 
includes the ensembles with unphysical sea-quark masses.  In this second analysis, referred to 
as the ``chiral analysis,'' we analyze our complete data set within the framework of staggered chiral perturbation theory (\schpt) for all-staggered heavy-light mesons~\cite{Lee-Sharpe,Aubin:2003mg,Bernard-Komijani}.
The  inclusion of the unphysical-mass ensembles gives us tighter control on discretization effects because \schpt\ connects the quark-mass and lattice-spacing dependence of the data, reducing the statistical errors on the decay constants significantly, and allowing us to make more refined adjustments for mistuning of masses.  We therefore take our final central values for $f_{D^+}$, $f_{D_s}$, and $f_{D_s}/f_{D^+}$ from the chiral analysis.  The physical-mass analysis provides a cross check of the chiral analysis and is used in our final estimate of systematic uncertainties.   

An earlier result for $f_{K^+}/f_{\pi^+}$ was presented in Ref.~\cite{FKPRL}.  Here we update this analysis
with slightly more statistics and improved estimates for the systematic errors.
Preliminary results for the charm decay constants and quark masses were presented in Ref.~\cite{LATTICE13_FD}.

\bigskip
This paper is organized as follows.
\Secref{params} gives details about the lattice ensembles used in our calculation and the method for 
extracting the decay constants from two-point correlation functions.  As discussed in
\secref{twopointfits}, the first stage in our analysis is to fit the two-point correlators to determine
the meson masses and decay amplitudes for each pair of valence-quark masses.  \Secref{analysis} presents
the main body of our analysis, which proceeds in two stages.  In the first stage, described in
\secref{physical-mass-analysis}, we use the physical-mass  ensembles to compute quark-mass ratios and
$f_{K^+}/f_{\pi^+}$, as well as some additional intermediate quantities required for the later chiral
analysis of the $D$-meson decay constants.  In the first part of the physical-mass analysis,
\secref{valencemassfits}, we fit the valence-quark mass dependence of the masses and amplitudes, and
evaluate the decay amplitudes at the resulting tuned valence masses.  Next, in \secref{seacontinuumfits},
we adjust the quark-mass ratios and decay amplitudes to account for the slight sea-quark mass mistuning
and extrapolate these results to the continuum.  In the last part of the physical-mass analysis,
\secref{FV-EM}, we consider systematic errors from finite-volume and electromagnetic effects.  In the
second analysis stage described in \secref{chiral-analysis}, we use heavy-light staggered chiral
perturbation theory to combine the unphysical light- and strange-quark mass ensembles with the
nearly-physical quark mass ensembles to obtain the charm-meson decay constants.  We first present the
chiral perturbation theory for all-staggered heavy-light mesons in \secref{CHIPT}.  We then discuss the
required mass-independent scale setting in \secref{LAT-SCALE}, where we take care to correct for effects
on the scale and quark-mass estimates of mistunings of the sea-quark masses. We present the
chiral-continuum fits in \secref{CHIPT-fits}, and  discuss the systematic errors from the continuum
extrapolation, as well as from other sources, in \secref{SYSERRS}.  We present our final results for the
decay constants and quark-mass ratios with error budgets in \secref{conclusions}, in which we also
compare our results to other unquenched lattice calculations.  Finally, we discuss the impact of our
results on CKM phenomenology in \secref{CKM}. 
The Appendix gives details about the inclusion of nonleading heavy-quark effects in our chiral formulas. 

\section{Lattice simulation parameters and methods}
\label{sec:params}

Table~\ref{tab:ensembles} summarizes the lattice ensembles used in this calculation.  Discussion of the parameters
relevant to the lattice generation, such as integration step sizes and acceptance rates, choice of the RHMC or RHMD
algorithm, and autocorrelations of various quantities can be found in Ref.~\cite{HISQ_CONFIGS}.
In particular, we find that the effects of using the RHMD algorithm
rather than the RHMC algorithm in some of our ensembles are negligible.
The dependence of error estimates for the decay constants in this
work on the jackknife block size is consistent with the more general
results on autocorrelations in Ref.~\cite{HISQ_CONFIGS}.
Reference~\cite{HISQ_CONFIGS} also shows the molecular dynamics time evolution of
the topological charge for many of these ensembles and histograms of the topological
charge.  We have since also verified that on the
$a\approx0.06$ fm physical quark mass ensemble the autocorrelation time for the topological charge
is much shorter than the topological charge autocorrelation time on the $a\approx0.06$ fm
$m_l^\prime = m_s^\prime/5$ ensemble shown in Fig.~2 of Ref.~\cite{HISQ_CONFIGS}.  The dependence
on the light-quark mass can be understood by thinking of the decorrelation process as a random walk in
the topological charge.

\begin{table} \renewcommand{\arraystretch}{0.85}
\caption{
\label{tab:ensembles}
Ensembles used in this calculation. The first column is the gauge coupling $\beta=10/g^2$, and the
next three columns are the sea-quark masses in lattice units.  The primes on the masses indicate that they
are the values used in the runs, and in general differ from the physical values either by choice, or because of
tuning errors.
The lattice spacings in this table are obtained separately on each ensemble using $f_{\pi^+}$ as the length
standard, following the procedure described in Sec.~\protect\ref{sec:valencemassfits}.
(In Sec.~\ref{sec:chiral-analysis} we use a mass-independent lattice spacing, described there.)
The lattice spacings here differ slightly from
those in Ref.~\cite{HISQ_CONFIGS} since we use $f_{\pi^+}$ as the length scale,
while those in \rcite{HISQ_CONFIGS} were determined using $F_{p4s}$ (discussed
at the beginning of \protect{\secref{physical-mass-analysis}}). 
Values of the strange quark mass chosen to be unphysical are marked with a dagger ($\dagger$); while the
asterisk~(*) marks an ensemble that we expect to extend in the future.
}
\begin{tabular}{|l|lll|l|l|llll|}
\hline
$\beta$ & $am'_l$ & $am'_s$ & $am'_c$ & $(L/a)^3\times (T/a)$      & $N_{lats}$ & $a$ (fm)& $L$ (fm) & $M_\pi L$ & $M_\pi$ (MeV) \\ \hline
\hline
5.80  & 0.013   & 0.065   & 0.838 & $16^3\times 48$  & 1020  & 0.14985(38) & 2.38 & 3.8 & 314 \\
5.80  & 0.0064  & 0.064   & 0.828 & $24^3\times 48$  & 1000  & 0.15303(19) & 3.67 & 4.0 & 214 \\
5.80  & 0.00235 & 0.0647  & 0.831 & $32^3\times 48$  & 1000  & 0.15089(17) & 4.83 & 3.2 & 130 \\
\hline
6.00  & 0.0102  & 0.0509  & 0.635 & $24^3\times 64$  & 1040  & 0.12520(22) & 3.00 & 4.5 & 299 \\
6.00  & 0.0102  & 0.03054$\null^\dagger$  & 0.635 & $24^3\times 64$  & 1020  & 0.12104(26) & 2.90 & 4.5 & 307 \\
6.00  & 0.00507 & 0.0507  & 0.628 & $24^3\times 64$  & 1020  & 0.12085(28) & 2.89 & 3.2 & 221 \\
6.00  & 0.00507 & 0.0507  & 0.628 & $32^3\times 64$  & 1000  & 0.12307(16) & 3.93 & 4.3 & 216 \\
6.00  & 0.00507 & 0.0507  & 0.628 & $40^3\times 64$  & 1028  & 0.12388(10) & 4.95 & 5.4 & 214 \\
6.00  & 0.01275 & 0.01275$\null^\dagger$     & 0.640 & $24^3\times 64$  & 1020  & 0.11848(26) & 2.84 & 5.0 & 349 \\
6.00  & 0.00507 & 0.0304$\null^\dagger$      & 0.628 & $32^3\times 64$  & 1020  & 0.12014(16) & 3.84 & 4.3 & 219 \\
6.00  & 0.00507 & 0.022815$\null^\dagger$    & 0.628 & $32^3\times 64$  & 1020  & 0.11853(16) & 3.79 & 4.2 & 221 \\
6.00  & 0.00507 & 0.012675$\null^\dagger$    & 0.628 & $32^3\times 64$  & 1020  & 0.11562(14) & 3.70 & 4.2 & 226 \\
6.00  & 0.00507 & 0.00507$\null^\dagger$     & 0.628 & $32^3\times 64$  & 1020  & 0.11311(19) & 3.62 & 4.2 & 230 \\
6.00  & 0.0088725 & 0.022815$\null^\dagger$  & 0.628 & $32^3\times 64$  & 1020  & 0.12083(17) & 3.87 & 5.6 & 286 \\
6.00  & 0.00184 & 0.0507  & 0.628 & $48^3\times 64$  & 999   & 0.12121(10) & 5.82 & 3.9 & 133 \\
\hline
6.30  & 0.0074  & 0.037   & 0.440 & $32^3\times 96$  & 1011  & 0.09242(21) & 2.95 & 4.5 & 301 \\
6.30  & 0.00363 & 0.0363  & 0.430 & $48^3\times 96$  & 1000  & 0.09030(13) & 4.33 & 4.7 & 215 \\ 
6.30  & 0.0012  & 0.0363  & 0.432 & $64^3\times 96$  & 1031  & 0.08779(08) & 5.62 & 3.7 & 130 \\
\hline
6.72  & 0.0048  & 0.024   & 0.286 & $48^3\times 144$ & 1016  & 0.06132(22) & 2.94 & 4.5 & 304 \\
6.72  & 0.0024  & 0.024   & 0.286 & $64^3\times 144$ & 1166  & 0.05937(10) & 3.79 & 4.3 & 224 \\
6.72  & 0.0008  & 0.022   & 0.260 & $96^3\times 192$ & 583*  & 0.05676(06) & 5.44 & 3.7 & 135 \\
\hline
\end{tabular} 
\end{table}

Our extraction of the pseudoscalar decay constants with staggered quarks
follows that used for asqtad quarks \cite{FPI04,milc_rmp} and for
$f_{K^+}$ with the HISQ action \cite{FNAL11,FKPRL}.
The decay constant $f_{PS}$ is given by the matrix element
of $\bar \psi \gamma_5 \psi$ between the vacuum and the pseudoscalar meson.
For staggered fermions, using the pion taste corresponding to the axial
symmetry broken only by quark masses, this becomes the operator
\BNE {\cal O}_P (\vec x, t) = \bar\chi^a(\vec x,t) (-1)^{ x + y + z + t} \chi^a(\vec x,t) \ \ \ ,\ENE
where $a$ is a color index.
The desired matrix element can be obtained from the amplitude of a
correlator using this operator at the source and sink,
\BNE P_{PP}(t) = \frac{1}{V_s}\sum_{\vec y} \langle {\cal O}_P(\vec y,0) {\cal O}_P(\vec 0,t) \rangle
= C_{PP} e^{-M t}
+ {\rm excited\ state\ contributions}\ \ \ , \ENE
where $V_s$ is the spatial volume, $M$ is the pseudoscalar meson mass
 and the sum over $\vec y$ isolates the zero spatial momentum states.
Then the decay constant is given by \cite{JLQCD_FPI,TOOLKIT}
\BNE f_{PS} = (m_A+m_B) \sqrt{\frac{V_s}{4} } \sqrt{\frac{C_{PP}}{M^3}} \ \ \ ,\ENE
where $m_A$ and $m_B$ are valence quark masses and $M$ is the pseudoscalar meson mass.

In our computations, we use a ``random-wall'' source for the quark propagators,
where a randomly oriented unit vector in color space is placed on each spatial
site at the source time.  Then quark and antiquark propagators originating on
different lattice sites are zero when averaged over the sources.  We use three
such source vectors for each source time slice.

We also compute pion correlators using a ``Coulomb-wall'' source, where
the gauge field is fixed to the lattice Coulomb gauge, and then a uniform color
vector source is used at each spatial site.   In practice these vectors are
the ``red'', ``green,'' and ``blue'' color axes.   The Coulomb-wall source
correlators are somewhat less contaminated by excited states than the random
wall source correlators, so by simultaneously fitting the correlators with
common masses we are able to determine the masses better, and hence get a
better determined amplitude for the random-wall source correlator.

Four source time slices are used on each lattice, with the exception of
the $0.06$ fm physical quark-mass ensemble where, because these lattices
are longer in the Euclidean time direction, six source time slices are
used.  The location of the source time slices on successive lattices
is advanced by an amount close to one half of the spacing between sources,
but incommensurate with the lattice time size, so that the source
location cycles among all possible values.

In each lattice ensemble, two-point correlators are computed for
a range of valence-quark masses.  The complete set of valence-quark
masses is given in Table~\ref{tab:valencemasses}.  The lightest valence
mass used is one-tenth the strange quark mass for the coarser ensembles
with heavier sea-quark masses, 1/20 the strange quark mass for the
$a\approx0.06$ fm ensembles with heavier than physical sea-quark
mass, and the physical light-quark  mass for
the ensembles with physical sea-quark mass.  The valence masses chosen
then cover the range from this lightest mass up to the estimated strange-quark mass. 
We then choose additional masses at the estimated charm-quark
mass  (the same as the  charm-quark mass
in the sea), as well as nine-tenths of that value, so that we can make adjustments for mistuning of the
charm-quark mass.  For these last two quarks, the coefficient of the
three-link term in the fermion action (the ``Naik term'') is 
adjusted to improve the quark's dispersion relation \cite{hpqcd_hisqprd}.
Specifically, the expansion resulting from combining Eqs. (24) and (26) of Ref.~\cite{hpqcd_hisqprd}
is used; the improvement has been checked in HISQ simulations \cite{hpqcd_hisqprd,scaling09}.

\begin{table}
\caption{
\label{tab:valencemasses}
Valence-quark masses used in this project.  Correlators with random
wall and Coulomb-wall sources are computed for each possible pair
of valence-quark masses.   Light valence masses $m_\text{v}$ are given in units of
the (ensemble value of the) sea strange quark mass $m'_s$.
Note that for the four ensembles with near-physical sea-quark mass, the lightest
valence mass is the same as the light sea mass.
 The two heavy valence masses are in
units of the charm sea-quark mass $m'_c$.  For the ensembles with unphysical strange
quark mass (included in ``All'' at $\beta=6.0$), the valence masses are given in units of
the approximate physical strange quark mass, $0.0507$.
}
\begin{tabular}{|l|lll|r|r|}
\hline
$\beta$ & \multicolumn{3}{c|}{sea quark masses} & light valence masses & charm valence masses     \\
        & $am'_l$ & $am'_s$ & $am'_c$ & $m_\text{v}/m'_s$ & $m_\text{v}/m'_c$ \\
\hline
5.80  & 0.013   & 0.065   & 0.838 & 0.1,0.15,0.2,0.3,0.4,0.6,0.8,1.0 & 0.9,1.0 \\
5.80  & 0.0064  & 0.064   & 0.828 & 0.1,0.15,0.2,0.3,0.4,0.6,0.8,1.0 & 0.9,1.0 \\
5.80  & 0.00235 & 0.0647  & 0.831 & 0.036,0.07,0.1,0.15,0.2,0.3,0.4,0.6,0.8,1.0 & 0.9,1.0 \\
\hline
6.00  & 0.0102  & All  & 0.635 & 0.1,0.15,0.2,0.3,0.4,0.6,0.8,1.0 & 0.9,1.0 \\
6.00  & 0.00507 & All  & 0.628 & 0.1,0.15,0.2,0.3,0.4,0.6,0.8,1.0 & 0.9,1.0 \\
6.00  & 0.00184 & 0.0507  & 0.628 & 0.036,0.073,0.1,0.15,0.2,0.3,0.4,0.6,0.8,1.0 & 0.9,1.0 \\
\hline
6.30  & 0.0074  & 0.037   & 0.440 & 0.1,0.15,0.2,0.3,0.4,0.6,0.8,1.0 & 0.9,1.0 \\
6.30  & 0.00363 & 0.0363  & 0.430 & 0.1,0.15,0.2,0.3,0.4,0.6,0.8,1.0 & 0.9,1.0 \\
6.30  & 0.0012  & 0.0363  & 0.432 & 0.033,0.066,0.1,0.15,0.2,0.3,0.4,0.6,0.8,1.0 & 0.9,1.0 \\
\hline
6.72  & 0.0048  & 0.024   & 0.286 & 0.05,0.1,0.15,0.2,0.3,0.4,0.6,0.8,1.0 & 0.9,1.0 \\
6.72  & 0.0024  & 0.024   & 0.286 & 0.05,0.1,0.15,0.2,0.3,0.4,0.6,0.8,1.0 & 0.9,1.0 \\
6.72  & 0.0008  & 0.022   & 0.260 & 0.036,0.068,0.1,0.15,0.2,0.3,0.4,0.6,0.8,1.0 & 0.9,1.0 \\
\hline

\end{tabular}
\end{table}

\section{Two-point correlator fits}
\label{sec:twopointfits}

To find the pseudoscalar masses and decay amplitudes, the random-wall and
Coulomb-wall correlators are fitted to common masses but independent amplitudes.
With staggered quarks the Goldstone-taste pseudoscalar correlators with
unequal quark masses contain contributions from opposite-parity states, which
show up as exponentials multiplied by an alternating sign, $(-1)^t$.  For
valence-quark masses up to and including the strange quark mass these
contributions are small, and good fits can be obtained while neglecting them.
In fact, in our previous analyses with the asqtad quark action, these states
were not included in the two-point fits.
However, with these data sets, slightly better fits are obtained when an opposite-parity state
is included in the light-light fits, and so we include such a state in the
unequal quark mass correlators.

The light-charm correlators (where ``light'' here includes masses up to the physical strange quark mass $m_s$)
are more difficult to fit than the light-light correlators for several reasons.
First, because the difference in the valence-quark mass is large,
the amplitude of the opposite-parity states is not small.  Second, the mass
splitting between the ground state and the lowest excited single particle
state is smaller.  For the light-light correlators, the approximate chiral
symmetry makes the ground state mass smaller than typical hadronic scales, which
has the side effect of making the mass gap to the excited single particle
states large, and these excited states can be suppressed by simply taking a large
enough minimum distance.  For the charm-light correlators we include an
excited state in the fit function.  (In principle, multiparticle states also
appear in these correlators.  For example, the lowest excited state in the
pion correlator would be a three-pion state.  Empirically these states do not
enter with large amplitudes, and the important excited states correspond more
closely to single particle states.)

To make the fits converge reliably, it is necessary to loosely constrain
the masses of the opposite-parity and excited states
by Gaussian priors.  The central value of the gap between the ground state and opposite
parity states is taken to be 400 MeV, motivated by the 450 MeV gap between
the $D$ mass and the $0^+$ light-charm mass, and the 350 MeV gap between the
$D_s$ mass and a poorly established $0^+$ strange-charm meson~\cite{PDG}.
The central value for the gap between the ground state and excited state
masses is taken to be 700 MeV, motivated by the 660 MeV gap between the
$\eta_c$ and the corresponding 2S state.
In most cases the widths of the priors for the opposite-parity and excited
state gaps are taken to be 200 MeV and 140 MeV respectively, although in
some cases these need to be adjusted to get all of the jackknife fits to
converge.

Another factor that makes the light-charm correlators more difficult to fit is
the faster growth of the statistical error.  The time dependence of the
variance of a correlator is expected to depend on time as $e^{-E_2 t}$, where
$E_2$ is the energy of the lowest lying state created by ${\cal O}{\cal O^\dagger}$,
where ${\cal O}$ is the source operator for the correlator itself, with the
proviso that quark and antiquark lines all go from source to sink, rather than
coming back to the source~\cite{LEPAGE_TASI}.
For the pion correlator, the state created by ${\cal O}{\cal O^\dagger}$ is just
the two pion state, leading to the expectation that the fractional statistical
error on the pion correlator is roughly independent of distance.  However, for
the light-charm correlator, the quarks and antiquarks created by ${\cal O}{\cal O^\dagger}$
can pair up to form an $\eta_c$ and a pion.  Then, the reduction of the pion's mass
from chiral symmetry makes this state much lighter than $2M_D$, so the
fractional error of the propagator grows rapidly with distance.  This makes it
essential to use smaller minimum distances in the fit range for the
light-charm correlators, which of course makes the problem of excited states
discussed in the previous paragraph even more serious.

\begin{figure}
\vspace{-1.0in}
\centerline{\includegraphics[width=0.9\textwidth]{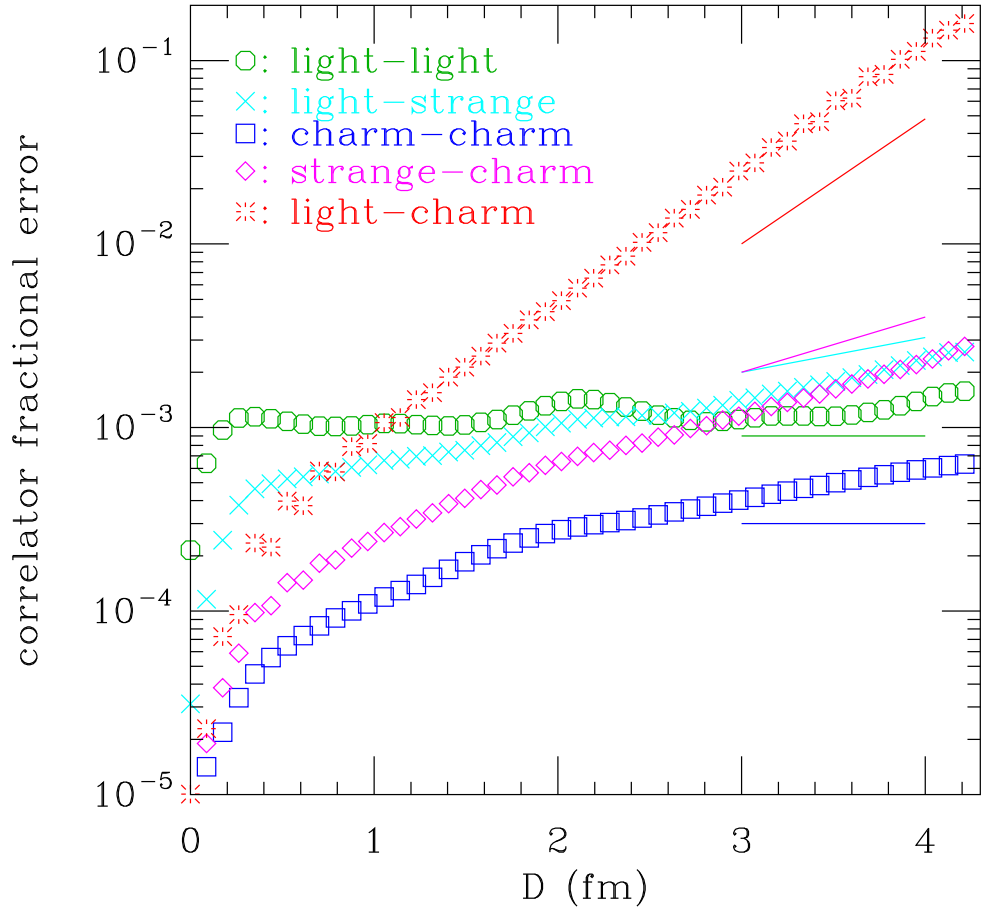}}
\vspace{-1.0in}
\caption{ \label{fig:fracerr_m0012}
Fractional errors for pseudoscalar correlators as a function of distance from the 0.09~fm physical quark-mass ensemble.  The line segments
show the slope expected from the states in Table~\protect\ref{tab:errorstates},
which give a good approximation to the observed growth of the errors with
the exception of the charm-charm correlator.
}
\end{figure}

Table~\ref{tab:errorstates} shows our expectations for the
states controlling the growth of statistical errors for the various
pseudoscalar correlators. Figure~\ref{fig:fracerr_m0012} shows the fractional
errors for the random-wall correlators for the $0.09$ fm physical quark-mass
ensemble, with comparison to the slopes expected from Table~\ref{tab:errorstates}.
With the exception of the charm-charm correlator, the behavior of the statistical
error agrees with our theoretical expectations.

\begin{figure}
\vspace{-1.0in}
\centerline{\includegraphics[width=0.9\textwidth]{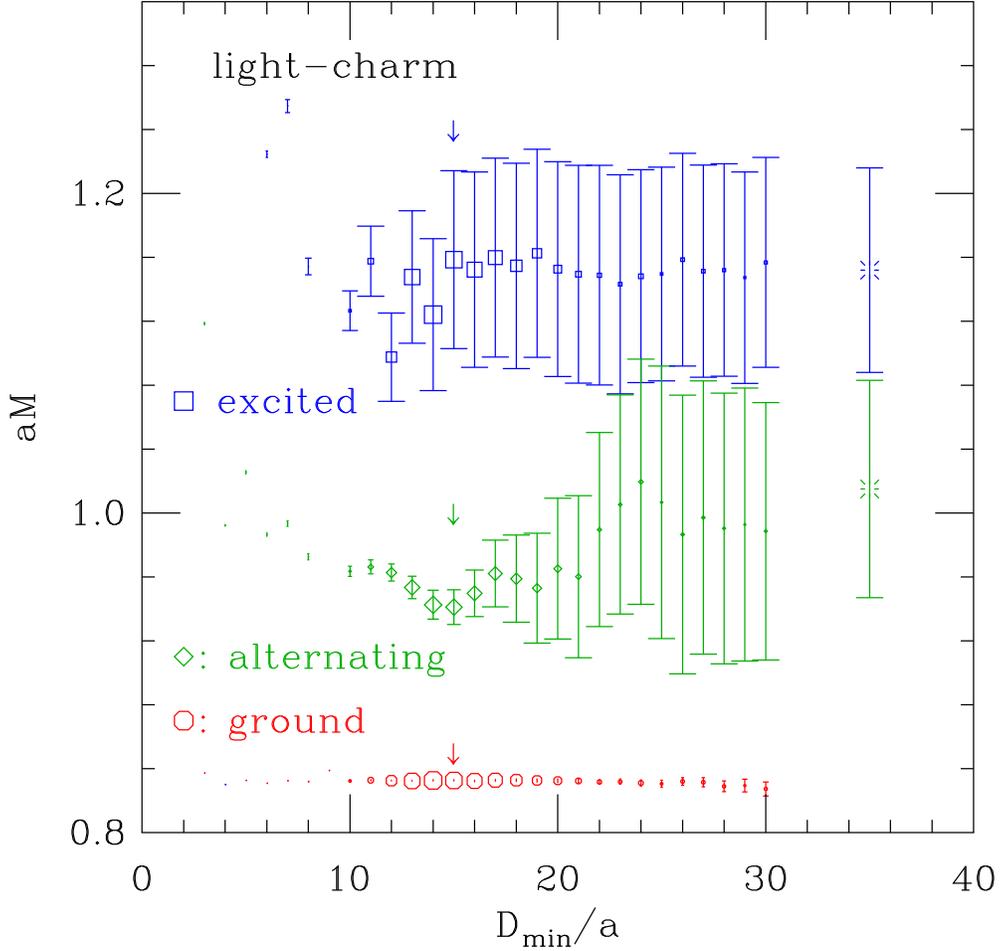}}
\vspace{-1.0in}
\caption{ \label{fig:fitchoice_lc}
Fits for the light-charm pseudoscalar correlator (mass $M$) in the ensemble with $a\approx 0.09$ fm and physical 
sea-quark masses.   We plot the ground state, alternating state (opposite parity) and excited state
masses as a function of minimum distance included in the fit.  The size of the symbols is 
proportional to the $p$~value of the fit, with the size of the symbols in the legend corresponding
to 0.5.  The two bursts on the right show the priors and their errors for the alternating
and excited masses.  The vertical arrows at $D\sb{min}=15$ indicate the fit that is chosen.
Further discussion is in the text. Here the masses and distance are in units of the
lattice spacing.
}
\end{figure}

\begin{figure}
\vspace{-1.0in}
\centerline{\includegraphics[width=0.9\textwidth]{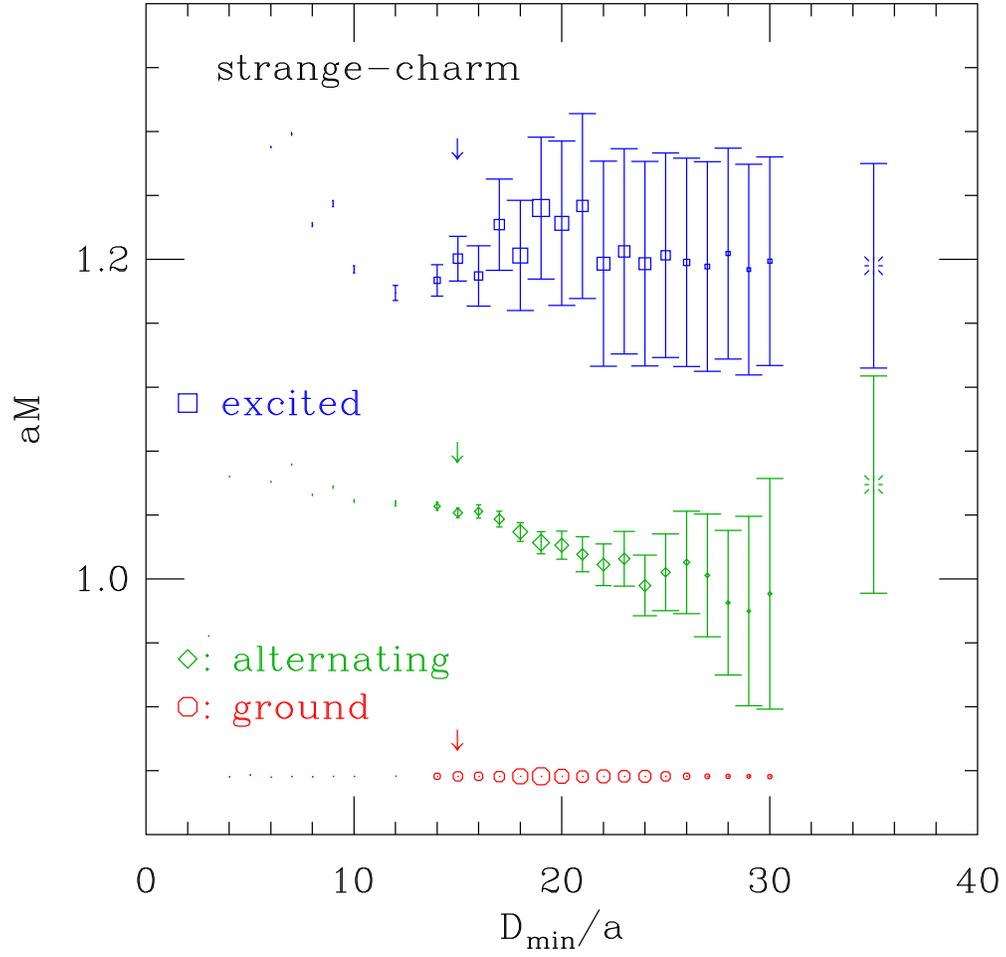}}
\vspace{-1.0in}
\caption{ \label{fig:fitchoice_sc}
Fits for the strange-charm correlator in the ensemble with $a\approx 0.09$ fm and physical sea-quark 
masses.   The format and symbols are the same as in Fig.~\protect\ref{fig:fitchoice_lc}.
}
\end{figure}

Figures~\ref{fig:fitchoice_lc} and \ref{fig:fitchoice_sc} show the masses in
the 2+1 state fits for the light-charm correlators in the
$a\approx0.09$ fm physical quark-mass ensemble as a function of the minimum distance included
in the fit, where the light-quark  mass is the
physical $(m_u+m_d)/2$ (\figref{fitchoice_lc}) and $m_s$ (\figref{fitchoice_sc}).
Fit ranges are chosen from graphs like
this for all the ensembles, and analogous graphs for the light-light and charm-charm
correlators.  We show this ensemble because it, together  with the $a\approx0.06$ fm
physical mass ensemble, is the most important to the final results.  In these
graphs the error bars on the right show the central values and widths of the
priors used for the opposite-parity and excited masses.  At short distances,
these masses are more accurately determined by the data, while at larger
$D_{min}$ the input prior controls the mass.  The linear sizes of the symbols in
these figures are proportional to the $p$~value of the fit, with the size of the
symbols in the legend corresponding to 50\%.
In the two-point correlator fits used to choose the fit types and ranges, as in
\figrefs{fitchoice_lc}{fitchoice_sc}, autocorrelations among the
lattices are minimized by first blocking the data in blocks of four lattices,
or 10 to 24 molecular dynamics time units.
However, statistical errors on results in
later sections are obtained from the jackknife procedures described in
\secrefs{physical-mass-analysis}{chiral-analysis}.  In these
analyses the two-point fits are repeated in each jackknife resampling. From 
these and similar graphs for other ensembles and different numbers of
excited states, 
keeping the minimum distance in physical units reasonably constant,
the minimum distances and fit forms in Table~\ref{tab:twopointfitforms} are chosen.
The need for using a smaller minimum distance and including an
   excited state in the heavy-light fits is consistent with our expectations
   from Table~\ref{tab:errorstates} and Fig.~\ref{fig:fracerr_m0012}.
Because the statistical errors increase with distance from the source, the fits are
much less sensitive to the choice of maximum distance.  In most cases the maximum
distance is taken to be one less than the midpoint of the lattice.  However, in
the $a\approx0.09$ and $0.06$ fm ensembles, the light-charm and charm-charm fits used a
smaller maximum distance because having fewer points in the fit gave a better
conditioned covariance matrix.  These maximum distances are also included
in Table~\ref{tab:twopointfitforms}.

\begin{table}
\caption{ \label{tab:errorstates}
States expected to control the
statistical errors on the correlators, for the pseudoscalars with physical
valence-quark masses.
The second column shows the state expected to control the growth of the
statistical error on the correlator, the third column the mass gap between
half the mass of the error state and the particle mass, and the fourth column the length
scale for the growth of the fractional statistical error.
Here $\bar s s$ is the unphysical flavor nonsinglet state, with mass
680 MeV.
}
\begin{tabular}{|llll|}
\hline
State\ \           & Error         	& $\mathrm{{}^{Energy}_{gap\;(MeV)}}$	& $\mathrm{{}^{Growth}_{length\; (fm)}}$ \\
\hline
$\pi$           & $2\pi$        	& 0 		& $\infty$  \\
$K$             & $\pi+\bar s s\ \ \ $        & 90 	& 2.26  \\
$\eta_c$        & $2\eta_c$     	& 0 		& $\infty$  \\
$D_s$           & $\eta_c+\bar s s$     & 140 	& 1.42  \\
$D$             & $\eta_c+\pi$  	& 310 	& 0.64  \\
	\hline
\end{tabular}
\end{table}

\begin{table}
\caption{ \label{tab:twopointfitforms}
Fit forms and minimum distance included for the two-point correlator
fits.  Here the fit form is the number of negative parity
({\it i.e.}, pseudoscalar) states ``plus'' the number of positive parity
states.  When the valence quarks have equal masses, the
opposite-parity states are not included.
In this work the charm-charm fits are needed only for computing the
mass of the $\eta_c$ meson, used as a check on the quality of our
charm physics.
}
\begin{tabular}{|l|lll|lll|lll|}
\hline
		& \multicolumn{3}{c|}{light-light} & \multicolumn{3}{c|}{light-charm} &
\multicolumn{3}{c|}{charm-charm} \\
\hline
		& form & $D_{min}$ & $D_{max}$ & form & $D_{min}$ & $D_{max}$ & form & $D_{min}$ & $D_{max}$ \\
\hline
$a\approx 0.15$ fm & 1+1 & 16 & 23 & 2+1 & 8 & 23 & 2+0 & 9 & 23 \\
$a\approx 0.12$ fm & 1+1 & 20 & 31 & 2+1 & 10 & 31 & 2+0 & 12 & 23 \\
$a\approx 0.09$ fm & 1+1 & 30 & 47 & 2+1 & 15 & 37 & 2+0 & 18 & 35 \\
$a\approx 0.06$ fm & 1+1 & 40 & 71 & 2+1 & 20 & 51 & 2+0 & 21 & 50 \\
\hline
\end{tabular}
\end{table}

\clearpage
\section{Determination of decay constants and quark-mass ratios}
\label{sec:analysis}
This section describes the details of the analyses that produce our results for light-light and
heavy-light decay constants and the ratios of quark masses.  We perform two versions of the analysis. The
first, the ``physical-mass analysis'' described in \secref{physical-mass-analysis}, is a straightforward procedure that essentially uses
only the physical-quark mass ensembles.  On these ensembles, a chiral extrapolation is not needed: only
interpolations are required in order to find the physical quark-mass point.  The physical-mass analysis
produces our results for quark-mass ratios and $f_{K^+}/f_{\pi^+}$, as well as some additional
intermediate quantities required for the chiral analysis of the $D$ meson decay constants, which follows.
The second analysis of charm decay constants, described in
\secref{chiral-analysis},  uses chiral perturbation theory to perform a combined fit to all of our
physical-mass and unphysical-mass data, and to thereby significantly reduce the statistical uncertainties
of the results.  We take the more precise values of $f_{D^+}$, $f_{D_s}$, and their ratio from the chiral
analysis as our final results, and use those from the simpler physical-mass analysis only as a
consistency check, and to aid in the estimation of systematic errors.
 
In the physical-mass analysis of \secref{physical-mass-analysis}, we  first determine the lattice spacing
and quark masses separately for each ensemble, using, in essence, the five experimental values of
$f_{\pi^+}$,  $M_{\pi^0}$, $M_{K^0}$, $M_{K^+}$ and $M_{D_s}$, as explained in \secref{valencemassfits}.
In order to adjust for mistuning of the sea-quark masses, we perform a parallel scale-setting and
quark-mass determination on the unphysical-mass ensembles; there, however, an extrapolation in the
valence-quark mass is generally required.  We extrapolate the quark-mass ratios to the continuum, after
small sea-quark mistuning adjustments, in \secref{seacontinuumfits}.  We follow the same procedure on the
physical-mass ensembles to also obtain values for decay constants.  In particular, we update our result
for $f_{K^+}/f_{\pi^+}$ from \rcite{FKPRL}.  Although the results for charm decay constants from the
physical-mass analysis are not taken as our final values, they are used as additional inputs in the
estimation of systematic errors from the continuum extrapolation.  Finally, the physical-mass analysis
allows us to make straightforward estimates of systematic errors coming from finite-volume and
electromagnetic (EM) effects on the decay constants and quark-mass ratios, as described in
\secref{FV-EM}.

The values of the physical quark-mass ratios $m_c/m_s$, $m_s/m_l$, and (to a lesser extent, in order to
take into account isospin-violating effects) $m_u/m_d$ obtained in \secref{physical-mass-analysis} are
used in the subsequent chiral analysis in \secref{chiral-analysis}.  Further, in the physical-mass
analysis, we determine the useful quantity $F_{p4s}$ \cite{HISQ_CONFIGS}, which is the light-light
pseudoscalar decay constant $F$ evaluated at a fiducial point with both valence masses equal to
$m_{p4s}\equiv0.4m_s$ and physical sea-quark masses.  The meson mass at the same fiducial point,
$M_{p4s}$, as well as the ratio $R_{p4s}\equiv F_{p4s}/M_{p4s}$, are similarly determined. The unphysical
decay constant $F_{p4s}$ provides an extremely precise and convenient quantity  to set the relative scale
in the chiral analysis (see \secref{LAT-SCALE}), while we use $R_{p4s}$ to tune the strange sea-quark
mass.

The chiral analysis of the decay constants of charm mesons  is described in detail 
in \secref{chiral-analysis}.  With chiral perturbation theory, one can 
take advantage of all our data by including both the physical-mass and unphysical-mass ensembles
in a unified procedure.
In particular, the statistical error in $\Phi_{D^+} $ is slightly more than a factor of
two smaller with the chiral analysis than in the physical-mass analysis of \secref{physical-mass-analysis}. 
In addition, the use of the relevant form of staggered chiral perturbation theory for this case, heavy-meson, 
rooted, all-staggered chiral perturbation theory (\aschpt) 
\cite{Bernard-Komijani}, allows us to relate the quark-mass and lattice-spacing dependence of the data, and thereby
use the unphysical-mass ensembles to tighten the control of the continuum extrapolation.  Our final central values for the charm decay constants given in the conclusions are taken from the chiral analysis.  We
increase some of the systematic uncertainties, however, 
to take into account differences with the results of the physical-mass analysis.

\subsection{Simple analysis  from physical quark-mass ensembles}
\label{sec:physical-mass-analysis}

Here we determine the quark-mass ratios and decay constants employing primarily the physical quark-mass 
ensembles.  First, in \secref{valencemassfits}, we determine the lattice spacing, quark masses, and decay
constants separately for each ensemble.  Next, in \secref{seacontinuumfits}, we adjust the quark masses and 
decay constants for slight sea-quark mass mistuning, and extrapolate to the continuum.  Finally, we estimate 
the systematic uncertainties in the quark-mass ratios and decay constants in \secref{FV-EM}.  We present 
results and error budgets for these quantities obtained from the physical mass analysis in Table~
\ref{tab:error_budget_1}.

\subsubsection{Valence-quark mass interpolation}
\label{sec:valencemassfits}

\vspace{2.5mm}

In this stage of the analysis we determine tuned quark masses and the
lattice spacing (using $f_{\pi^+}$ to fix the scale) for each ensemble, and then find the decay constants by
interpolation or extrapolation in valence-quark mass to these corrected
quark masses.  There are a number of possible choices for the procedure
used, and we include the differences among a few sets of choices in
our systematic error estimate.  It is important to remember that there is
inherent ambiguity in defining a lattice spacing for ensembles with
unphysical sea-quark masses, but all sensible
choices should have the same limit at zero lattice spacing and physical
sea-quark masses.  For example, in the ensemble-by-ensemble fitting procedure
described in this section, we take the value of $f_{\pi^+}$ on each ensemble to be
$130.41$ MeV, independent of sea-quark masses, while for the chiral
perturbation theory analysis we take the lattice spacing to be independent
of the sea-quark masses.
\vspace{1.5mm}

\begin{figure}[p]
\vspace{-1.0in}
\centerline{\includegraphics[width=1.00\textwidth]{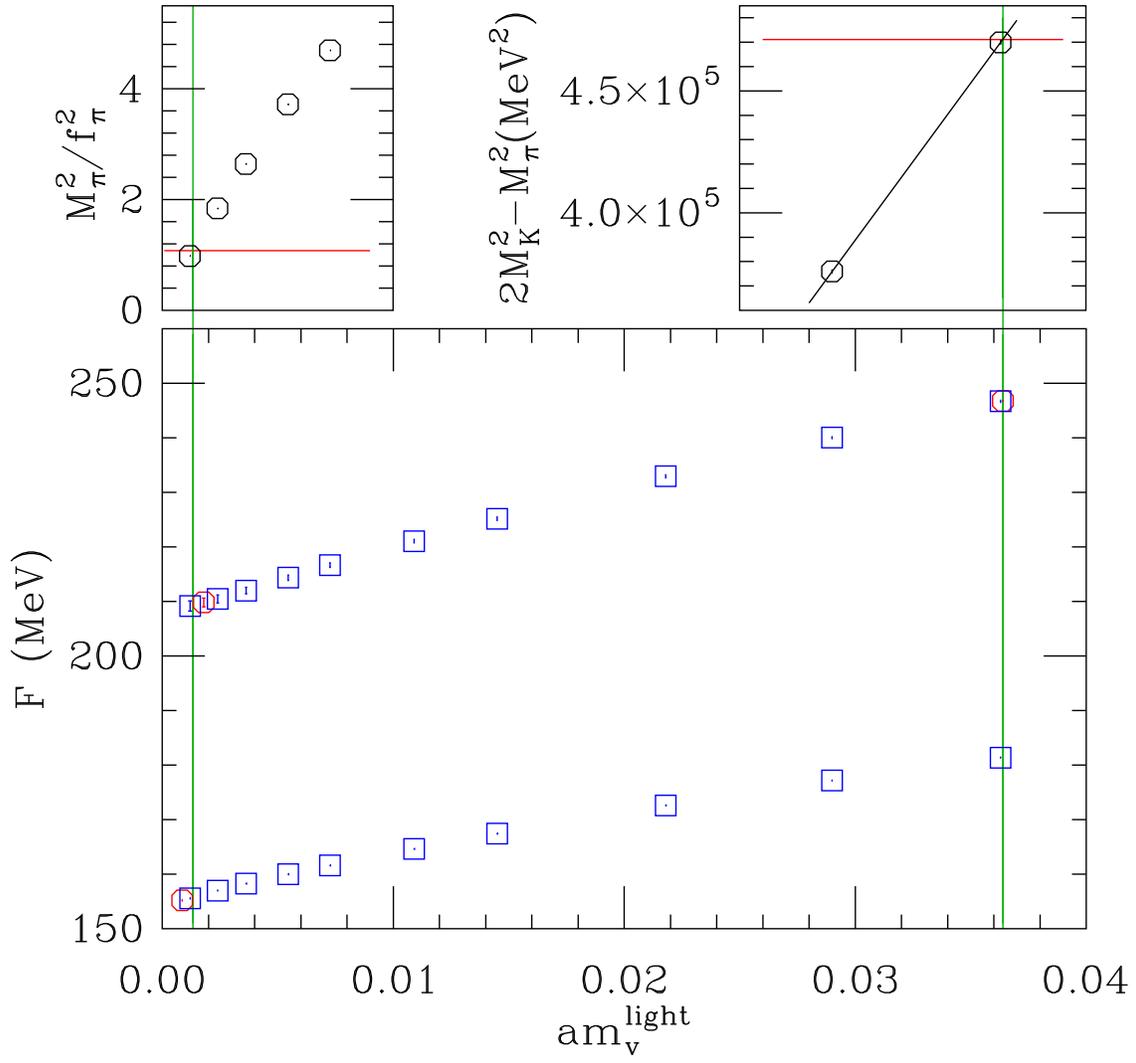}}
\vspace{-1.0in}
\caption{ \label{fig:tuning_m0012}
Illustration of the ``$f_\pi$'' tuning for the $a\approx 0.09$ fm physical quark mass
ensemble.  $F$ is the decay constant of a generic pseudoscalar meson. The procedure illustrated is described in the text.
}
\end{figure}

Figure~\ref{fig:tuning_m0012} illustrates some of the features of our
procedure, and referring to it may help clarify the following description.
Since the decay amplitude $F$ depends on valence-quark mass, and
we wish to use $f_{\pi^+}=130.41$ MeV to set the lattice scale, we must determine
the lattice spacing and tuned light-quark  mass simultaneously.
To do so, we find the light valence-quark mass where the mass and amplitude
of the pseudoscalar meson with degenerate valence quarks have the physical
ratio of $M_\pi^2/f_{\pi^+}^2$.  
(Actually we adjust this ratio for finite size effects, using the
pion mass and decay constant in a 5.5 fm box.  This correction is discussed 
 in  \secref{FV-EM}.)
This light-quark  mass is the average of the up and down quark masses, $m_l=(m_u+m_d)/2$.
Here we use the mass of the $\pi^0$, since it
is less affected by electromagnetic corrections than the $\pi^+$.   Since
the $\pi^+$ contains one up and one down quark, the error in $f_{\pi^+}$ from using
degenerate light valence quarks is negligible.  
This tuning is illustrated in the upper left panel of \figref{tuning_m0012}, which
shows this ratio as a function of light valence mass for the $0.09$ fm physical
quark-mass ensemble, one of the two ensembles that are most important in our analysis.
The octagons in this panel are the ratio at the valence-quark masses where we
calculated correlators, with error bars that are too small to be visible.
The horizontal red line is the desired value of this ratio, and the green vertical
line shows the light-quark  mass where the ratio has its desired value.  With the
tuned light-quark  mass determined, we use the decay amplitude at this mass, $f_{\pi^+}$,
to fix the lattice spacing.
In performing the interpolation or extrapolation of $M_\pi^2/f_\pi^2$ we use points with degenerate light valence-quark 
mass $m_\text{v}$ and employ
a continuum, partially quenched,  SU(2) \chpt\ form \cite{Sharpe:1997by,Aubin:2003mg},
\BEA \label{eq:lightchipt}
\frac{M_\pi^2}{f_\pi^2} &=& \frac{B2m_\text{v}}{f^2} \left\{ 1+\frac{1}{16\pi^2 f^2}\left[
B(4m_\text{v}-2m_l^\prime)\log(2Bm_\text{v}/\Lambda_\chi^2) \right.\right. \EL
   &&\hspace{7mm}   \left. \phantom{\frac{1}{16\pi^2 f^2}}\left.+4B(m_\text{v}+m_l^\prime)\log(B(m_\text{v}+m_l^\prime)/ \Lambda_\chi^2) \right] + Cm_\text{v} \right\} \EL
f_\pi &=& f \LB 1 - \frac{2B(m_\text{v}+m_l^\prime)}{16 \pi^2 f^2} \log(B(m_\text{v}+m_l^\prime)/\Lambda_\chi^2) + Cm_\text{v} + D m_\text{v}^2 \RB \  , \\
\EEA
where $m_l^\prime$ is the light sea-quark mass and $\Lambda_\chi$ is the chiral scale.   In applying \eq{lightchipt}, we fix the low energy constants
$B$ and $f$ in the coefficients of the logarithms to values determined from 
lowest order \chpt\ using the smallest valence-quark mass.  
We then fix the coefficients of $m_\text{v}$ and $m_\text{v}^2$ in $M_\pi^2/f_\pi^2$
using the smallest two valence-quark masses available, and we fix the analytic coefficients
in $f_\pi$ using the three smallest valence-quark masses.
In the physical quark-mass ensembles, such as
the one shown in \figref{tuning_m0012}, this is only a small correction to
the quark mass.  On the other hand, in most of the ensembles 
with $m'_l/m'_s=0.1$ or $0.2$, 
the lightest valence-quark mass is $0.05m'_s$ or $0.1m'_s$, and a significant extrapolation
is made.
However, these unphysical-mass
ensembles are used only in the analysis of this section to correct the results of the physical-mass ensembles for small mistunings of the 
sea masses in the physical-mass ensembles.

We then fix the tuned strange quark mass to the mass that gives the
correct $2M_K^2-M_\pi^2$.  This is illustrated in the upper right panel
of \figref{tuning_m0012}.  In all of our ensembles, we use valence
``strange'' quark masses at the expected strange quark mass and at $0.8$ times
this mass. The two data points shown in the figure have these strange masses and
the lightest available light-quark  valence mass. A linear
interpolation or extrapolation is performed through these two points.  Again, the horizontal
red line shows the desired value of this mass difference, and the vertical
green line the resulting value of $m_s$.
In this stage of the tuning the kaon mass is corrected for finite volume effects,
electromagnetic effects and isospin breaking effects, where again we defer
the details to the discussion of systematic errors in \secref{FV-EM}.

Next we determine the up-down quark mass difference, and hence the up
and down quark masses.
We use the difference in $K^0$ and $K^+$ masses,
\begin{equation} m_d-m_u = \frac{ M_{K_\text{adj}^0}^2 - M_{K_\text{adj}^+}^2  }
                   { \PAR{M_K^2}{m_l} }         \ \ \ \ .
\end{equation}
Here the kaon masses are adjusted for finite volume and electromagnetic
effects, and again we defer the details to \secref{FV-EM}.  We note that the
electromagnetic corrections are a small effect on the strange quark mass
tuning, but are absolutely crucial in the determination of $m_d-m_u$.
To estimate the derivative $ \partial M_K^2/\partial {m_l}$, we use the masses of kaons
containing a valence quark near the strange quark mass and a second valence
quark that is one of the two lightest valence quarks we have.

Then the tuned charm quark mass is determined from the experimental
value of $M_{D_s}$.  We use $M_{D_s}$ rather than $M_D$ because it has
much smaller statistical errors.  In all of our ensembles we have correlators with
valence-quark masses at the expected charm quark mass and at $0.9$ times
this mass.  Using linear interpolations in $m_s$ of the $D_s$ meson mass at these
two ``charm'' masses to the strange quark mass found earlier, and a linear
interpolation in $m_c$ between these, we find a tuned charm quark mass.

Now that we have found the lattice spacing and tuned quark masses, we
can find decay constants and masses of other mesons by interpolating
or extrapolating to these quark masses.  The bottom panel of 
\figref{tuning_m0012} illustrates this process.  The lower set
of points in this graph are the decay constants at each light
valence mass, interpolated using the two ``strange'' valence masses
to the tuned strange quark mass.  Then $f_{K^+}$ is found by extrapolating
these points to the tuned $m_u$, illustrated by the red octagon
at the lower left.  Similarly, the upper set of data points is the
decay constant at each light-quark  mass, linearly interpolated or
extrapolated using the two ``charm'' valence masses to the tuned $m_c$.
This graph is then interpolated or extrapolated to the tuned $m_d$ to
find $f_{D^+}$, shown in the red octagon at the upper left, or to
the tuned $m_s$ to find $f_{D_s}$, shown by the red octagon at the
upper right.

As checks on our procedure, we also similarly interpolate or extrapolate in
the meson masses to find $M_{D^0}$, $M_{D^+}$ and $M_{\eta_c}$.

\subsubsection{Sea-quark mass adjustment and continuum extrapolation}
\label{sec:seacontinuumfits}

In this stage we combine the results from the
individual ensembles and fit to a function of the lattice spacing to
find the continuum limit.  We  use the ensembles with unphysical
sea-quark masses to make small adjustments for the fact that the
sea-quark masses in the physical quark-mass ensembles were fixed
after short tuning runs, and inevitably turned out to be slightly
mistuned when the full runs are done.  
The amount of mistuning is shown in Table~\ref{tab:tunedmasses},
which gives the sea-quark masses and
the tuned quark masses for the physical quark-mass ensembles.

\begin{table}
\caption{\label{tab:tunedmasses} Tuned lattice spacings (using $f_{\pi^+}$ to set the scale) and quark masses for the
physical quark-mass ensembles. The quark mass entries show the
light, strange and charm quark masses in units of the lattice
spacing. The column labeled $am'$ gives the run values of the sea quark masses.
}
\begin{tabular}{|ll|l|l|}
\hline
$a_{approx}$(fm) & $a_{tuned}$(fm) & $am' $ & $am_{tuned}$ \\
\hline
0.15 & 0.15089(17) & 0.00235/0.0647/0.831 & 0.002426(8)/0.06730(16)/0.8447(15) \\
0.12 & 0.12121(10) & 0.00184/0.0507/0.628 & 0.001907(5)/0.05252(10)/0.6382(8)  \\
0.09 & 0.08779(8)  & 0.0012/0.0363/0.432  & 0.001326(4)/0.03636(9)/0.4313(6) \\
0.06 & 0.05676(6)  & 0.0008/0.0220/0.260  & 0.000799(3)/0.02186(6)/0.2579(4) \\
\hline
\end{tabular} \end{table}

\begin{figure}
\centerline{\includegraphics[width=0.9\textwidth]{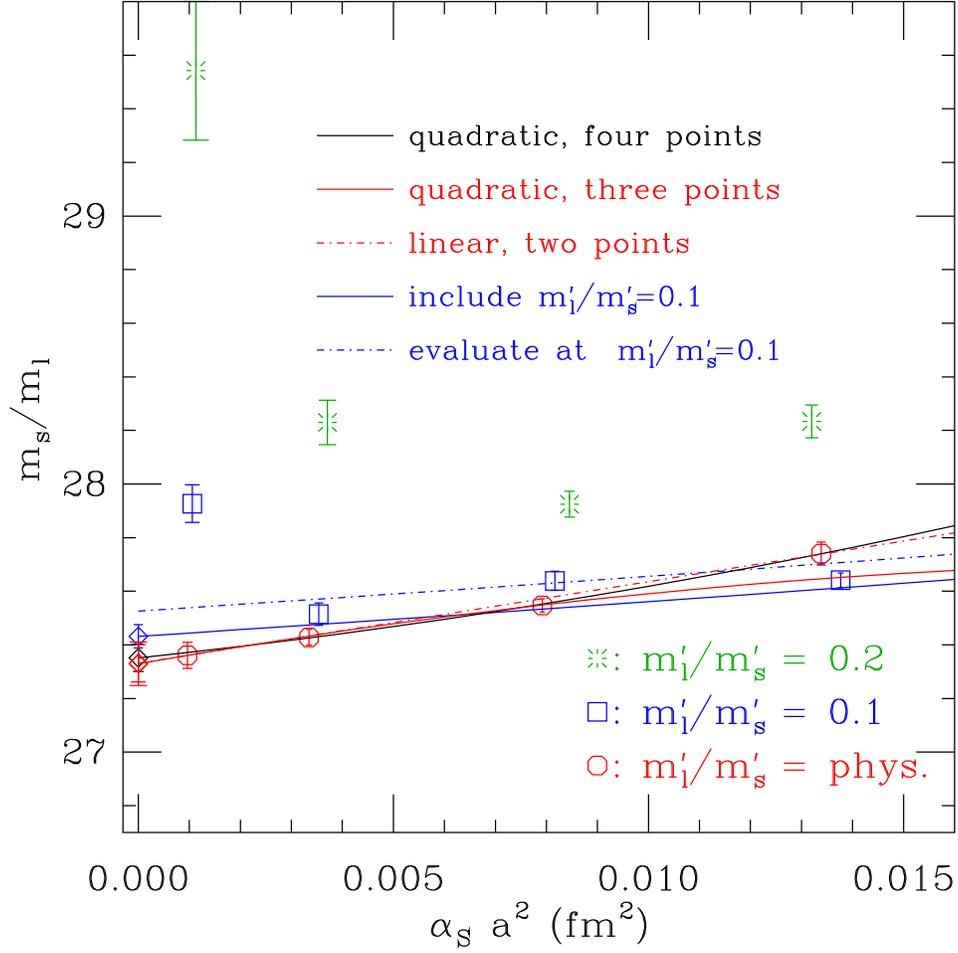}}
\caption{\label{fig:slratio}
The tuned ratio of strange quark mass to light-quark  mass, $m_s/m_l$, on each ensemble,
for the physical quark-mass ensembles (red octagons), for $m'_l/m'_s =0.1$
(blue squares) and for $m'_l/m'_s=0.2$ (green bursts).
The fits shown in this and subsequent figures are described in the text.
The diamonds at the left indicate the continuum extrapolations of the various fits.
}
\end{figure}

Fitting to the lattice spacing dependence is straightforward,
because the results from each ensemble are statistically independent.
We have performed continuum extrapolations for the ratios of quark
masses, $m_u/m_d$, $m_s/m_l$, and $m_c/m_s$, which come automatically
from the fitting for each ensemble described in \secref{valencemassfits}.
\Figrefthree{slratio}{csratio}{udratio}
show the results for each ensemble, together
with fits to the lattice spacing dependence.
In these plots the abscissa is $a^2 \alpha_S$, where $\alpha_S$ is an
effective coupling constant determined from taste violations in the pion masses.
The relative value of $\alpha_S$ at a given coupling $\beta$, compared to its value at a fixed, fiducial coupling $\beta_0$,
is given by
\begin{equation}\eqn{alphas-taste}
\frac{\alpha_S(\beta)}{\alpha_S(\beta_0)}  =  \sqrt{\frac{(a^2\bar \Delta)_\beta\; a^2(\beta_0)}
{(a^2\bar \Delta)_{\beta_0}\; a^2(\beta)}}\;,
\end{equation}
where $(a^2\bar \Delta)_\beta$ is the mean squared taste splitting at coupling $\beta$,
and $a(\beta)$ is the lattice spacing given below in \tabref{results_a}. 
\Equation{alphas-taste} assumes that $a^2\bar\Delta$  is proportional to 
$\alpha^2_S a^2$, its leading behavior.  
We use $\beta_0=5.8$
in these plots, and scale $\alpha_S$ to agree with the coupling $\alpha_V$ at $\beta_0=5.8$,
which in turn may be determined from the plaquette \cite{Davies:2002mv} as explained after Eq.~(9) of 
\rcite{HISQ_CONFIGS}.

\begin{figure}
\centerline{\includegraphics[width=0.9\textwidth]{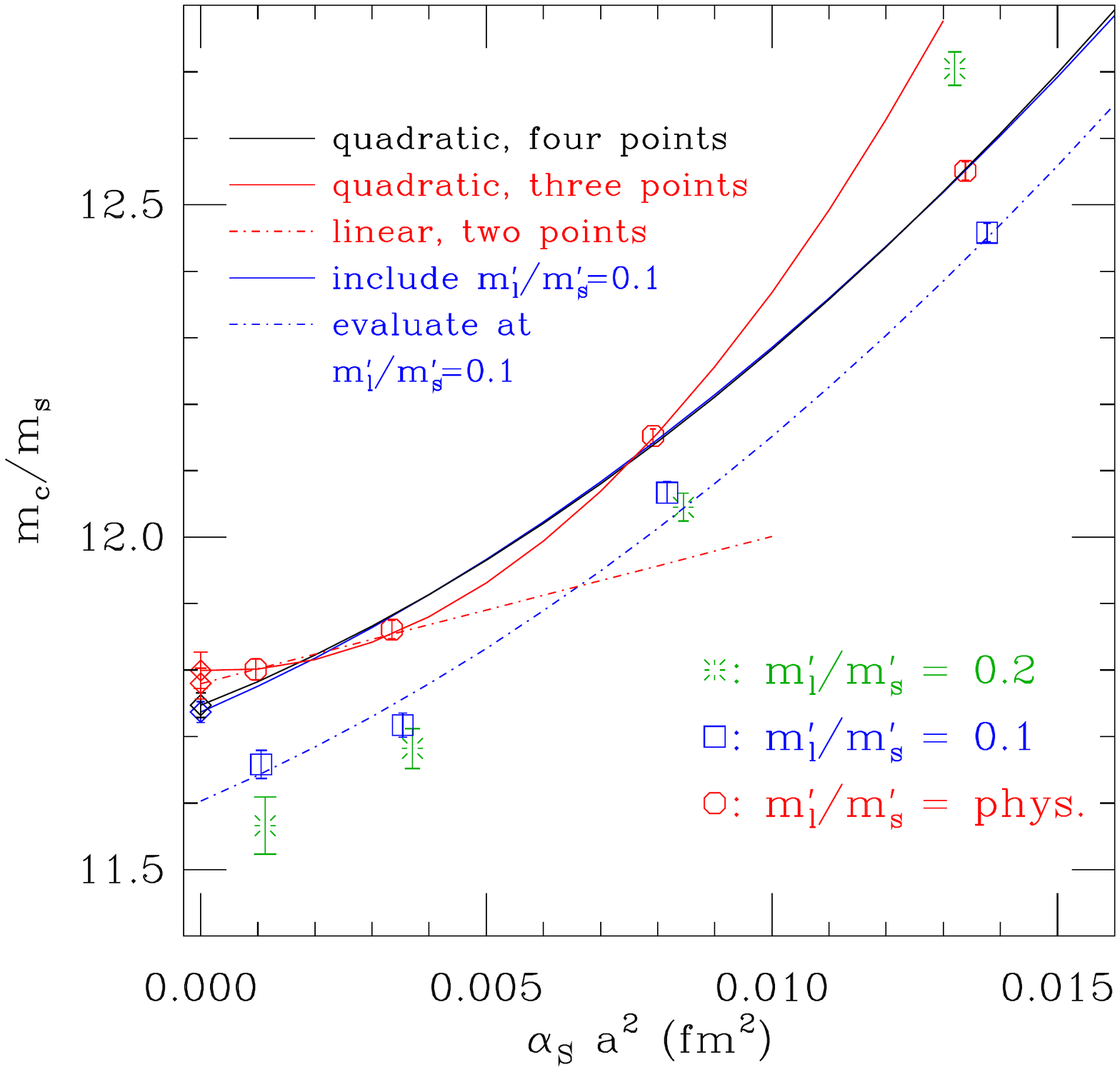}}
\caption{\label{fig:csratio}
The tuned ratio of charm quark mass to strange quark mass, $m_c/m_s$, on each ensemble.
The notation and choice of fits is the same as in Fig.~\protect\ref{fig:slratio}.
}
\end{figure}

\begin{figure}
\centerline{\includegraphics[width=0.9\textwidth]{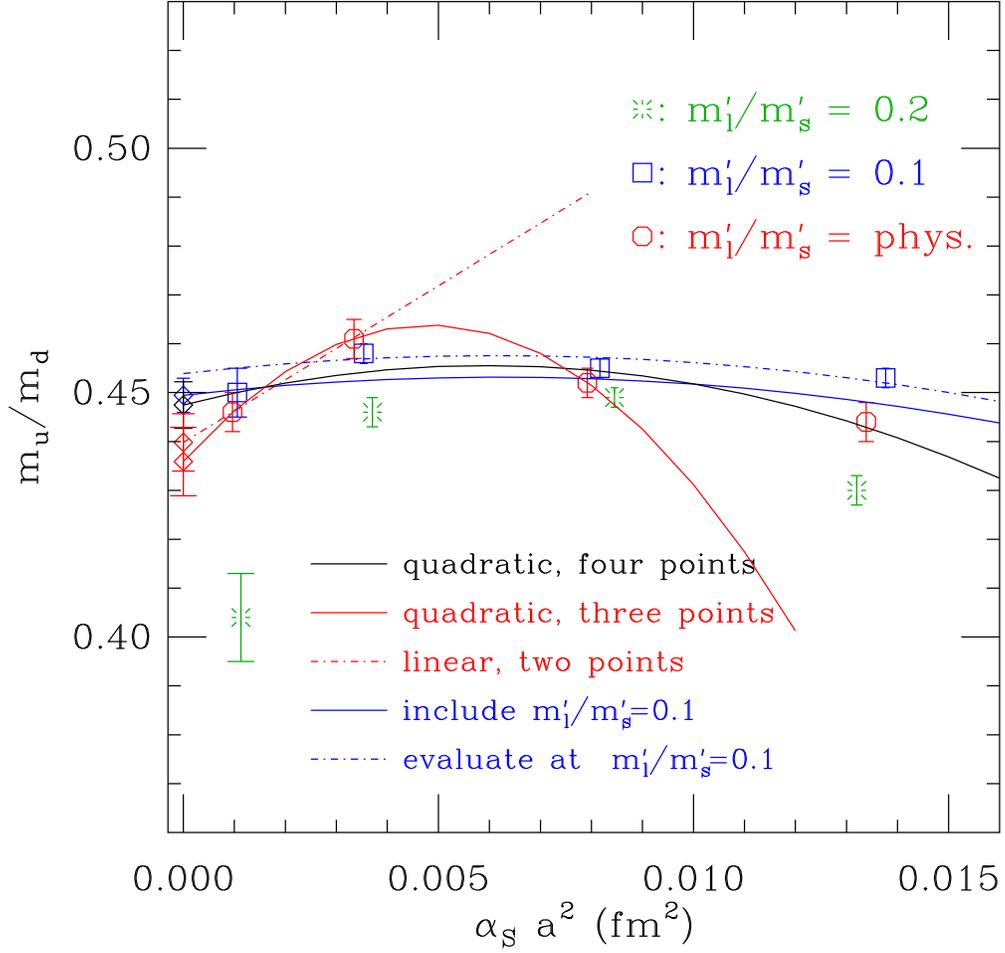}}
\caption{\label{fig:udratio}
The ratio of up quark mass to down quark mass, $m_u/m_d$, on each ensemble.
The notation and choice of fits is the same as in Fig.~\protect\ref{fig:slratio}.
}
\end{figure}

In these figures the fit used to determine the central value is shown in black.
This is a quadratic
polynomial fit through the four physical quark-mass points.  In this fit, small
adjustments have been made to compensate for sea-quark mass mistuning.  To make these
adjustments, the derivative of each quantity with respect to sea-quark mass is found
from a fit including both the physical quark-mass ensembles and the $0.1 m'_s$ ensembles,
and this derivative is used to adjust each point in the fit.  The resulting adjustments
are too small to be visible in
\figrefthree{slratio}{csratio}{udratio}.
Other fits shown in these figures are used in estimating the systematic error
resulting from our choice of fitting forms.
The blue lines in each figure show the fit including the $0.1 m'_s$ points, where the fit is
quadratic in $a^2$ and linear in $m_l^\prime/m_s^\prime$.
Here the solid line is the fit evaluated at the physical sea-quark mass, and the dashed line
is the fit evaluated at $m_l^\prime=0.1 m_s^\prime$.
The red lines are extrapolations using only the finer lattice spacings: the curved solid line
is a quadratic through the $0.06$, $0.09$ and $0.12$ fm ensembles, and the dashed straight
line is a line through the finest two points.
The diamonds at $\alpha_Sa^2=0$ indicate the continuum extrapolations of the various fits.
It is clear from the curvature in \figrefthree{slratio}{csratio}{udratio} that
a quadratic term is needed.   However, it makes only a negligible difference
whether this quadratic term is taken to be $(\alpha_Sa^2)^2$, as is done here for
convenience, or simply $(a^2)^2$.
Other continuum extrapolations not shown here use $\alpha_V a^2$, where
 $\alpha_V$ is the
 strong coupling constant computed from
 the plaquette, or simply $a^2$ as the abscissa.
 
 The four extrapolations in \figrefthree{slratio}{csratio}{udratio},
 together with quadratic
 fits to the physical mass points using $\alpha_V a^2$ or $a^2$ as the
 abscissa,
 make a set of six continuum extrapolations for these and other quantities.  
 The six versions are used to estimate the systematic errors of the quark mass ratios and light-meson
 decay constants, and  to inform the systematic error analysis of \secref{SYSERRS}.

In \figref{slratio} and, to a lesser extent in \figrefs{csratio}{udratio}, the points at small
lattice spacing with unphysical light sea quark masses deviate strongly from the physical
sea quark mass points.  This is mostly a partial quenching effect that shows up for valence
quark masses small compared to the light sea quark mass.  In particular, the squared pseudoscalar
meson mass is increased by a partially quenched chiral log, which means that a smaller
tuned light valence quark mass is needed to give the desired $M^2/F^2$.
This has the direct effect of increasing $m_s/m_l$, with smaller effects on all other quantities.
This is mostly seen at the smallest lattice spacing because at larger lattice spacings
taste violations smear out the chiral logs.  Note that this
partial quenching effect has negligible effect on our results for $m_s/m_l$ and $m_c/m_s$, which depend
almost exclusively on the data from the physical-mass ensembles.

\begin{figure}
\centerline{\includegraphics[width=0.9\textwidth]{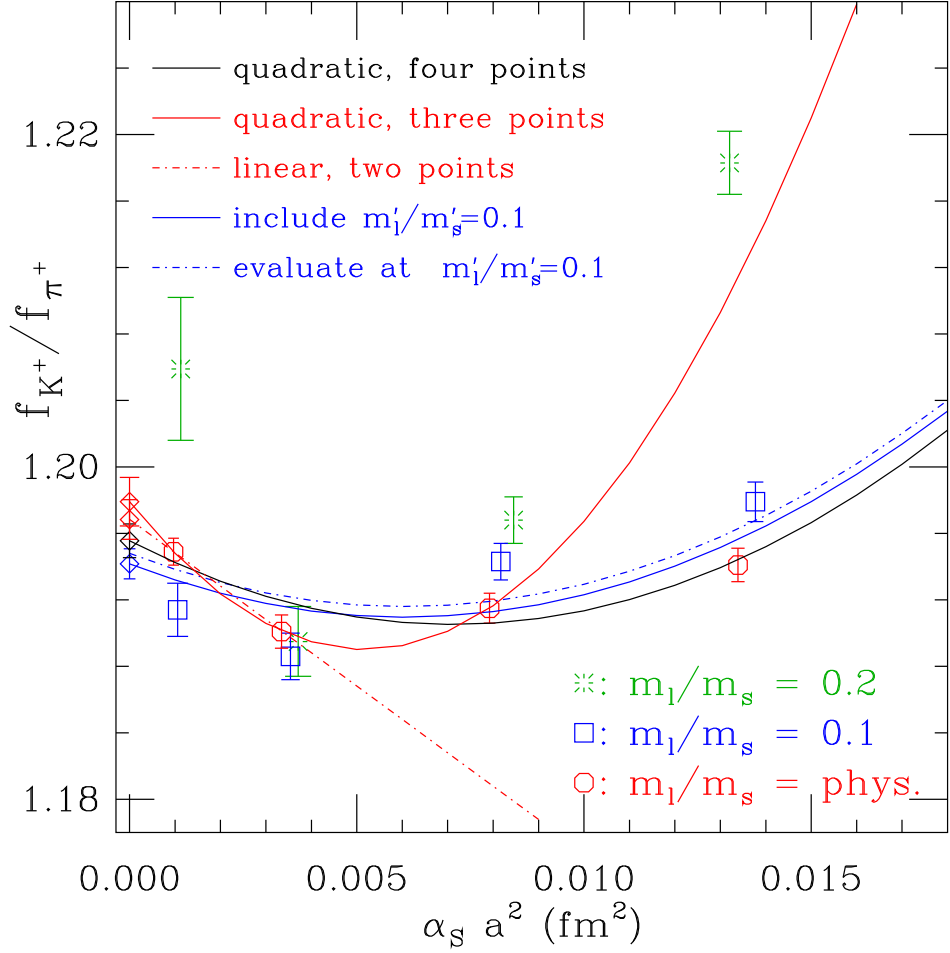}}
\caption{\label{fig:fkpiratio}
The ratio $f_{K^+}/f_{\pi^+}$ on each ensemble,
The notation and choice of fits is the same as in Fig.~\protect\ref{fig:slratio}.
}
\end{figure}

\begin{figure}
\centerline{\includegraphics[width=0.9\textwidth]{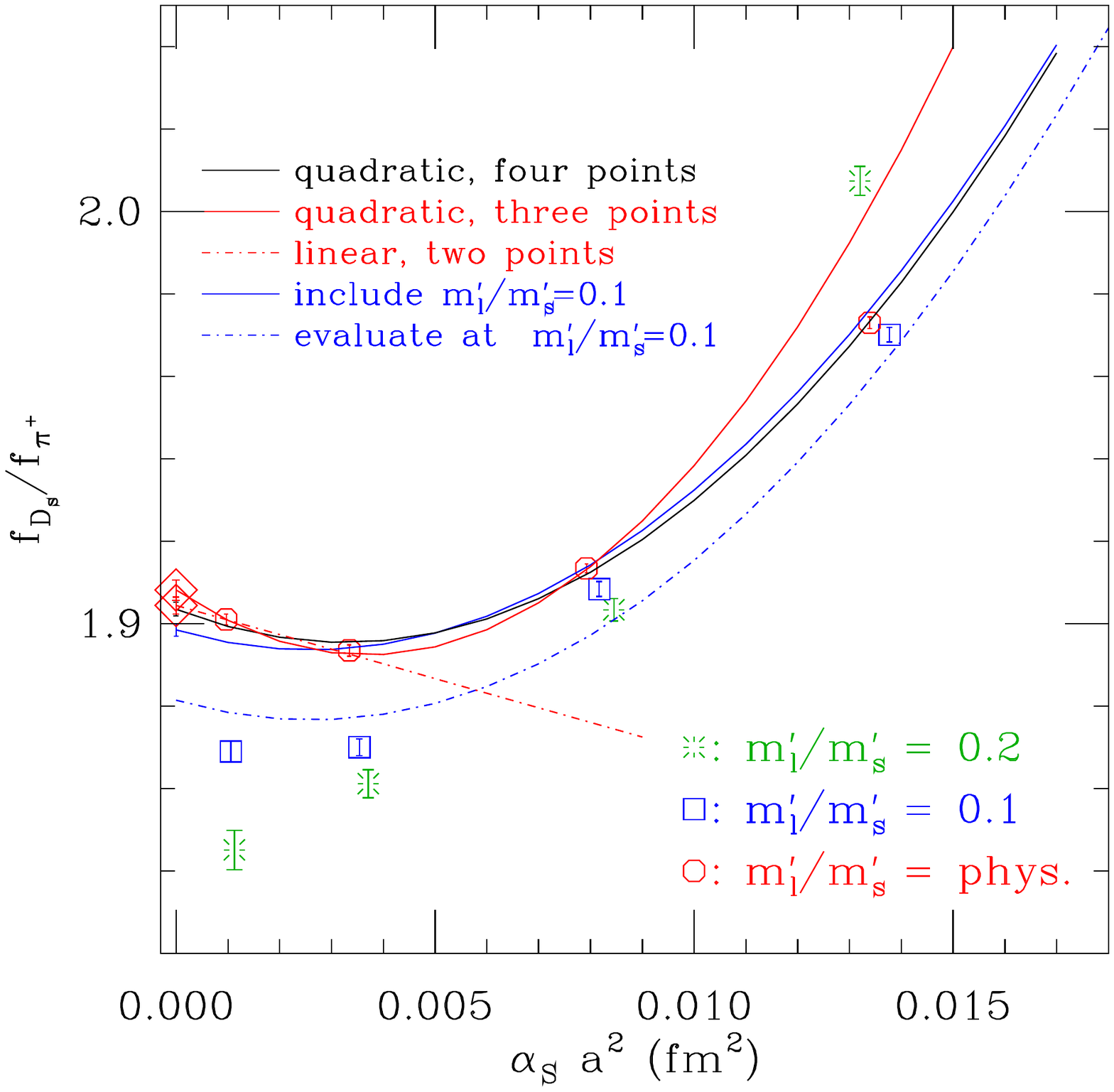}}
\caption{\label{fig:fdspiratio}
The ratio $f_{D_s}/f_{\pi^+}$ on each ensemble.
The notation and choice of fits is the same as in Fig.~\protect\ref{fig:slratio}.
}
\end{figure}

\begin{figure}
\centerline{
   \includegraphics[width=0.57\textwidth]{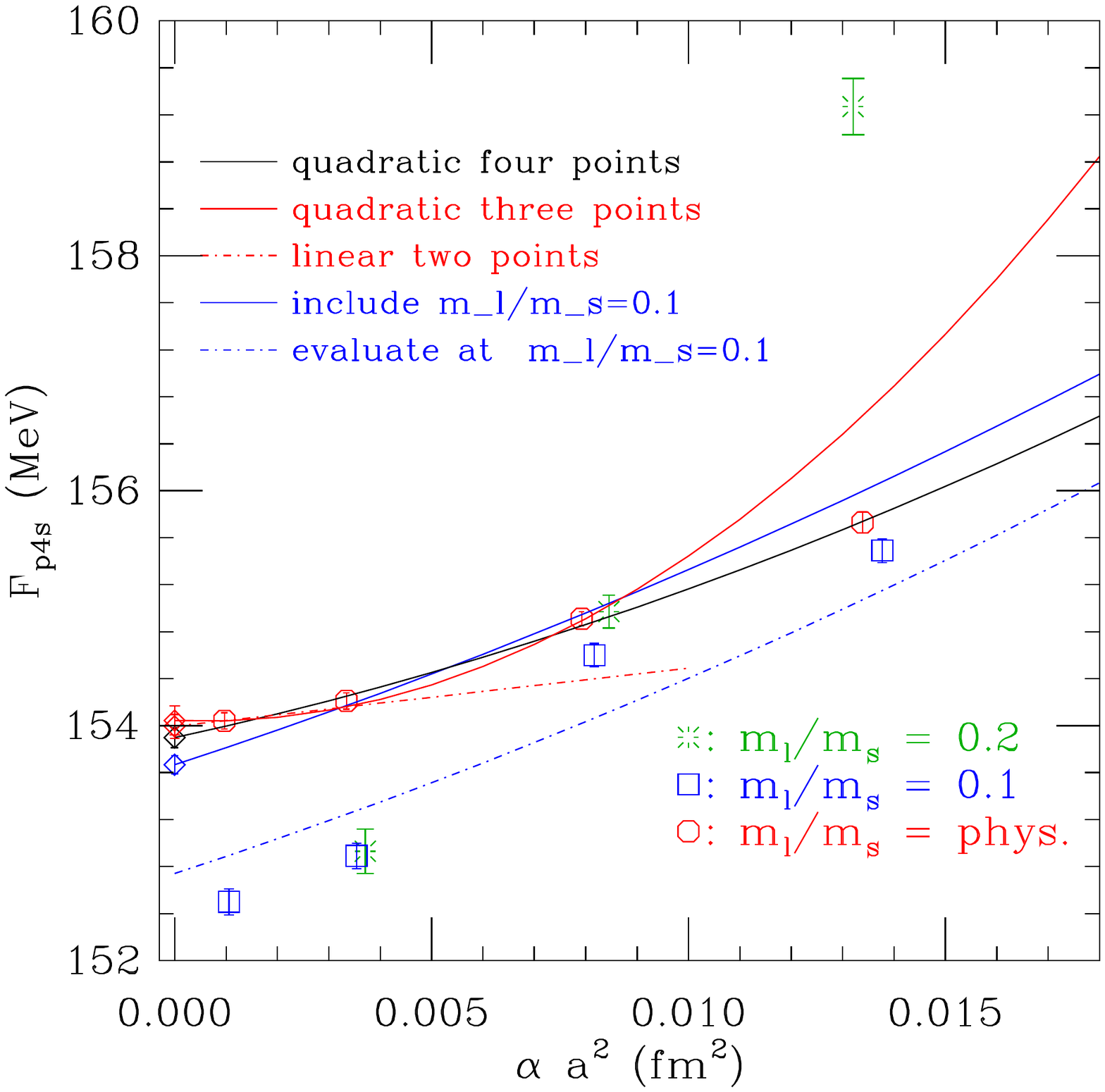}
   \hspace{-0.06\textwidth}
   \includegraphics[width=0.57\textwidth]{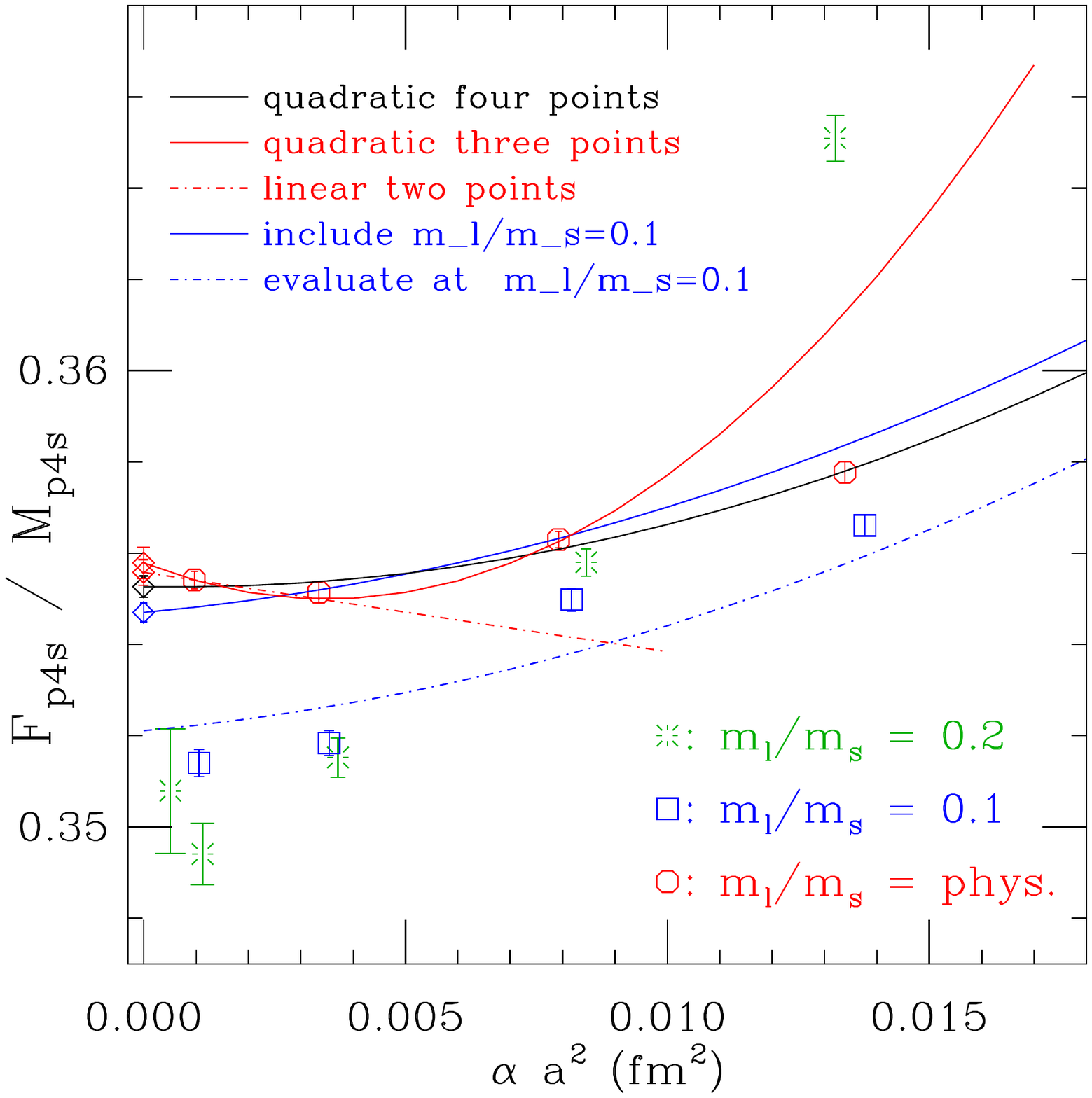}
}
\caption{\label{fig:fp4s}
$F_{p4s}$ and the ratio $F_{p4s}/M_{p4s}$ on each ensemble.  Here $f_\pi=130.41$ MeV
was used to set the scale to express $F_{p4s}$ in MeV.
The notation and choice of fits is the same as in Fig.~\protect\ref{fig:slratio}.
}
\end{figure}

We perform similar continuum extrapolations for the ratios of decay constants
$F_{p4s}/f_{\pi^+}$, $f_{K^+}/f_{\pi^+}$, $f_{D^+}/f_{\pi^+}$,  $f_{D_s}/f_{\pi^+}$, and $f_{D_s}/f_{D^+}$,
and for $M_{p4s}$ and $R_{p4s} = F_{p4s}/M_{p4s}$.
\Figref{fkpiratio} shows the individual ensemble values and the same
set of continuum extrapolations for the ratio $f_{K^+}/f_{\pi^+}$.
As an example of a quantity involving a charm quark, \figref{fdspiratio} shows values and
continuum extrapolations for the ratio $f_{D_s}/f_{\pi^+}$.
The extrapolated value for $f_{K^+}/f_{\pi^+}$ is our result for this quantity.
\Figref{fp4s} shows the continuum extrapolations for $F_{p4s}$ and  $R_{p4s} \equiv F_{p4s}/M_{p4s}$.
The resulting continuum values for  $F_{p4s}$ and $R_{p4s}$
are used in the later analysis in \secref{chiral-analysis}.
The values for the charm-meson decay constants provide consistency checks
on the analysis in \secref{chiral-analysis}, and the spread in
continuum values among the different extrapolations is included in
our estimates of the systematic uncertainty from the continuum extrapolation.
Finally, as a check, we extrapolate the mass of the $\eta_c$ meson.
These continuum extrapolations and their statistical errors are shown
in Table~\ref{tab:error_budget_1}.

Statistical errors on these quark mass ratios and decay constants are estimated with a jackknife method,
where for each ensemble we perform
the entire fitting procedure eliminating one configuration at a time.  Autocorrelations
are handled by estimating the final error from the variance of the
jackknife resamples, after first blocking the jackknife results in blocks
of 20 (eliminated) lattices,
which corresponds to 50 molecular dynamics time units for
the $a \approx 0.15$ fm physical quark mass ensemble, 100 molecular dynamics time
units for the other  $a \approx 0.15$ fm and the $0.12$ fm ensembles and 120 time units
for the $a \approx 0.09$ and $0.06$ fm ensembles.

\subsubsection{Finite volume and electromagnetic uncertainties}
\label{sec:FV-EM}

\begin{figure}
\centerline{\includegraphics[width=0.9\textwidth]{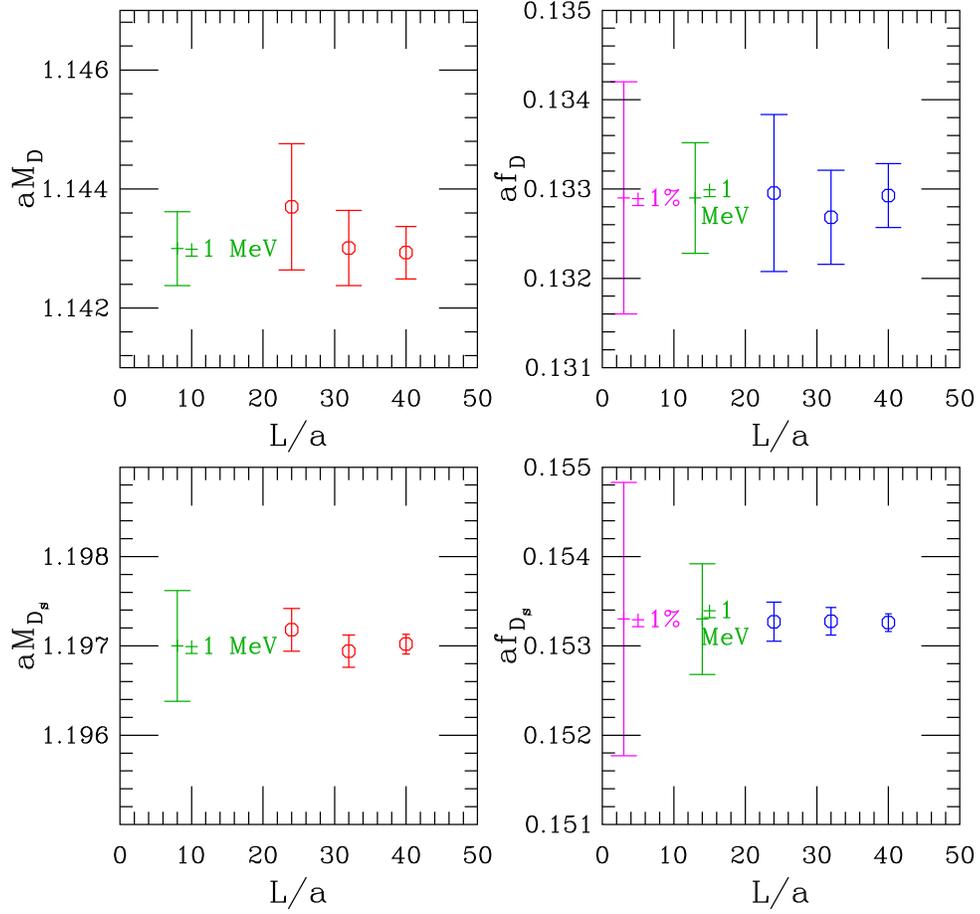}}
\caption{\label{fig:FV_check3}
Spatial size effects on 
$M_D$, $M_{D_s}$, $f_D$ and $f_{D_s}$, as determined by comparison of ensembles
with $L/a=24$, 32, and 40 at $\beta=6.0$ ($a\approx0.12$ fm).
To show the magnitude of the effects, green error bars show
an arbitrary value $\pm 1$ MeV, and magenta error bars $\pm 1$\%.
}
\end{figure}

Our treatment of finite volume effects on the pion and kaon masses and
decay constants is the same as described in Ref.~\cite{FKPRL}, and we refer the reader to the discussion
there.  To summarize very briefly, we adjust these masses and decay constants to their
values in a 5.5~fm box, the size of our physical quark mass lattices, and use these
adjusted values in the tuning procedure described above.
After the tuning and continuum extrapolation, at which point we have determined $f_{K^+}$ in a 5.5~fm box, the adjustment is removed to get our result for $f_{K^+}$ in infinite volume.
As an estimate of the remaining finite size uncertainty we
use the difference between results using staggered chiral
perturbation theory and continuum chiral perturbation theory (NNLO for $M_\pi$ and
$f_{\pi^+}$, NLO for $M_K$ and $f_{K^+}$) \protect\cite{FKPRL}.
This difference, along with other systematic effects, is tabulated in Table~\ref{tab:error_budget_1}.
Finite size effects on the charm-meson masses and decay constants are, as
expected, quite small.  \Figref{FV_check3} shows the charm-meson masses and
decay constants on the three ensembles differing only in spatial size, showing no
detectable finite size effects.

\begin{center}\begin{table}[t]
\caption{ \label{tab:error_budget_1}
Values for various physical quantities evaluated at zero lattice spacing, as
well as statistical and systematic errors, 
obtained from the simple physical-mass ensemble analysis.
Here $\Phi_{D^+} \equiv f_{D^+} \sqrt{M_{D^+}}$ etc.
We also include the $p$~value of the central fit of this analysis.
For the systematic errors, we tabulate the amount by which the central values change.
Finite size errors are the difference between results using staggered chiral
perturbation theory and continuum chiral perturbation theory (NNLO for $M_\pi$ and
$f_{\pi^+}$, NLO for $M_K$ and $f_{K^+}$) \protect\cite{FKPRL}.
``EM1'' is the effect of varying $\epsilon$ by 0.021, or one standard deviation.
``EM2'' is the effect of subtracting 450 MeV$\null^2$ from $M_K^2$.
``EM3'' is the effect of lowering the $D_s$ meson mass by 1 MeV.
``Cont. extrap.'' is the full amount of variation among the alternative continuum extrapolation
fits.
``Priors'' is the effect of using narrower priors for the mass gaps in the 0.09 and 0.06 fm physical quark mass correlator
fits.
More details on these systematic effects are in the text.
}
\smallskip
\begin{tabular}{|l|l|l|l|llllll|}
\hline
Quantity  & Central	& Stat.	& $p$ val. & Finite	& EM1 & EM2 & EM3	& Cont.   & Priors \vspace{-3.0mm}\\
	  & value	& 	&        & size  	&        &        &      	& extrap. &  \\
\hline
$M_{\eta_c}$ (MeV)    & 2982.33	& 0.35		& 0.18	&  \phantom{$+$}0.29   &  \phantom{$+$}0.11   &  \phantom{$+$}0.35   & $-$1.81   & ${}^{+1.41}_{-0.88}$ &  \phantom{$+$}0.01 \\
$f_{K^+}/f_{\pi^+}$     & 1.1956	& 0.0010	& 0.025	& $-$0.0010 & $-$0.0003 & $-$0.0004 &
\phantom{$+$}0.0000 & ${}^{+0.0023}_{-0.0014}$     & \phantom{-}0.0002 \\
$F_{p4s}$ (MeV)      & 153.90	& 0.09    	& 0.10	& $-$0.15   & $-$0.02   & $-$0.05   &  \phantom{$+$}0.00   & ${}^{+0.14}_{-0.23}$ &  \phantom{$+$}0.00 \\
$M_{p4s}$ (MeV)      & 433.24	& 0.17    	& 0.11	& $-$0.02   & $-$0.12   & $-$0.41   &  \phantom{$+$}0.00   & ${}^{+0.01}_{-0.33}$ & $-$0.01 \\
$R_{p4s}$       & 0.35527	& 0.00024    	& 0.035	& $-$0.00030&  \phantom{$+$}0.00007&  \phantom{$+$}0.00023&
\phantom{$+$}0.00000& ${}^{+0.00052}_{-0.00015}$ & \phantom{-}0.00001 \\
$m_u/m_d$       & 0.4482	& 0.0048	& 0.025	&  \phantom{$+$}0.0001 & $-$0.0156 &  \phantom{$+$}0.0000 &  \phantom{$+$}0.0000 & ${}^{+0.0021}_{-0.0115}$    &  \phantom{$+$}0.0000 \\
$m_s/m_l$       & 27.352	& 0.051		& 0.72	& $-$0.039  & $-$0.015  & $-$0.053  &  \phantom{$+$}0.000  & ${}^{+0.080}_{-0.020}$     & $-$0.001 \\
$m_c/m_s$       & 11.747	& 0.019		& 0.010	& $-$0.006  &  \phantom{$+$}0.009  &  \phantom{$+$}0.025  & $-$0.010  & ${}^{+0.052}_{-0.032}$     & \phantom{$+$}0.001 \\
$f_{D_s}/f_{D^+}$   & 1.1736	& 0.0036	& 0.97	&  \phantom{$+$}0.0003 & $-$0.0003 & $-$0.0003 &  \phantom{$+$}0.0000 & ${}^{+0.0004}_{-0.0015}$      & $-$0.0002 \\
$f_{D^+}/f_{\pi^+}$     & 1.6232	& 0.0057	& 0.59	& $-$0.0016 &  \phantom{$+$}0.0003 &  \phantom{$+$}0.0000 & $-$0.0001 & ${}^{+0.0097}_{-0.0034}$     & \phantom{$+$}0.0006 \\
$f_{D_s}/f_{\pi^+}$ & 1.9035	& 0.0017	& 0.010	& $-$0.0015 & $-$0.0001 & $-$0.0004 & $-$0.0001 & ${}^{+0.0089}_{-0.0050}$     & $-$0.0001 \\
$\Phi_{D^+}$ (MeV${}^{3/2}$)        & 9161.5	& 33.7    	& 0.61	& $-$9.3    &  \phantom{$+$}1.6    &  \phantom{$+$}0.6    & $-$3.1    & ${}^{+16.1}_{-44.9}$ &  \phantom{$+$}3.0 \\
$\Phi_{D_s}$ (MeV${}^{3/2}$)    & 11012.9	& 9.7    	& 0.007	& $-$8.9    & $-$0.7    & $-$2.6    & $-$3.4    & ${}^{+51.6}_{-28.8}$ & $-$0.1 \\
\hline
\end{tabular}
\end{table}\end{center}

Our treatment of EM effects also follows Ref.~\cite{FKPRL}, which in turn
follows Ref.~\cite{FPI04}.  The current analysis uses updated inputs for the electromagnetic
effects, so we repeat some of the discussion.
Because our sea quarks are isospin symmetric, we adjust
the experimental inputs to what they would be in a world without electromagnetism or sea-quark isospin violation before
matching the simulation data to experiment to find the strange quark mass $m_s$ and the average light-quark  mass
$\hat m= (m_u + m_d)/2$.
Specifically, we do not adjust the neutral pion mass because the leading-order isospin
correction to $M_{\pi^0}^2$ is $\propto (m_u - m_d)^2/\Lambda^2_\chi$ in  \chpt\ and therefore small, and the
electromagnetic corrections vanish in the chiral limit for neutral mesons and are thus also small.
For the kaon, we consider the isospin-averaged mass $M_{\widehat{K}}^2 = (M_{K^+}^2 + M_{K^0}^2)_{\rm QCD}/2$,
where the subscript ``QCD" indicates that the leading EM effects in the masses are removed from the experimental
masses \cite{PDG}. To remove these effects we use results from our ongoing lattice QED+QCD simulations with asqtad 
sea quarks~\cite{Basak:2012zx,Basak:2013iw} for the parameter $\epsilon$ that characterizes violations
of Dashen's theorem:
\begin{equation}
\eqn{eps-def}
 (M^2_{K^\pm}-M^2_{K^0})^\gamma =(1+\epsilon) (M^2_{\pi^\pm}-M^2_{\pi^0})^\gamma\ ,
\end{equation}
where the superscript $\gamma$ denotes the EM contribution to the splittings.  In \rcites{Basak:2012zx,Basak:2013iw}, we found
$\epsilon=0.65(7)(14)(10)$, but this result did not yet adjust for finite volume effects on the photon field.  
A recent preliminary result \cite{CB-Lat14} including finite volume effects is  $\epsilon=0.84(21)$, and we use that here.

We estimate the uncertainty due to EM effects by
varying the values of the EM-subtracted meson masses used in the quark-mass tuning; this affects $m_u$ the most.
We vary the parameter $\epsilon$ by its error.  We also
consider possible EM effects on the neutral kaon mass itself, which are less well understood than
the EM effects on the $K^+$--$K^0$ splitting that are described by $\epsilon$. 
In \rcite{Basak:2013iw}, the EM contribution to the squared $K^0$ mass was estimated to be about $900$ MeV$^2$. However, this estimate did not take into account the effects of
EM quark mass renormalization, which should be subtracted from the result.  A rough calculation of the renormalization effect
(using one-loop perturbation theory) suggests it is of order of half the size of the contribution.  We thus include
as a systematic error the effect of shifting the squared $K^0$ mass by 450 MeV$^2$.
We do not consider direct EM effects on the weak matrix elements $f_{\pi^+}$, $f_{K^+}$, $f_{D^+}$ and $f_{D_s}$, which are by definition pure QCD quantities~\cite{PDG}.  Such direct EM effects, however, are relevant in the extraction of CKM elements by comparison with experimental rates, as described in \secref{CKM}.

The shifts in various quantities resulting from these electromagnetic
uncertainties are also tabulated in Table~\ref{tab:error_budget_1}.
The two effects labeled ``EM1'' and ``EM2'' are combined in quadrature to give our
quoted EM systematic errors for $m_s/m_l$ and $f_{K^+}/f_{\pi^+}$.  
The  ``EM3'' column in Table~\ref{tab:error_budget_1} shows the effect of lowering the
input $D_s$ meson mass by 1 MeV, an order-of-magnitude estimate for the electromagnetic
effect on this mass, which affects the tuning of the charm-quark mass.
This effect has not been directly determined in QCD+QED simulations.
Assuming that the EM effect on  $M_{D^+}$ is approximately the same as on  $M_{D_s}$, since the
two mesons have the same charge, the EM3 error on the decay constants
of these mesons is negligible:  To very good approximation, the changes in
$\Phi_{D^+}$ and $\Phi_{D_s}$ due to the change in the estimate of the charm-quark mass, are canceled by
the changes in the factors of
$M_{D^+}^{1/2}$ or $M_{D_s}^{1/2}$ in these quantities. The fact that the decay constants themselves are 
only mildly dependent on the heavy-quark mass  (for example, the difference
between $f_{D_s}$ and $f_{B_s}$ is only about 10\% \cite{Aoki:2013ldr}) indicates that such cancellations must take place. 
The EM3 error does  lead to a significant uncertainty on $m_c/m_s$, and we include it in our
systematic error estimate for that quantity.

\subsection{Chiral perturbation theory analysis of $f_D$ and $f_{D_s}$ including unphysical
quark-mass ensembles} \label{sec:chiral-analysis}

In this section, we present the combined chiral extrapolation/interpolation and continuum
extrapolations used to obtain the physical values of the $D^+$ and $D_{s}$ meson decay constants.
We first discuss chiral perturbation theory for all-staggered heavy-light mesons in
\secref{CHIPT}, giving the formulas used for the chiral fits and describing our method for
incorporating discretization effects into the extrapolation.   An explanation of our method for
setting the lattice scale follows in \secref{LAT-SCALE}.   Chiral perturbation theory assumes a
mass-independent scale-setting procedure.  In practice, we use $F_{p4s}$ to set the scale and
$F_{p4s}/M_{p4s}$ to tune the strange sea-quark mass.  We take these values from the physical
quark-mass analysis in \secref{physical-mass-analysis}.  This means that the absolute scale comes
ultimately from $f_{\pi^+}$, which is used to set the scale in \secref{physical-mass-analysis}.

The chiral fits themselves are presented in \secref{CHIPT-fits}, while systematic errors in the chiral analysis are described in \secref{SYSERRS}.  
Chiral/continuum extrapolation errors are found by considering a large number (18) of alternative chiral fits, as well as six versions of the continuum extrapolation of the inputs, resulting in 108 possibilities.  We also estimate finite volume and EM errors within the chiral analysis by propagating the errors in the corresponding inputs through the chiral fits.
Equations~(\ref{eq:fD-chiral-result})--(\ref{eq:ratio-chiral-result}) show our results for the charm decay constants from the self-contained chiral analysis with complete systematic error budgets.

\subsubsection{Chiral perturbation theory for $f_{D^+}$ and $f_{D_s}$}  \label{sec:CHIPT}

The quark-mass and lattice-spacing dependence of the decay constant has been derived at one loop in  heavy-meson, 
rooted, all-staggered chiral perturbation theory (\aschpt) in \rcite{Bernard-Komijani}. 
At fixed heavy-quark mass $m_Q$,  one may argue  following Ref.~\cite{Bazavov:2011aa} that 
inclusion of  hyperfine splittings (\eg $M^*_D-M_D$) and flavor splittings (\eg $M_{D_s}-M_D$), but no 
other $1/m_Q$ effects,
constitutes a systematic approximation at NLO in \aschpt.  The argument is based on the
power counting introduced by Boyd and Grinstein \cite{BoydGrinstein}.
With ${\rm v}$ denoting the light valence quark, $Y$ the ${\rm v}\bar {\rm v}$ valence meson, and
$\Phi_{D_{\rm v}}\equiv f_{D_{\rm v}} \sqrt{ M_{D_{\rm v}}}$,
\rcite{Bernard-Komijani} obtains for the pseudoscalar-taste heavy-light meson:
\begin{eqnarray}\label{eq:chpt-form}
 \Phi_{D_{\rm v}}
      & = &     \Phi_0 \Biggl\{  1 + \frac{1}{16\pi^2f^2} \frac{1}{2}
      \Biggl(-\frac{1}{16}\sum_{\mathscr{S},\Xi} \ell(M_{\mathscr{S}{\rm v},\Xi}^2)
          - \frac{1}{3}
          \sum_{j\in \cM_I^{(3,{\rm v})}} 
          \frac{\partial}{\partial M^2_{Y,I}}\left[ 
                R^{[3,3]}_{j}(
              \cM_I^{(3,{\rm v})};  \mu^{(3)}_I) \ell(M_{j}^2) \right]
              \nonumber \\*&&\hspace{-5mm}{} 
              -   \Bigl( a^2\delta'_V \sum_{j\in \cM_V^{(4,{\rm v})}}
              \frac{\partial}{\partial M^2_{Y,V}}\left[ 
                R^{[4,3]}_{j}( \cM_V^{(4,{\rm v})}; \mu^{(3)}_V)
              \ell(M_{j}^2)\right]
                  + [V\to A]\Bigr) \nonumber \\*&&\hspace{-5mm}{}
            -3g_\pi^2\frac{1}{16}\sum_{\mathscr{S},\Xi} J(M_{\mathscr{S}{\rm v},\Xi},\Delta^*+\delta_{\mathscr{S}{\rm v}})
            - g_\pi^2
          \sum_{j\in \cM_I^{(3,{\rm v})}} 
          \frac{\partial}{\partial M^2_{Y,I}}\left[ 
                R^{[3,3]}_{j}(
              \cM_I^{(3,{\rm v})};  \mu^{(3)}_I) J(M_{j},\Delta^*) \right]
              \nonumber \\*&&\hspace{-5mm}{} 
              \hspace{0cm} -3g_\pi^2 \Bigl( a^2\delta'_V \sum_{j\in \cM_V^{(4,{\rm v})}}
              \frac{\partial}{\partial M^2_{Y,V}}\left[ 
                R^{[4,3]}_{j}( \cM_V^{(4,{\rm v})}; \mu^{(3)}_V)
              J(M_{j},\Delta^*)\right]
                  + [V\to A]\Bigr)  
            \Biggr)\ \nonumber \\*&&\hspace{-5mm}{}+
      L_\text{s} (x_u + x_d + x_s) + L_{\rm v} x_{\rm v} + L_{a} \frac{x_{\bar\Delta}}{2}\Biggr\} \ , \eqn{chiral-form}
\end{eqnarray}
where $\Phi_0$, $L_\text{s}$, $L_{\rm v}$, and $L_a$ are low-energy constants (LECs); the indices $\mathscr{S}$ and $\Xi$ run over sea-quark flavors and meson tastes, respectively;
$\Delta^*$ is the lowest-order hyperfine splitting;  $\delta_{\mathscr{S}{\rm v}}$  is the
flavor splitting between a heavy-light meson with  light quark of flavor $\mathscr{S}$ and one
of flavor ${\rm v}$; and $g_\pi$ is the $D$-$D^*$-$\pi$ coupling.
In infinite volume, the chiral logarithm functions $\ell$ and $J$ are defined 
by \cite{Aubin:2003mg,Bazavov:2011aa}
\begin{eqnarray}
        \ell(m^2) &= & m^2 \ln \frac{m^2}{\Lambda_\chi^2}
\qquad \textrm{[infinite volume]} , \label{eq:chiral_log_infinitev} \\
    J(M,\Delta) &=& (M^2-2\Delta^2)\log(M^2/\Lambda^2) +2\Delta^2 -4\Delta^2 F(M/\Delta)  \qquad \textrm{[infinite volume]},  \eqn{Jdef}
\end{eqnarray}
with \cite{Stewart:1998ke}
\begin{equation}
F(1/x) \equiv
    \begin{cases}
     -\frac{\sqrt{1-x^2}}{x_{\phantom{g}}}\left[\frac{\pi}{2} - \tan^{-1}\frac{x}
    {\sqrt{1-x^2}}\right], & \text{if $ |x|\le 1$,} \\
    \frac{\sqrt{x^2-1}}{x}\ln(x + \sqrt{x^2-1}), & \text {if $ |x|\ge 1$.}
    \end{cases}
    \eqn{Fdef}
\end{equation}
The residue functions $R_j^{[n,k]}$ are given by
\begin{eqnarray}\eqn{R-def}
        R_j^{[n,k]}\left(\left\{m\right\}\!;\!\left\{\mu\right\}\right)
                  &\equiv & \frac{\prod_{i=1}^k (\mu^2_i- m^2_j)}
		   {\prod_{r\not=j}^n (m^2_r - m^2_j)}\ .
\end{eqnarray}  
The sets of masses in the residues are  
\begin{eqnarray}
  \mu^{(3)} & = & \{m^2_U,m^2_D,m^2_S\}\ ,\\*
  \cM^{(3,\text{v})} & = & \{m_Y^2,m_{\pi^0}^2, m_{\eta}^2\}\ ,\\*
  \cM^{(4,\text{v})} & = & \{m_Y^2,m_{\pi^0}^2, m_{\eta}^2, m_{\eta'}^2\}\ .
\end{eqnarray}
Here taste labels (\eg $I$ or $V$ for the masses) are implicit.
We define dimensionless quark 
masses and a measure of the taste splitting by
\begin{eqnarray}\eqn{x-defs}
x_{u,d,s,{\rm v}}  \equiv \frac{4B}{16 \pi^2 f_{\pi}^2} m_{u,d,s,{\rm v}}\;,\ \mathrm{and} &\hspace{10mm}&
x_{\bar\Delta} \equiv \frac{2}{16 \pi^2 f_{\pi}^2}a^2\bar\Delta \;,
\end{eqnarray}
where $B$ is the LEC that gives the Goldstone pion mass $M_\pi^2 = B(m_u+m_d)$, and
$a^2\bar\Delta$ is the mean-squared pion taste splitting.  The $x_i$ are  natural variables of \aschpt; 
the LECs $L_\text{s}$, $L_{\rm v}$, and $L_a$ are therefore expected to be $\cO(1)$. 
All ensembles in the current analysis have degenerate light sea quarks: $x_u=x_d\equiv x_l$. 
The taste splittings have been determined
to $\sim\!1$--10\% precision \cite{HISQ_CONFIGS} and are used as input to \eq{chiral-form}, as are
the taste-breaking hairpin parameters $\delta'_A$ and $\delta'_V$, whose ranges
are taken from chiral fits to light pseudoscalar mesons~\cite{Bazavov:2011fh}. 

To include the finite-volume effects for a spatial volume~$L^3$ in \eq{chiral-form}, we  replace~\cite{Bazavov:2011aa}
\begin{eqnarray}
      \ell( m^2)  &\to & \ell( m^2)  + m^2 \delta_1(mL) \qquad \textrm{[finite volume]},\\
      J(m,\Delta) &\to & J(m,\Delta) + \delta J(m,\Delta,L) \qquad \textrm{[finite volume]},
\end{eqnarray}
where
\begin{equation}
\delta J(m,\Delta,L)  = \frac{m^2}{3}\delta_1(mL) - 16\pi^2\left[\frac{2\Delta}{3}J_{FV}(m,\Delta,L)
+\frac{\Delta^2-m^2}{3} K_{FV}(m,\Delta,L)\right] \ ,
\end{equation}
with
\begin{equation}
    K_{FV}(m,\Delta,L) \equiv \frac{\partial}{\partial \Delta} J_{FV}(m,\Delta,L) ,
\end{equation}
and with $\delta_1(mL)$ and $J_{FV}(m,\Delta,L)$ defined in Refs.~\cite{Aubin:StagHL2007,Arndt:2004}.

Because we have data with $\sim\!1\%$ to less than $0.1\%$ statistical errors and 314 to 366 data points 
(depending on whether $a\approx0.15\;$fm is included), NLO \aschpt\ is not adequate to describe
fully the quark-mass dependence, 
in  particular for masses near $m_s$.  We therefore include all NNLO and NNNLO mass-dependent analytic terms.  
There are four independent  functions of $x_{\rm v}$, $x_l$ and $x_s$ at NNLO and seven at NNNLO, for a total of eleven additional fit parameters. 
It is not necessary to keep all the seven terms appearing at NNNLO to get a good fit, nevertheless we include all of them
to make it a systematic approximation at the level of analytic terms.

While \eq{chiral-form} is a systematic NLO approximation for the decay constant at fixed $m_Q$,
we have data on each ensemble with two different values of the valence charm mass: $m'_c$ and $0.9 m'_c$,
where $m'_c$ is the value of the charm sea mass of the ensembles, and is itself not precisely equal to 
the physical charm mass $m_c$ because of tuning errors, which are in some cases as large as 
this difference (\ie 10\% of $m'_c$).  Since
such changes in the value of the charm mass lead to corrections to decay constants that  
are comparable in size to those from the  pion masses at NLO,
\eq{chiral-form} needs to be modified in order to fit the data. We therefore allow the 
LEC $\Phi_0$ to depend on $m_Q$ as suggested by HQET. For 
acceptable fits to the highly correlated data at valence charm masses $m'_c$ and $0.9 m'_c$, we 
need to introduce both $1/m_Q$ and $1/m_Q^2$ terms.
(For more details see the Appendix.) 
Furthermore, $\Phi_0$ has generic lattice-spacing dependence
that must be included to obtain good fits.  With HISQ quarks,
the leading generic discretization errors are $\cO(\alpha_S a^2)$.  But because the high degree of 
improvement in the HISQ action drastically reduces the coefficient of these leading errors, formally higher  $\cO(a^4)$ errors
are also apparent, as can be seen from the curvature in \figrefto{slratio}{fp4s}.
In \eq{chiral-form}, we thus 
replace 
\begin{equation}\eqn{Phi0-form}
\Phi_0\to \Phi_0 \left(1 + k_1\frac{\Lambda_{\rm HQET}}{m_Q}+ k_2\frac{\Lambda_{\rm HQET}^2}{m_Q^2}\right)\Big(1+ c_1\alpha_S (a\Lambda)^2 +c_2  (a\Lambda)^4 
\Big)\;,
\end{equation}
where the $k_i$ are new physical LECs, $c_i$ are additional fit parameters,
$\Lambda_{\rm HQET}$ is a physical scale for HQET effects, and $\Lambda$ is
the scale of discretization effects.
 
In cases where the valence and sea values of the charm quark mass differ, $m_Q$ in \eq{Phi0-form} is 
taken equal to the valence mass.   This is based on the expectation from 
decoupling \cite{Appelquist:1974tg} that effects due to
variations in the charm sea mass on low-energy physical 
quantities are small.
Note that HQET tells us that heavy-light decay constants come from  the physics of the light-quark
at scale $\Lambda_{\rm QCD}$,
despite the presence of the heavy valence quark.
Thus we do not introduce extra terms corresponding 
to the charm sea mass here. As discussed in \secref{SYSERRS}, however, such terms are included in alternative fits
used to estimate systematic errors.

Generic dependence on $a$ is also allowed 
for the physical LECs $L_\text{s}$, $L_{\rm v}$,   $k_1$ and $k_2$.  However, because these parameters first 
appear at  NLO in the chiral or HQET expansions, it is sufficient to include at most the leading $a$-dependence, for example:
\begin{equation}\eqn{L-form}
L_{\rm v} \to L_{\rm v} + L_{\rm v\delta} \;\alpha_S (a\Lambda)^2 
\end{equation}
Thus we add 4 fit parameters related to generic discretization effects: 
$L_{\rm v\delta} $,   $L_{\rm s\delta} $,    $k_{1\delta} $,  and $k_{2\delta}$. 
There are also 3 parameters related to taste-violation effects: $L_a$, $\delta'_A$ and $\delta'_V$.  These parameters are
taken proportional to the measured average taste splitting $a^2 \bar \Delta$,
which depends on $a$ approximately as
$\alpha_S^2 a^2$  \cite{HISQ_CONFIGS}. In addition, we find that $m_Q$-dependent discretization errors 
must  be considered if data at the coarsest lattice spacing ($a\approx0.15\;$fm) is included in the fits.
This is not surprising because $am^{\rm phys}_c\approx0.84$ at this lattice spacing, which by the power counting
estimates of \rcite{HPQCD_HISQ} suggests $\sim\!5\%$ discretization errors 
(although this may be reduced  by dimensionless factors).  
We therefore add $c_3\alpha_S (am_Q)^2 +c_4(am_Q)^4$ to the analytic terms in \eq{chiral-form}, where $m_Q$ is taken to denote the valence charm mass.
If the $a\approx0.15\;$fm data are omitted, good fits may be obtained with $c_3$ and $c_4$ set to zero. 
 As discussed below, one can also add similar terms for the charm sea mass.

For the LEC $g_\pi$, a reasonable range is $g_\pi=0.53(8)$, which comes from recent lattice calculations 
\cite{Becirevic,Detmold}.
When this central value and range are included as Bayesian priors, fits to our full data set
tend to pull  $g_\pi$ low, several sigma below 0.53. 
Hence, we simply fix $g_\pi=0.45$, 1-sigma below its nominal value, in our central fit.  
This problem is ameliorated for 
alternative  fits, used to estimate the systematic errors, that drop the data at  $a\approx0.15$ fm or that
use the experimental value of $f_{K^+}$, rather than that of $f_{\pi^+}$,  for $f$ in \eq{chpt-form}.
Other alternatives considered in the systematic error estimates are to allow $g_\pi$ to be a free parameter, 
or to keep it fixed at its nominal value. We give more details about fits with varying treatments of $g_\pi$ in \secref{SYSERRS}.

\subsubsection{Setting the relative lattice scale}  \label{sec:LAT-SCALE}

Relative scale setting in the combined chiral analysis is done using $F_{p4s}$.  The value of
$F_{p4s}$ in physical units, which is  only needed at the end of this analysis, has been
obtained by comparison with $f_{\pi^+}$ in
\secref{physical-mass-analysis}, as are the other needed inputs:  $R_{p4s}\equiv F_{p4s}/M_{p4s}$ and the
quark-mass ratios $m_c/m_s$, $m_s/m_l$ and $m_u/m_d$.
All those quantities are listed in \tabref{error_budget_1}, and
\figref{fp4s} shows the data and continuum extrapolations used to
determine $F_{p4s}$ and $R_{p4s}$.

We use $F_{p4s}$ in the chiral analysis, rather than $f_{\pi^+}$ itself, for several reasons.  First of
all, $F_{p4s}$ gives highly-precise relative lattice spacings between ensembles.  Precision scale
setting is required  in order to get good chiral fits to our large partially-quenched data
set (366 points) with large correlations of the points within each ensemble.  Second,  $F_{p4s}$ can
be accurately adjusted for mistunings in the sea-quark masses using unphysical-mass ensembles for
which the physical valence-quark mass values needed to find $f_{\pi^+}$ can only be reached by
extrapolation.  Finally, and perhaps most importantly, there are no logarithms of light pseudoscalar
masses ($\sim\!m_\pi$) in the \schpt\ expression for the decay constant \cite{Aubin:2003mg} evaluated
at the relevant quark masses for $F_{p4s}$. The lightest meson that enters is a valence-sea meson for
quark masses $0.4m_s$ and $m_l$, which has mass $\sim\!325$ MeV (for the Goldstone taste).  This means
that $F_{p4s}$ should be well approximated by its Taylor series in $a^2$, and we do not need to modify
\eq{chpt-form} to take into account chiral logarithms that enter through the scale-setting procedure.
We have checked this assumption by performing a more complicated three-step analysis: (1) The
degenerate light-light decay-constant data for all ensembles are fit to the NLO \schpt\ form of
\rcite{Aubin:2003mg}. (2) From the fit, we determine $F_{p4s}$ as a function of $a^2$. (3) The data
for $\Phi_{D_\text{v}}/F_{p4s}^{3/2}$ are fit to \eq{chpt-form} divided by the 3/2 power of
$F_{p4s}(a^2)$.  The results of this procedure differ from the results reported  in
\tabref{results_Phi} below by less than half of the statistical errors, and the systematic errors
are essentially the same in both approaches.  

We use a mass-independent scale-setting scheme.  We first determine $aF_{p4s}$ and $am_{p4s}$ on the
physical-mass ensembles; then, by definition, all ensembles at the same $\beta$ as a given
physical-mass ensemble have a lattice spacing $a$ and value of $am_{p4s}$ equal to those of the
physical-mass ensemble.  Since we do not know the correct strange-quark mass until after the lattice
spacing is fixed, $aF_{p4s}$ and $am_{p4s}$ must be determined self-consistently.  
We find $am_{p4s}$ and $aF_{p4s}$ on a given physical-mass ensemble by
adjusting $am_\text{v}$ until $aF/(aM)$ has the expected physical ratio $R_{p4s}$.

To determine $aF_{p4s}$ and $am_{p4s}$ accurately, data must be adjusted for mistunings in the sea-quark
masses. The sea-quark masses of the physical-mass ensembles are tuned relatively well (especially at
0.09 and 0.06 fm), and adjustments are  small. Nevertheless, the adjustments may change the final
results of $f_{D^+}$ and $f_{D_s}$ by more than the size of the statistical errors. 
 
To make these adjustments, we first find an approximate value of $am_{p4s}$ on each physical-mass
ensemble by passing a parabola through $(M/F)^2$ as a function of $m_\text{v}$, for the three values of $m_\text{v}$
closest to  $m_{p4s}$.  The sea-quark masses are kept fixed (initially, to their values in the run) in
this process.  We use $(M/F)^2$  here rather than $F/M$, since we expect $M^2$ to be approximately
linear in ${\rm m_\text{v}}$, and $F^2$ to be approximately constant.  The value of $am_\text{v}$ where the ratio
takes its expected value $1/R^2_{p4s}$ is the tentative value of $am_{p4s}$, and the  corresponding
value of $aF$ is the tentative value of $a F_{p4s}$. The procedure also gives tentative values of the
physical sea-quark masses in lattice units: $am_s\cong2.5\,am_{p4s}$, $am_l\cong2.5\,am_{p4s}/ (m_s/m_l)$, and $am_c\cong2.5\,am_{p4s} (m_c/m_s)$.  We then adjust the data for $aF$ and $aM$ to the values they would have at the tentative new
sea-quark masses, and iterate the whole process until it converges.
 
The adjustment of the data requires a determination of the following derivatives
\begin{equation}\eqn{derivatives} \frac{\partial F^2 }{\partial m'_l},\ \frac{\partial F^2 }{\partial
m'_s},\ \frac{\partial F^2 }{\partial m'_c},\ \frac{\partial M^2 }{\partial m'_l},\ \frac{\partial M^2
}{\partial m'_s},\ \frac{\partial M^2 }{\partial m'_c},\ \frac{\partial^2 M^2 }{\partial m'_l \partial
m_\text{v}}, \ \frac{\partial^2 M^2 }{\partial m'_s \partial m_\text{v}}, \frac{\partial^2 M^2 }{\partial
m'_c \partial m_\text{v}}, \end{equation} where the derivatives should be evaluated at $m_\text{v}=m_{p4s}$,
and with $m'_l$, $m'_s$ and $m'_c$ at their physical values.  All quantities here are in ``$p4s$ units'',
which are (semi-) physical units in which $aF$ and $aM$ have been divided by (the tentative value of)
$aF_{p4s}$, and quark masses in lattice units have been divided by (the tentative value of) $am_{p4s}$ (and
therefore do not require renormalization).  The mixed partial derivatives with $m_\text{v}$ are needed
because we must adjust the data at different values of $m_\text{v}$ in order to iterate the process. Because
$M^2$ is approximately linear in $m_\text{v}$, the effect of the mixed partials in \eq{derivatives} is
non-negligible, while mixed partials of   $F^2$ may be neglected.
Since the effects of mistunings are already not much larger than our statistical errors, we expect
that we may neglect discretization errors and any mistuning effects in the derivatives themselves.
This means that we may use,  at all lattice spacings, the values determined for the derivatives in
\eq{derivatives} at any one lattice spacing.  This expectation is confirmed by alternative
determinations of the derivatives, which give results in agreement with the method we now describe.

\begin{figure}[t] 
\begin{center}
\null\vspace{-6.5cm}
\null\hspace{-15mm}\includegraphics[width=14cm]{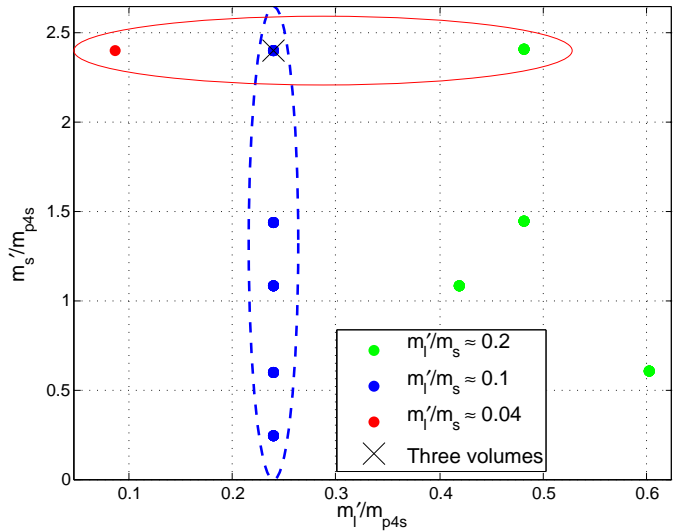} \end{center} \vspace{-6cm} \caption{
Values of $m'_s$ and $m'_l$ of the ensembles at $\beta=6.0$. At one value of $m'_s$ and $m'_l$, indicated
by the black cross, we have three ensembles with different volumes; the intermediate volume ensemble,
which is equal in volume to all the other ensembles shown here, is used in our calculation of the
derivatives.  Five ensembles inside the blue ellipse are used to calculate ${\partial F^2 }/{\partial
m'_s}$, ${\partial M^2 }/{\partial m'_s}$, and ${\partial ^2M^2 }/{\partial m'_s\partial m_\text{v}}$.
These five ensembles have the same charm sea masses.  Three ensembles inside the red ellipse are used
to calculate ${\partial F^2 }/{\partial m'_l}$, ${\partial M^2 }/{\partial m'_l}$, and ${\partial ^2M^2
}/{\partial m'_l\partial m_\text{v}}$. One of these ensembles has a slightly different charm sea mass,
which is adjusted before calculating the derivatives.  \label{fig:ms_ml_beta6}} \end{figure}

\begin{figure}[t] \vspace{-10cm} \begin{center} \begin{tabular}{l l}
\null\hspace{-40mm}\includegraphics[width=17cm]{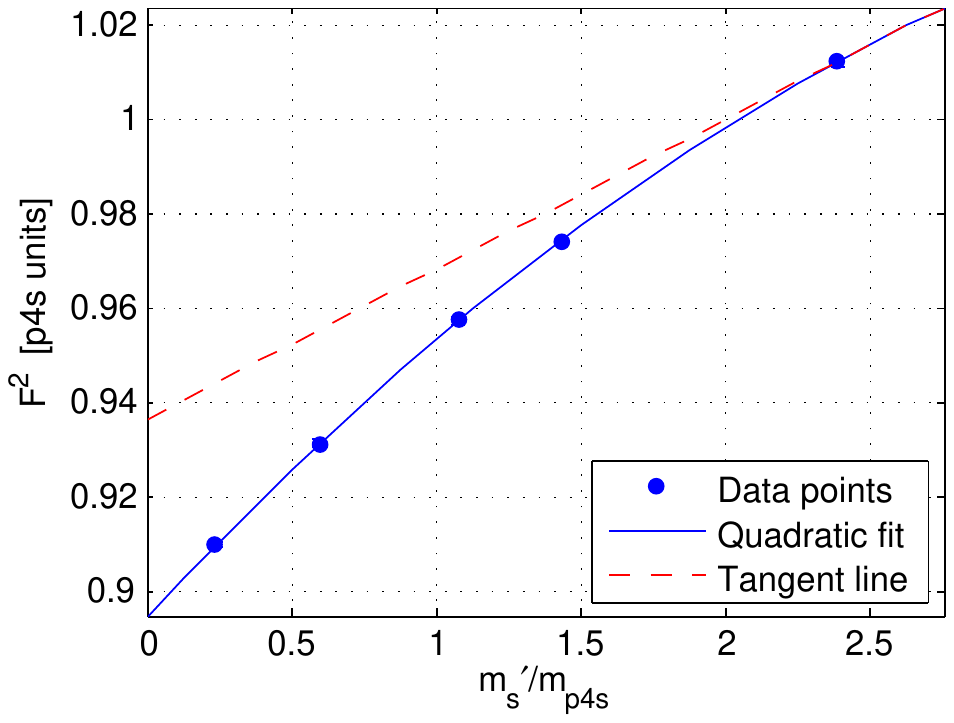} &\hspace{-90mm}
\includegraphics[width=17cm]{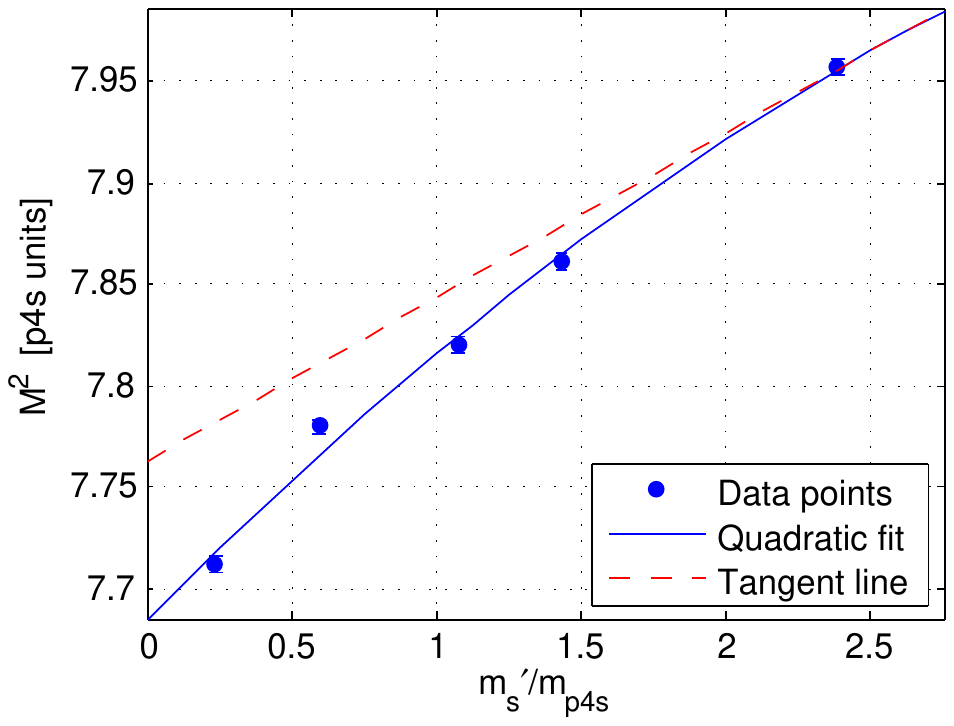} \end{tabular} \end{center} \vspace{-8.5cm} \caption{Data from
the $a\approx 0.12$ fm,  $m'_l/m_s\approx0.1$ ensembles, which are shown inside the blue
ellipse in \figref{ms_ml_beta6}. \label{fig:dFp4sandMp4s_dms} $F_{p4s}$ and $M_{p4s}$ are the
light-light pseudoscalar decay constant and mass for $m_{\rm v} = m_{p4s}$; quantities are expressed
in $p4s$ units, as described in the text.  The needed derivatives are given by
the slope of the tangent line at $m'_s/m_{p4s}$=2.5 } \end{figure}

Many of the derivatives may be calculated using the twelve ensembles that we have at $a \approx 0.12\
{\rm fm}$. \Figref{ms_ml_beta6} shows the light and strange sea  masses of these ensembles.  Most of
the ensembles have the same charm sea masses, which allows us to determine the derivatives with
respect to $m'_l$ and $m'_s$ accurately.  We first convert the lattice data to ${p4s}$ units using
(tentative values of) $a m_{p4s}$ and $a F_{p4s}$.  Ensembles in which the light sea mass is tuned
close to $0.1m'_s$, shown  inside the dashed blue ellipse in \figref{ms_ml_beta6}, are then
used to determine ${\partial F^2 }/{\partial m'_s},\ {\partial M^2 }/{\partial m'_s}$ and ${\partial^2
M^2 }/{\partial m'_s \partial m_\text{v}}$.  The three derivatives with respect to $m'_s$ are found by
fitting a quadratic function to the corresponding quantities of these ensembles, as shown in
\figref{dFp4sandMp4s_dms}.

\begin{figure}[t] \vspace{-8cm} \begin{center} \begin{tabular}{l l}
\null\hspace{-40mm}\includegraphics[width=17cm]{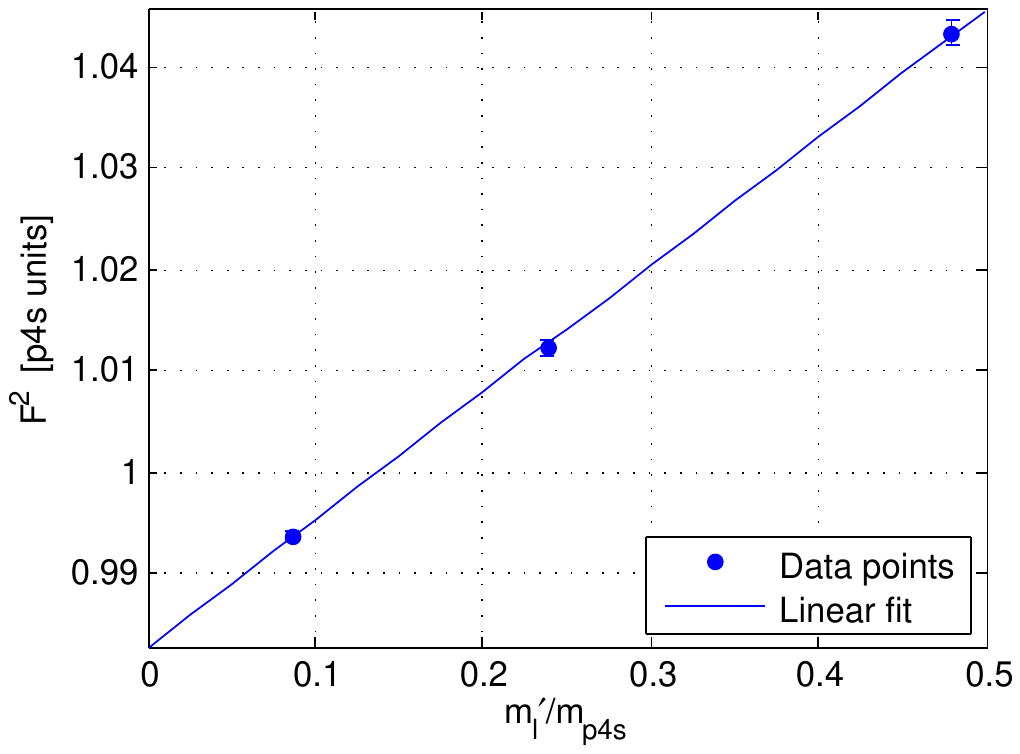} &\hspace{-90mm}
\includegraphics[width=17cm]{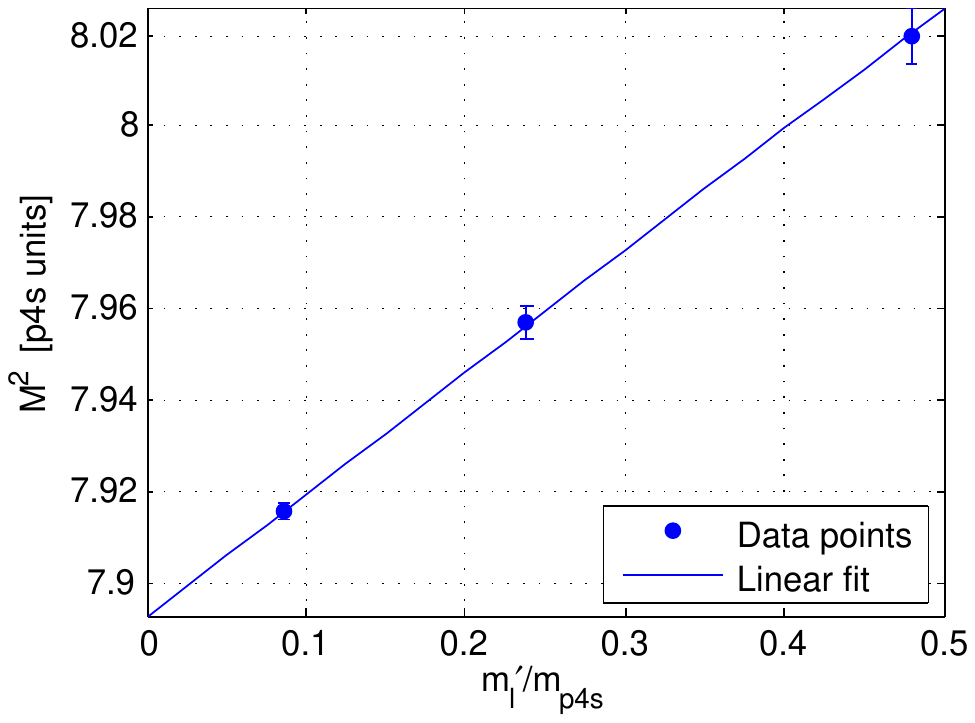} \end{tabular} \end{center} \vspace{-8.7cm} 
\caption{Data from
three ensembles with strange sea masses tuned close to $m_s$, the ensembles inside the red ellipse
in \figref{ms_ml_beta6}.   \label{fig:dFp4sandMp4s_dml} 
} \end{figure}

To calculate ${\partial F^2 }/{\partial m'_l},\ {\partial M^2 }/{\partial m'_l}$ and ${\partial^2 M^2
}/{\partial m'_l \partial m_\text{v}}$, we use the three ensembles with strange sea mass close to its
physical value, the ensembles inside the red ellipse  in \figref{ms_ml_beta6}.  We fit straight lines
to the corresponding data, as shown in \figref{dFp4sandMp4s_dml}.  Note that there are small
differences in the charm and strange sea masses of these ensembles, but they are taken into account by
a small adjustment using the derivatives with respect to $m'_s$ and $m'_c$.

The derivatives with respect to $m'_c$ cannot be calculated directly, because we do not have a group
of ensembles with different charm sea masses but equal light and strange sea masses.  So we have to
determine the charm-mass derivatives indirectly, by investigating ensembles with different charm sea
masses after adjusting for their differences in strange and light sea masses. This procedure can be
carried out using
the three ensembles available at $\approx\!0.06$ fm.  Since $m'_s$ and $m'_c$ vary by about 10\% on
these three ensembles, the lever arm is large enough to calculate the derivatives with respect
to $m'_c$.  We first use the derivatives with respect to $m'_s$ obtained at $\approx\!0.12$ fm to adjust
the data at $\approx\!0.06$ fm for mistuning of the strange sea masses, so only $m'_l$ and $m'_c$
dependence remains.  Then we calculate the $m'_c$ derivatives by passing a  function linear in both
$m'_l$ and $m'_c$ through the three data points for each quantity.  The $m'_c$ derivatives thus found
feed back into the small adjustments needed at $a\!\approx\!0.12$ fm in order to calculate $m'_l$
derivatives, as discussed in the preceding paragraph.  Our estimates of all the needed derivatives are
tabulated in Table~\ref{tab:Derivatives_mc}. 

\begin{table} \caption{ \label{tab:Derivatives_mc} The values of derivatives needed for adjusting the
data for mistunings.  All the derivatives are in $p4s$ units, and are evaluated at the valence mass
$m_\text{v}=m_{p4s}$ and at physical values of sea masses $m_l$, $m_s$, and $m_c$.  Derivatives are found
using 0.12 fm and 0.06 fm ensembles, as described in the text.}
\begin{tabular}{|l|l|l|l|l|l|} \hline
$\frac{\partial F^2 }{\partial m'_l}$    & 0.1255(32) & $\frac{\partial M^2 }{\partial m'_l}$    &
0.266(15)  & $\frac{\partial^2 M^2 }{\partial m'_l \partial m_{\rm v}}$ & 0.182(55)  \\ \hline
$\frac{\partial F^2 }{\partial m'_s}$    &0.0318(17) & $\frac{\partial M^2 }{\partial m'_s}$    &
0.0810(85)&		$\frac{\partial^2 M^2 }{\partial m'_s \partial m_{\rm v}}$ & 0.060(30) \\
\hline $\frac{\partial F^2 }{\partial m'_c}$    &0.00554(85)& $\frac{\partial M^2 }{\partial m'_c}$
&0.0209(41) & $\frac{\partial^2 M^2 }{\partial m'_c \partial m_{\rm v}}$ & 0.023(13) \\ \hline
\end{tabular} \end{table}

It is noteworthy that we can analytically determine the first order derivatives with respect to $m'_c$
by integrating out the charm quark for  processes that occur at energies well below its mass.  By
decoupling \cite{Appelquist:1974tg}, the effect of a heavy (enough) sea quark on low-energy quantities
occurs only through  the change it produces in the effective value of $\Lambda_{\text{QCD}}$ in the
low-energy (three-flavor) theory~\cite{Bernreuther:1981sg}.   (For a pedagogical discussion see Sec.~1.5 of
\rcite{MAN_WISE}.) 
Thus, assuming $m'_c$ is heavy enough, we may calculate the $m'_c$ derivatives
of  any quantity that is proportional to $\Lambda_{\text{QCD}}$, where the proportionality constant is some
pure number, independent of the light quark masses.  Examples of such quantities are the LEC $B$ in
\eq{x-defs} and the light-light decay constant in the chiral limit, $f$.  At leading order in
weak-coupling perturbation theory,  one then obtains (see Eq.(1.114) in \rcite{MAN_WISE}),
\begin{equation} \frac{\partial B}{\partial m'_c} = \frac{2}{27}\frac{B}{m'_c}\ , \qquad \frac{\partial
f}{\partial m'_c} = \frac{2}{27}\frac{f}{m'_c}\ .  \end{equation} At the nonzero values of $m_\text{v}$, $m'_l$,
and $m'_s$ at which we need to evaluate the derivatives in \eq{derivatives}, there are corrections to
these expressions.  However, chiral perturbation theory suggests that such corrections are relatively
small.   At the relevant light masses, we therefore expect \begin{eqnarray} \frac{\partial
F^2}{\partial m'_c} = 2F \frac{\partial F }{\partial m'_c} \approx \frac{4}{27}\frac{ F^2 }{ m'_c} =
0.00504\ \ [p4s\ {\rm  units}],  \eqn{mc-deriv-F}\\ \frac{\partial M^2 }{\partial m'_c} \approx
2m_{p4s}\frac{\partial B}{\partial m'_c} \approx \frac{2}{27}\frac{ M^2 }{ m'_c}=  0.01998\ \ [p4s\ {\rm
units}], \eqn{mc-deriv-M} \end{eqnarray} which agree with our numerical results within 10\%; see
Table~\ref{tab:Derivatives_mc}.  Indeed, the fact that the agreement is this close is probably due to
chance, especially for the derivative of the decay constant:  Our argument has neglected the
difference between $f$ and $F_{p4s}$, but that difference is  $\sim\!40\%$.  

Having the required derivatives, we now iteratively adjust for mistunings.  We first compute $a
m_{p4s}$ and $a F_{p4s}$, then adjust the data, and repeat the entire process two more times.
The values of $a
m_{p4s}$ and $a F_{p4s}$ have then converged to well within their statistical errors.
The
results for the lattice spacing $a$ and $am_s$ are listed in Table~\ref{tab:results_a}.  The error
estimates  of these quantities will be discussed below.  Our investigation shows that the errors in
the derivatives change $a$ and $am_s$ by less than their statistical errors, so those errors are not
included in the analysis.  

Comparing \tabref{results_a} with \tabref{tunedmasses}, which uses $f_{\pi^+}$
to set the scale, we see significant differences
at the coarser lattice spacings, but not at the finest spacing.  This is as expected for two different schemes, 
which should only agree exactly in the continuum limit.

\begin{table} \caption{ \label{tab:results_a} Lattice spacing $a$ and $am_s$, as 
a function $\beta$, in the $p4s$ mass-independent scale-setting scheme.} 
\begin{tabular}{|c|l|} \hline $\beta$ = 5.8& $a = 0.15305(17)_{\rm
stat}({}^{+46}_{-23})_{a^2\,{\rm extrap}}(29)_{\rm FV}(4)_{\rm EM}$ fm \\ & $a m_s = 0.06863(16)_{\rm
stat}({}^{+43}_{-24})_{a^2\,{\rm extrap}}(26)_{\rm FV}(7)_{\rm EM}$ [lattice units] \\ \hline $\beta$
= 6.0& $a = 0.12232(14)_{\rm stat}({}^{+36}_{-19})_{a^2\,{\rm extrap}}(23)_{\rm FV}(3)_{\rm EM}$ fm \\
& $a m_s = 0.05304(13)_{\rm stat}({}^{+33}_{-18})_{a^2\,{\rm extrap}}(20)_{\rm FV}(6)_{\rm EM}$
[lattice units] \\ \hline $\beta$ = 6.3& $a = 0.08791(10)_{\rm stat}({}^{+26}_{-13})_{a^2\,{\rm
extrap}}(17)_{\rm FV}(2)_{\rm EM}$ fm \\ & $a m_s = 0.03631(9)_{\rm stat}({}^{+23}_{-13})_{a^2\,{\rm
extrap}}(14)_{\rm FV}(4)_{\rm EM}$ [lattice units] \\ \hline $\beta$ = 6.72& $a = 0.05672(7)_{\rm
stat}({}^{+17}_{-9})_{a^2\,{\rm extrap}}(11)_{\rm FV}(1)_{\rm EM}$ fm \\ & $a m_s = 0.02182(5)_{\rm
stat}({}^{+14}_{-8})_{a^2\,{\rm extrap}}(8)_{\rm FV}(2)_{\rm EM}$ [lattice units] \\ \hline
\end{tabular} \end{table}

\subsubsection{Chiral-continuum fits to D system}  \label{sec:CHIPT-fits}

So far, we have introduced eight fit parameters related to discretization effects
($c_1$, $c_2$, $c_3$, $c_4$, $L_{\rm v \delta}$, $L_{\rm s\delta}$, $k_{1\delta}$, and $k_{2\delta}$) 
and three parameters related to taste-violation effects ($L_a$, $\delta'_{\rm A}$, and $\delta'_{\rm V}$).
The latter parameters appear at NLO in \schpt\ and must be kept since our expansion is supposed to be 
completely systematic through NLO.
This is not the case for the former parameters;
several of them ($c_2$, $c_3$,  $c_4$, $L_{\rm v \delta}$,  $L_{\rm s\delta}$, and $k_{2\delta}$)  are 
formally NNLO and may be dropped. 
We indeed get acceptable fits when some of these parameters are dropped, especially if the $a
\approx0.15\;$fm data are omitted. 
In order to see the effects of these parameters, we present the results of two fits, with different sets of 
parameters,
to data at the three finer lattice spacings,
and we study the extrapolation of the chiral fit back to the coarsest lattice spacing ($a\approx0.15\;$fm,
$\beta=5.8$).

\begin{figure}[t]
\null\vspace{-5mm}
\hspace{-10mm}\includegraphics[width=17.2cm]{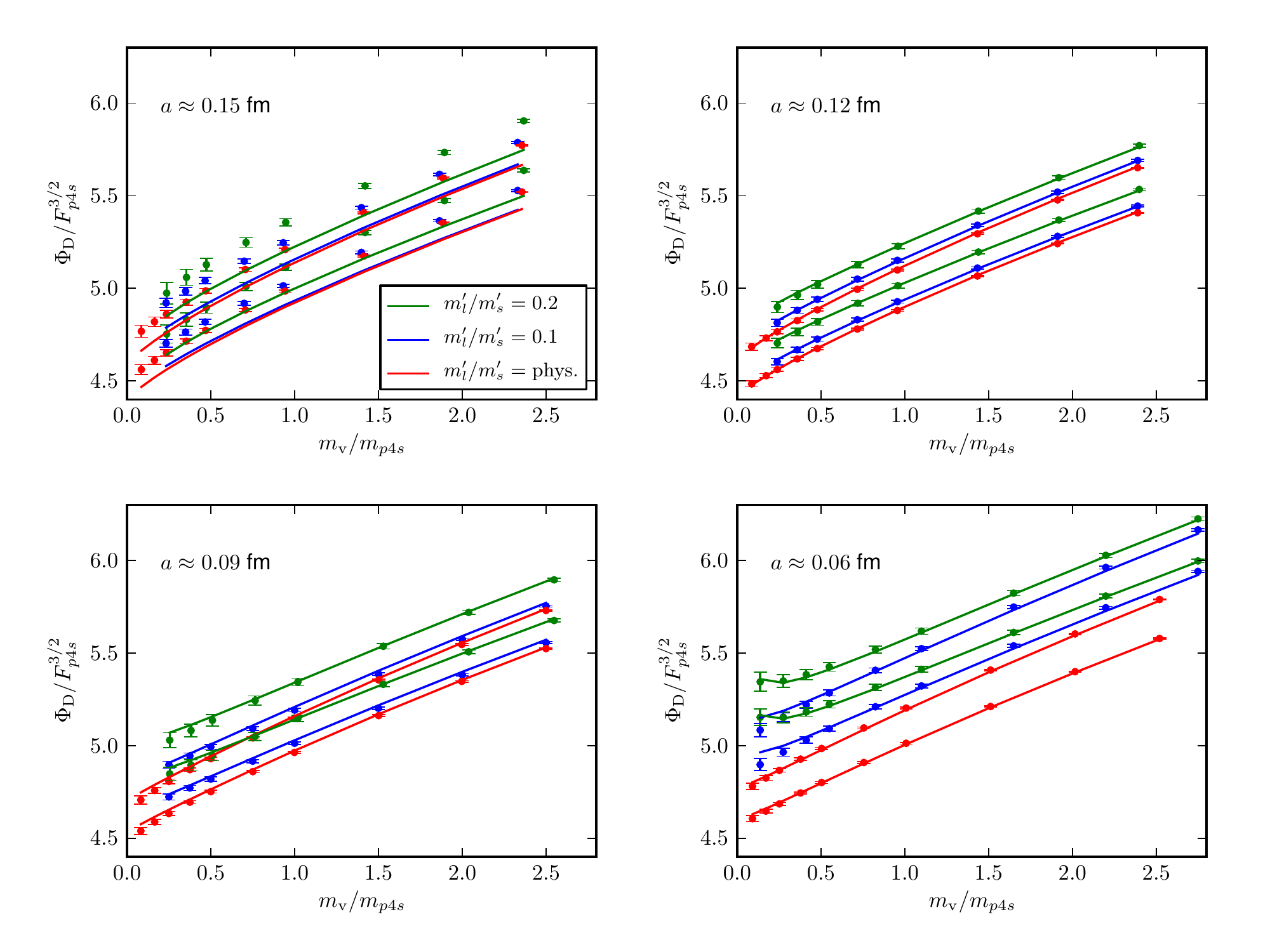}
\vspace{-5mm}
\caption{Simultaneous chiral fit to $\Phi_D$  as a function of $m_{\rm v}$, the valence-quark mass (in
units of $m_{p4s}$), at the three finer lattice spacings. The $a\approx\!0.15$ fm ($\beta=5.8$)
data is not included in the fit,
although the data  and the extrapolation of the chiral fit to it
are shown at the left in the top row. At the right of the top row we show the $a\approx0.12$ fm 
($\beta=6.0$) data,
and in the bottom row are 
$a\approx\!0.09$ fm ($\beta=6.3$, left) and $a\approx\!0.06$ fm ($\beta=6.72$, right).
The colors denote different light sea-quark masses, as indicated.  For each color
there are two lines, one for heavy valence-quark mass $\approx m'_c$ (higher line), and one for $\approx 0.9 m'_c$. 
In this fit, $g_\pi$ is fixed to $0.53$.   The fit has $\chi^2/{\rm dof}=339/293$, giving $p=0.033$.
\label{fig:chiral-no5p8-fit}}
\end{figure}

\Figref{chiral-no5p8-fit} shows a fit to partially quenched data at the three finer lattice spacings.   
(The $a\approx0.15\;$fm data are omitted.) Among the introduced fit parameters related to discretization effects,
only $c_1 $ in \eq{Phi0-form} and $k_{1\delta} $ in \eq{L-form} are taken as free parameters in this fit, and the 
others are set to zero.
This fit gives $p=0.033$, and as illustrated in \figref{chiral-no5p8-fit}, the extrapolation 
of the fit to the coarsest lattice spacing does not follow the corresponding data points.
We note that this fit and all other chiral fits in this paper include additional data (not shown) from ensembles
at $a\approx0.12\;$fm ($\beta=6.0$) either with $m'_s$ lighter than physical, or 
with volumes $24^3\times 64$ and $40^3\times64$, which were generated to check finite volume effects. (See \tabref
{ensembles}.)
Moreover, it is important to realize that the biggest source of variation in the data in the four plots 
shown in \figref{chiral-no5p8-fit} is not discretization errors,
but mistunings of the strange and, most importantly, charm-quark masses.

\begin{figure}[t]
\null\vspace{-5mm}
\hspace{-10mm}\includegraphics[width=17.2cm]{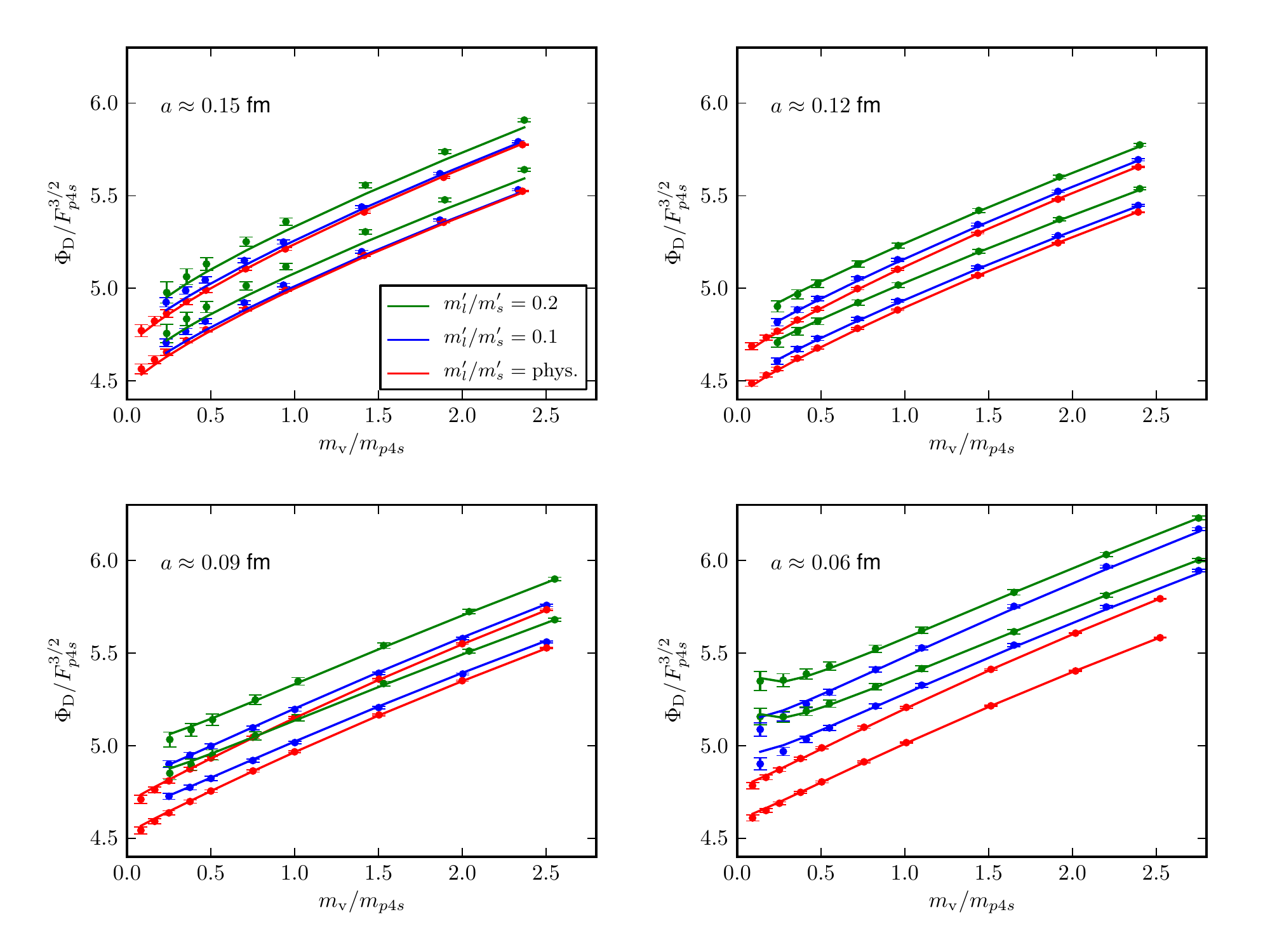}
\vspace{-5mm}
\caption{Simultaneous chiral fit to $\Phi_D$  as a function of $m_{\rm v}$ at the three finer lattice spacings.
Similar to the fit in \figref{chiral-no5p8-fit},  but with three extra fit parameters: $c_2$, $c_3$, and $c_4$.  
This fit has $\chi^2/{\rm dof}=239/290$, giving $p=0.986$. 
\label{fig:chiral-no5p8-w36-w23w25-fit}}
\end{figure}

Adding $c_3\alpha_S (am_Q)^2 +c_4(am_Q)^4$ to the analytic terms in \eq{chiral-form},
as well as including $c_2 $ in \eq{Phi0-form}, we get a new fit to the partially quenched data at the three finer lattice spacings.
By including these three extra parameters, an excellent fit is achieved, as shown in \figref{chiral-no5p8-w36-w23w25-fit}, 
and extrapolation of the fit to the coarsest lattice spacing gives lines
that pass relatively well through the corresponding data points. 
This comparison makes clear that higher-order discretization errors are important for the HISQ data, in 
which the leading-order discretization effects are suppressed. 

We have a total of 18 acceptable ($p>0.1$) versions of the continuum/chiral fits.
Five of the fits drop the $a\approx0.15$ fm ensembles; the rest keep those ensembles.
The chiral coupling $f$ is generally set to $f_{\pi^+}$, except for two fits with the coupling constant set to $f_{K^+}$. The LEC $g_\pi$ is usually fixed to either its nominal value or to $1\sigma$ below its nominal value, 
however it is allowed to be a free parameter in four of the fits.  
The LEC $B$ in \eq{x-defs} is generally
determined for each lattice spacing separately by fitting all data  for the squared meson mass $M^2$ vs.\ the sum of the valence masses to a straight line.
(At $a \approx 0.12\ {\rm fm}$ only the ensembles with strange sea masses close to its physical mass are included in the fit.) 
However, in two versions of the chiral fits, $B$ is determined from just the data on the physical-mass ensembles at each lattice spacing.

Another difference among the fits is how we determine the strong coupling $\alpha_S$ in discretization terms such as those
with coefficients $c_1$ and $c_3$. 
 Since the coefficients are free parameters, all that we actually need in the fits is the relative value of $\alpha_S$ at a given coupling $\beta$ compared to its value at a fixed, fiducial coupling $\beta_0$.  
In most of the fits, we have used measured light-light pseudoscalar taste splittings to fix this relative
value, as in \eq{alphas-taste}.
An alternative, which is used in two of our fits, is to use for $\alpha_S$ the coupling $\alpha_V$,  
determined from the plaquette \cite{Davies:2002mv}.  The scale for $\alpha_V$
is taken to be $q^*=2.0/a$.  Note that the NLO perturbative corrections to $\alpha_V$ 
have not
been calculated for the HISQ action, so we use the result for the asqtad action.  Since the
$n_f$ dependence of the NLO result is small, we expect the difference to have negligible effects
on the results of the fit. This expectation can be tested by, for example, flipping the sign of the
$n_f$ term in the asqtad result, which is likely a much bigger change than would actually come from
changing from asqtad to HISQ.  When we do this, we find that the results change by amounts comparable to or smaller than
the statistical errors, and significantly smaller than the total systematic errors.
Similar, but usually smaller, changes result from replacing  $q^*=2.0/a$ with $q^*=1.5/a$, which is another
reasonable choice, as discussed in \rcite{HISQ_CONFIGS}.

We have introduced eight fit parameters related to discretization effects 
($c_1$, $c_2$, $c_3$, $c_4$, $L_{\rm v \delta}$, $L_{\rm s\delta}$, $k_{1\delta}$, and $k_{2\delta}$), 
but it is not necessary to keep all of them to get an acceptable fit. 
Dropping some of these parameters, we have different continuum/chiral fits with the number of parameters 
ranging from 23 to 28.   We may also choose to 
constrain, with priors, the LECs in higher-order (NNLO and NNNLO) analytic terms to be $\cO(1)$ in 
natural units (as explained following \eq{x-defs}).  (Through NLO, where we have the complete chiral
expression, including logarithms, we always leave the LECs $\Phi_0$, $L_\text{s}$, $L_\text{v}$, and $L_a$ completely unconstrained,
while $g_\pi$, $\delta'_A$, and $\delta'_V$ are constrained by independent analyses as discussed above.)
We may similarly constrain the coefficients of
discretization terms to be $\cO(1)$ when the terms are written in terms of a reasonable
QCD scale (which we take, conservatively, to be 600 MeV). 
Among the 18 fits we consider, some have higher-order chiral terms and discretization terms completely
unconstrained, and others constrain either the chiral terms, or  the discretization terms, or both.

In \eq{Phi0-form}, $m_Q$ denotes the valence charm mass. To take into account the physical effects of the charm sea 
masses we can introduce a parameter $k'_1$
to \eq{Phi0-form}:
 \begin{equation}\eqn{Phi0-form-prime}
\Phi_0\to \Phi_0 \left(1 + k_1\frac{\Lambda_{\rm HQET}}{m_Q}+ k_2\frac{\Lambda_{\rm HQET}^2}{m_Q^2}+ k'_1\frac{\Lambda_{\rm HQET}}{m'_c}\right)\Big(1+ c_1\alpha_S (a\Lambda)^2 +c_2  (a\Lambda)^4 
\Big)\;,
\end{equation}
where $m'_c$ is the mass of the charm mass in the sea.
One of our 18 fits adds the  parameter  $k'_1$.
Further, discretization errors 
coming from the charm sea masses can be
included  by adding $c'_3\alpha_S (am'_c)^2 +c'_4(am'_c)^4$ to the analytic terms
in  \eq{chiral-form}, and one of the fits makes that addition.   It is interesting to note that it is possible
to obtain another acceptable fit in which $c_2$ in \eq{Phi0-form} is restricted by priors to be much 
smaller than its value in the central fit, but the $c'_3$ and $c'_4$ terms are added.  
This shows that our lattice data cannot distinguish in detail between various sources
of higher-order discretization effects.  However, the results in the continuum limit are rather insensitive to these differences. 

Since all 18 fits considered have acceptable $p$ values and give correction 
terms reasonably consistent with expectations from chiral perturbation theory and power counting,
whether or not such terms are constrained, we have no strong reason to choose one fit or groups of
fits as preferred in comparison to the rest. We therefore choose our ``central fit'' simply by requiring
that it be a fit to all ensembles and that it give results for $\Phi_{D+}$ and $\Phi_{D_s}$ that are as close as possible to the center of the
histograms for these quantities from all the fits and from all 
systematic variations in the inputs (\ie from the ``continuum extrapolation'' column in 
\tabref{error_budget_1}).
This central fit has  27 free parameters, with $g_\pi$ fixed to 1-sigma below its nominal value, and with
the $k'_1$, $c'_3$, and $c'_4$ terms discussed in the previous paragraph dropped, but all discretization terms aside from $c'_3$ and $c'_4$ kept. 
In the central fit, $c_2$ in \eq{Phi0-form} is equal to $1.3$  with $\Lambda=600\; {\rm MeV}$; while the HQET parameters
are 
$k_1=-1.0$\ and  $k_2=0.5$, with 
$\Lambda_\text{HQET}=600$ MeV.

\begin{figure}[t]
\null\vspace{-5mm}
\hspace{-8mm}\includegraphics[width=16.2cm]{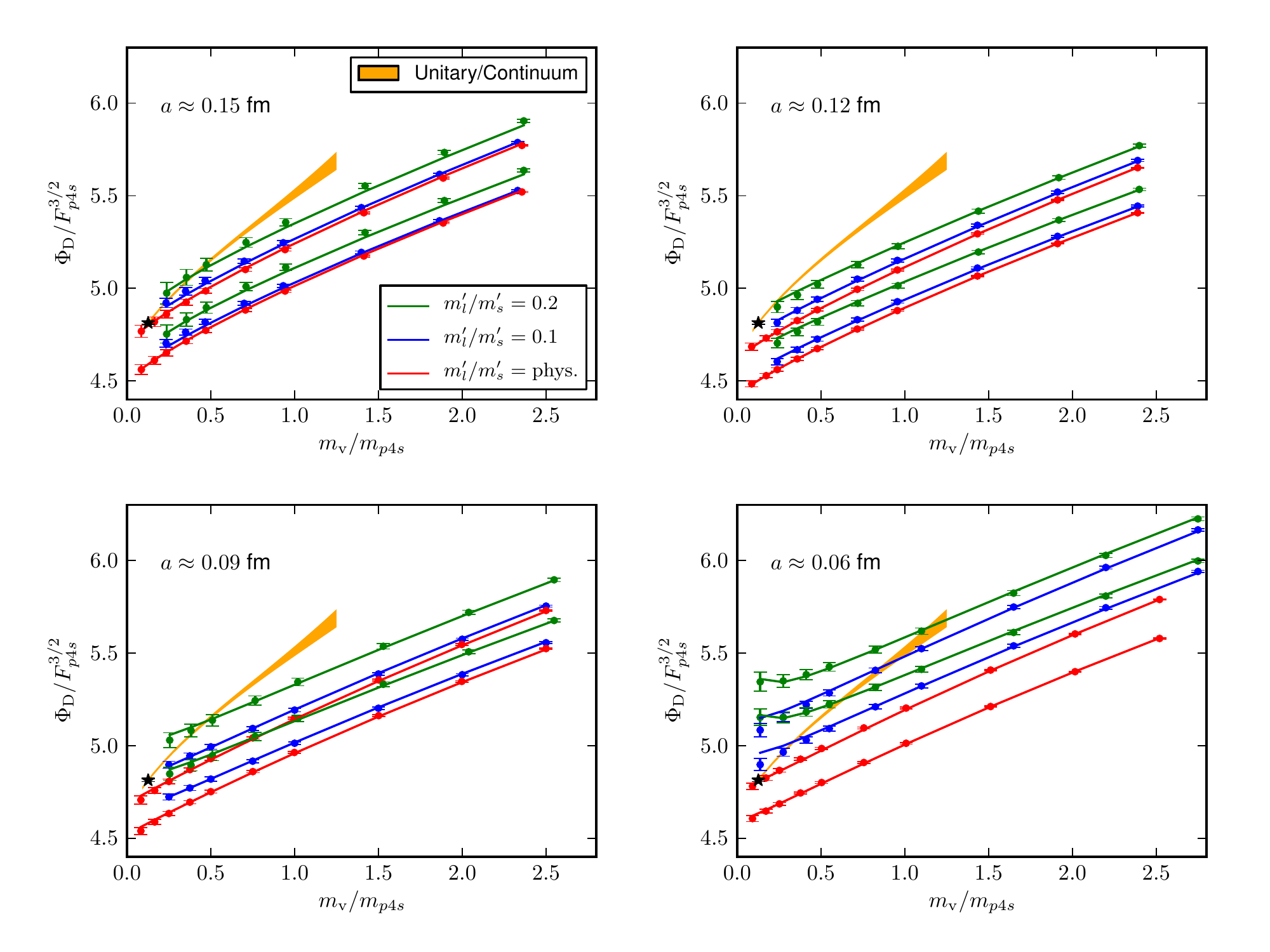}
\vspace{-5mm}
\caption{Simultaneous chiral fit to $\Phi_D$  as a function of $m_{\rm v}$, the valence-quark mass (in
units of $m_{p4s}$), at all four lattice spacings: $a\approx0.15$ fm and $0.12$ fm (top row), and
$0.09$ fm and $0.06$ fm (bottom row).  
This fit has $\chi^2/{\rm dof}=347/339$, giving $p=0.36$.  
In the fit lines for each ensemble, the light valence-quark mass varies, with all sea-quark masses held fixed.
The orange band, labeled as ``unitary/continuum,'' is identical in each panel. It
 gives the result after extrapolating to the continuum, setting the light valence-quark and sea-quark masses equal
(up to the small difference between $m_d$ and $m_l = (m_u+m_d)/2$),
and adjusting the strange and charm masses to their physical values.
The width of the band  shows the statistical error coming from the fit.
The black bursts indicate the value of $\Phi_{D^+}$ at the physical
light-quark mass point. 
\label{fig:chiral-fit}}
\end{figure}

\Figref{chiral-fit} shows our central fit to partially quenched data at all four lattice spacings. 
Extrapolating the parameters to the continuum, adjusting the  strange
sea-quark mass and charm valence- and sea-quark masses to their physical values,
and setting the light sea-quark mass equal to the light valence mass
(up to the small difference between $m_d$ and $m_l = (m_u+m_d)/2$)
gives the orange band.  Putting in the physical light-quark  mass then gives the black burst, which is 
the result for $\Phi_{D^+}$.  Note that the effect of isospin violation in the valence quarks is included in our
result. The effect of isospin violation in the sea has not been included, but we may easily estimate its size by 
putting in our values for $m_u$ and $m_d$ (instead of the average sea mass $m_l$) 
 in \eq{chpt-form} and in the
NNLO and NNNLO analytic terms.  This results in a change of only 0.01\% in $f_{D^+}$, and a still smaller change in $f_{D_s}$.

The width of the band shows the statistical error coming from the fit, which is only part of the total statistical
error, since it does not include the statistical errors in the inputs of the quark masses and the lattice scale.  To determine the
total statistical error of each output quantity, we divide the full data set into 100 jackknife resamples. 
The complete calculation, including the determination of the inputs, is performed on each resample, 
and the error is computed as usual from the variations over the resamples. 
(For convenience, we kept the covariance matrix fixed  to that from the full data set, rather than recomputing it for each resample.)
Each jackknife resample drops approximately ten consecutive stored configurations (50 to 60 trajectories) from each ensemble with $\approx\!1000$ configurations.
This procedure  controls for autocorrelations, since all our measures of the autocorrelations of these quantities indicate that 
they are negligible after four or eight consecutive configurations. For the 
physical-mass 0.06 fm ensemble with 583 configurations, we are forced to drop only about six
consecutive stored configurations at a time.
Our expectation is that the effect of any remaining autocorrelations, while perhaps not completely negligible, is small compared to other sources of error.    
The total statistical errors computed from the jackknife procedure are only about 10\% larger than the statistical error from the chiral/continuum fit, indicating that the inputs
are statistically quite well determined. The same procedure is performed to find the total statistical error of $a$ and $am_s$
at each lattice spacing.

\begin{figure}[t]
\begin{center}
\null\vspace{-08mm}
\includegraphics[width=16cm]{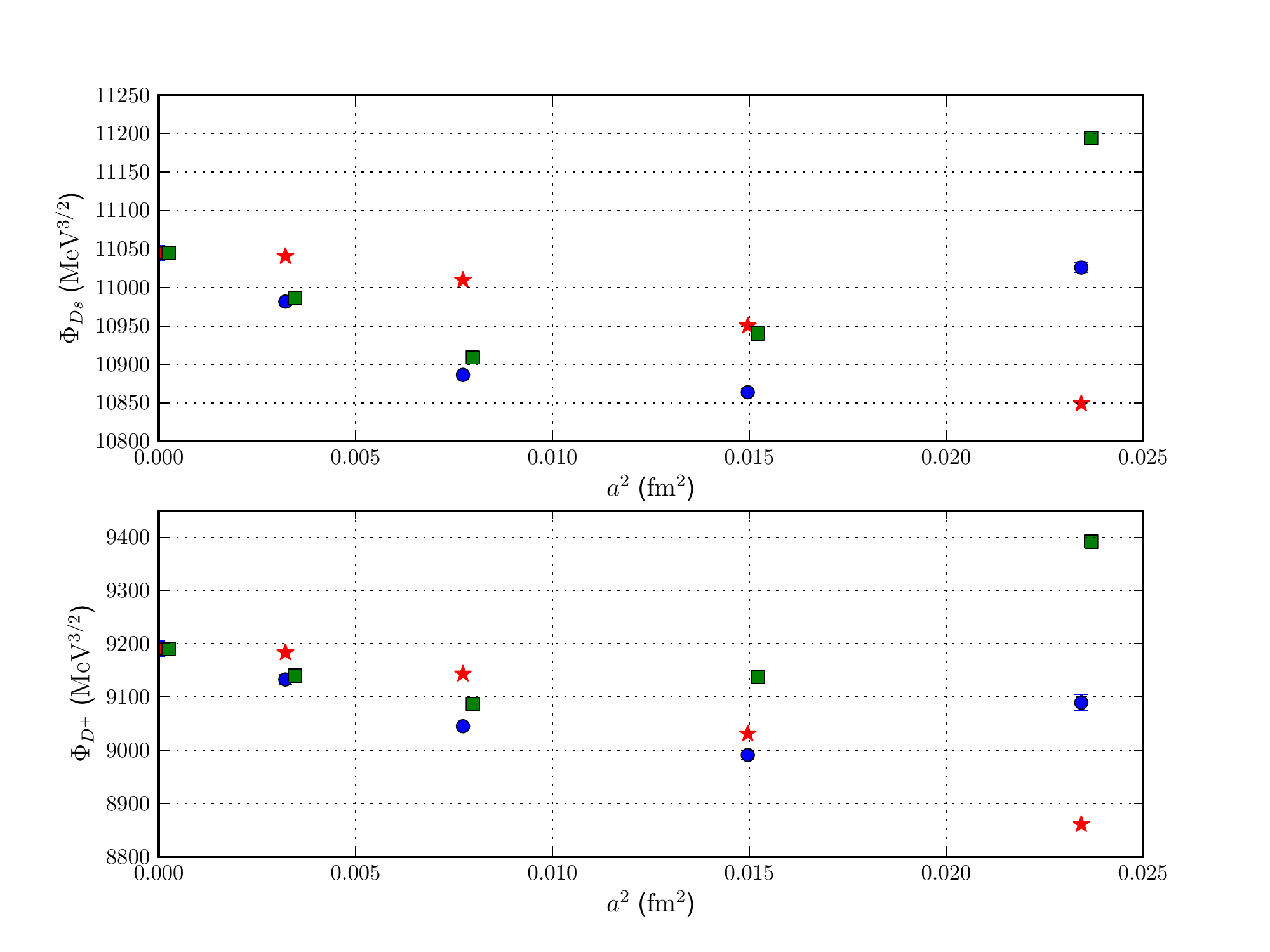}
\end{center}
\vspace{-8mm}
\caption{
Lattice spacing dependence of $\Phi_{D^+}$ and $\Phi_{D_s}$. The blue circles show
the lattice data, after adjustment for mistunings of valence- and sea-quark masses.  The red stars
show the modification of each continuum value by the $a^2$ dependence of the chiral logarithms, 
while the green squares show the corresponding modification by the $a^2$ dependence induced by the fit parameters. 
Red stars and green squares overlap at $a^2=0$ (only the green square is visible).  
Neglecting small cross terms, the deviation of the blue circles from the continuum value are given 
by the algebraic sum of the deviations of the red stars and the green squares.
\label{fig:a2dep}}
\end{figure}

\Figref{a2dep} illustrates how data for $\Phi_{D^+}$ and $\Phi_{D_s}$ depend on lattice spacing after 
adjustment to physical values of the quark masses (blue circles).  There is a 2--3\% variation between 
these points and the continuum value (green square at $a^2=0$).  Note that there is clear curvature
in the plot, evidence of significant $a^4$ terms in addition to the formally leading $\alpha_S a^2$ terms.
Both the small absolute size of the errors, and the competition between formally leading and subleading terms,
are typical of highly improved actions such as the HISQ action.
The red stars show the contribution from the chiral logarithms (with known taste splittings) to the $a^2$ dependence 
of the chiral fit function. The green squares show the corresponding contribution from the analytic fit parameters.
The two effects are of comparable magnitudes but the relative sign changes with lattice spacing; both are needed to describe 
the  $a^2$ dependence of the data.    

\subsubsection{Continuum extrapolation and systematic uncertainties}\vspace{-2.0mm} \label{sec:SYSERRS}

To determine the systematic error associated with the continuum extrapolation (and chiral interpolation) 
of the charm decay constants in the chiral perturbation theory analysis, we rerun the analysis
with alternative continuum/chiral fits, and with alternative inputs that come from different continuum extrapolations of the
physical-mass analysis, listed in the ``continuum extrapolation'' column in 
\tabref{error_budget_1}.

\begin{figure}[t]
\begin{center}
        \null\vspace{-08mm}
        \begin{tabular}{l l}
        \null\hspace{-2mm}\includegraphics[trim=0.5in 0 0.3in 0, clip, width=8.3cm]{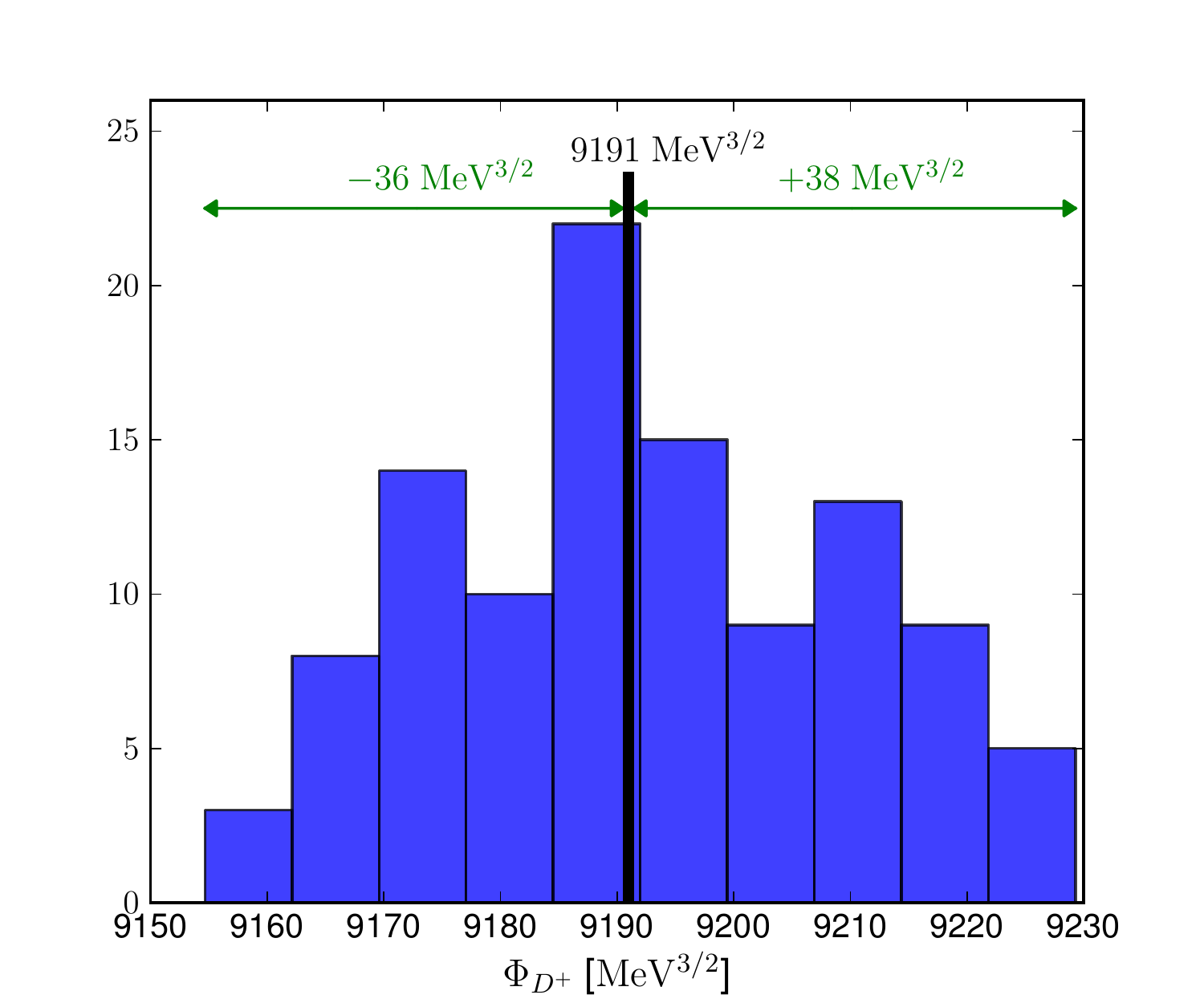}
        &\hspace{-2mm}
        \includegraphics[trim=0.5in 0 0.3in 0, clip, width=8.3cm]{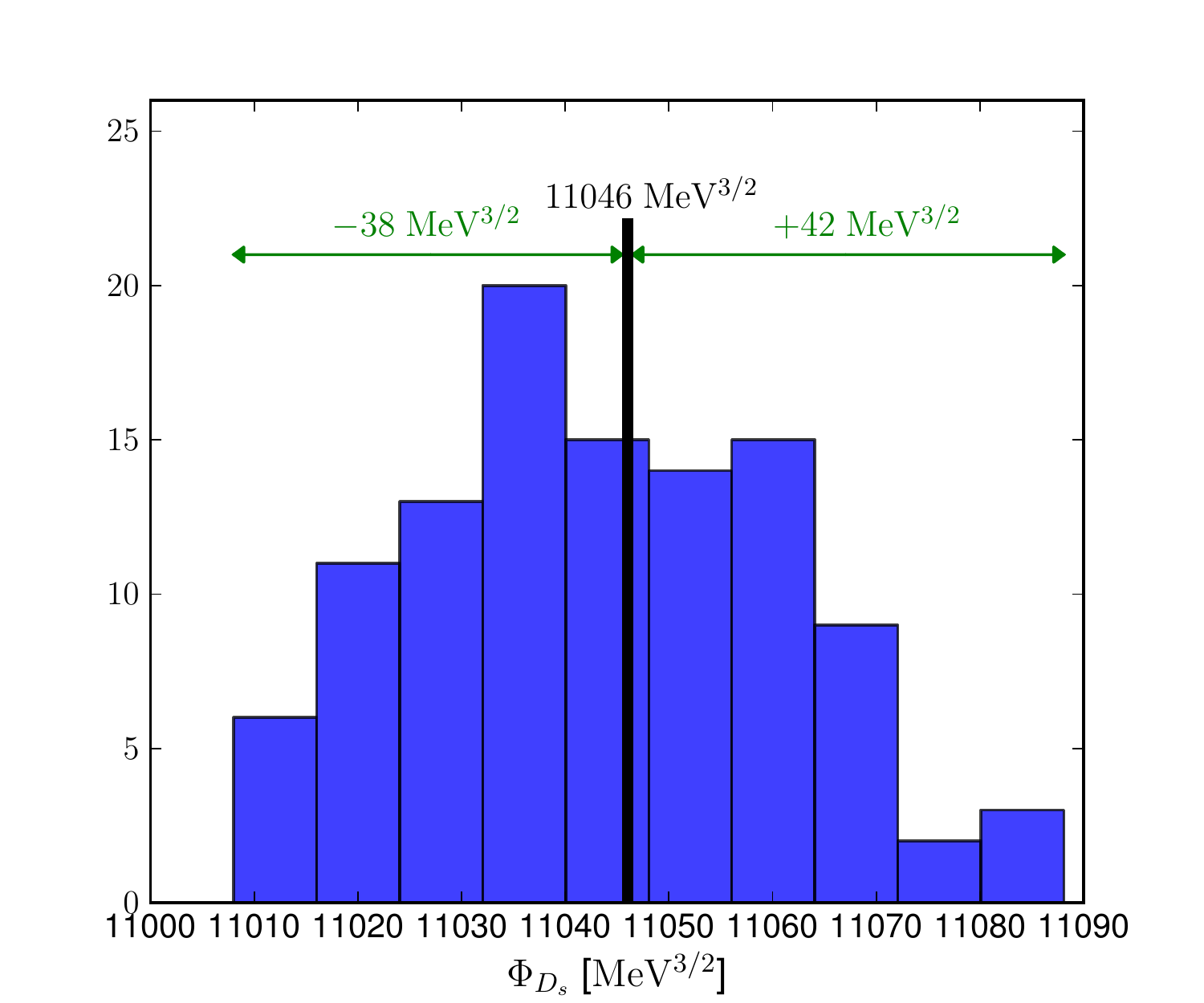}
        \end{tabular}
        \end{center}\vspace{-08mm}
\caption{
Histograms of $\Phi_{D^+}$ and $\Phi_{D_s}$ values obtained from various versions of the continuum/chiral extrapolation and various inputs of quark masses and scale values from the physical-mass analysis.
Our central fit gives $\Phi_{D^+}=9191\; {\rm MeV}^{3/2}$ and $\Phi_{D_s}=11046\; {\rm MeV}^{3/2}$; those values are marked
with vertical black lines.
At the top of each histogram, we show the range taken as the systematic error of the self-contained chiral analysis of the current section.
\label{fig:hist_phi}}
\end{figure}

As mentioned above, we have a total of 18 acceptable versions of the continuum/chiral fits.
We also have the six versions of the continuum extrapolations used in the physical-mass analysis
that leads to the inputs of quark 
masses and the lattice scale.
This gives a total of 108 versions of the analysis.  
Histograms of the 108 results for $\Phi_{D^+}$ and $\Phi_{D_s}$  are shown in \figref{hist_phi}.  Conservatively, 
we take the maximum difference seen in these results with our central values as the ``self-contained'' estimate of the continuum 
extrapolation errors within this chiral analysis. 
The central fit is chosen to give results that are close to the centers of the histograms, which results in more symmetrical 
error bars than in the preliminary analysis reported in \rcite{LATTICE13_FD}.  Note that the ``acceptable'' fits 
entering the histograms all have $p> 0.1$.  If the cutoff is instead taken to be $p> 0.05$, the additional fits 
allowed would not change the error estimates.  However a cutoff of $0.01$ or lower would give some 
additional outliers that would increase the width of the histograms.

As mentioned in \secref{CHIPT}, the chiral fits tend to pull $g_\pi$ to low values.  We can now look at this effect quantitatively.  The
central fit,  which has $g_\pi$ fixed to 0.45, $1\sigma$ below its nominal value of $0.53$, has $p=0.36$ and gives $\Phi_{D^+}=9191(14)\:\textrm{MeV}^{3/2}$, where the statistical error  comes only from the fit and not from the errors in the inputs. Allowing $g_\pi$ to be a free
parameter, with prior range $0.53(8)$, we find  $g_\pi=0.26(5)$, about  $3\sigma$ below its nominal value, and $p=0.71$.  However, 
 $\Phi_{D^+}$ then is $9184(15)\:\textrm{MeV}^{3/2}$, a change of only half the statistical error, and much less than the systematic error from the
range over the results of all chiral/continuum fits.  Alternatively, fixing $g_\pi$ to its nominal value gives $\Phi_{D^+}=9195(13)\:\textrm{MeV}
^{3/2}$,  $p=0.16$.  We can also consider the effect in fits that drop the data with $a\approx0.15$ fm and consequently use fewer lattice-spacing-dependent 
parameters. When $g_\pi$ is a free parameter with prior range $0.53(8)$, we find  $g_\pi=0.37(6)$,  $2\sigma$ below the nominal value, and 
 $\Phi_{D^+}=9189(12)\:\textrm{MeV}^{3/2}$, $p=0.37$.  The corresponding fits with $g_\pi$ fixed to its nominal value or one sigma below that value give
  $\Phi_{D^+}=9196(13)\:\textrm{MeV}^{3/2}$, $p=0.18$ and $\Phi_{D^+}=9192(12)\:\textrm{MeV}^{3/2}$,  $p=0.30$, respectively.  Thus,
 the systematic error on $\Phi_{D^+}$ associated with the value of $g_\pi$ is small compared to our other errors.  The systematic error 
 from $g_\pi$ on
 $\Phi_{D_s}$ is a factor of two smaller still.

The fact that a wide range of $g_\pi$ values give good fits indicates that our data has little to say about the physical value of that parameter.  
Indeed, even fits with $g_\pi$ set equal to zero have very good $p$ values, and do not change $\Phi_{D^+}$ by more than one statistical $\sigma$.
Such a fit that includes all data gives $\Phi_{D^+}=9180(13)\:\textrm{MeV}^{3/2}$, $p=0.83$, and one that drops the data with $a\approx0.15$ fm
gives $\Phi_{D^+}=9181(13)\:\textrm{MeV}^{3/2}$, $p=0.52$.

In practice, the NLO finite volume corrections are included in our fit function, \eq{chiral-form}, when it is applied to the data,
and the volume is sent to infinity when the continuum results are extracted.  We may conservatively
estimate the residual finite volume error
in the heavy-light data either by turning off all finite volume corrections and repeating the fit, or by using the current fit to
find the size of the NLO finite volume correction on our most-important, 0.06 fm physical-mass ensemble. Yet another way to
make the estimate is by direct comparison of our results on the $32^3\times 64$, $\beta=6.0$,
$m'_l/m'_s=0.1$ ensemble
(which is similar in physical size to our other $m'_l/m'_s=0.1$ ensembles) and the $40^3\times 64$, $\beta=6.0$,
$m'_l/m'_s=0.1$ ensemble. All three methods indicate that there are negligible direct finite volume effects in the heavy-light lattice data. Nevertheless, there are non-negligible finite volume effects in our final answers, which appear due to the scale setting in the light-quark sector through, ultimately, $f_{\pi^+}$.  (The value of $F_{p4s}$ in physical units that we use comes by comparison with
$f_{\pi^+}$.)  We then propagate the errors in  the inputs through our analysis.  Electromagnetic errors in the light quark masses are similarly propagated through our analysis.

Results for $\Phi_{D^+}$, $\Phi_{D_s}$ and their ratio at various values of the mass ratio of light to strange sea quarks are shown
in \tabref{results_Phi}; only the top subsection of the table gives physical results.  
Note that the valence masses  do not vary in the three different subsections of the table, so changes in results show
only the effects of the light sea mass. 
The EM error associated with the masses of 
the heavy-light mesons, which we call ``EM3,'' is
{\it not}\/ included in any of the quoted EM errors in the table.  As explained in \secref{FV-EM},
that is because the error cancels to good approximation when one extracts
the decay constants $f_{D^+}$, $f_{D_s}$ from $\Phi_{D^+}$, $\Phi_{D_s}$.  One should use the experimental masses
$M_{D^+}=1869.62$ MeV, $M_{D_s}=1968.50$  MeV \cite{PDG} in this extraction; the experimental errors in these masses are
negligible at the current level of precision.

 To quantify the effect of isospin violations, we also report $\Phi_{D}$ and $\Phi_{D^+}-\Phi_{D}$, where $
\Phi_{D}$ is the value 
of $\Phi$ in the isospin limit, when the light valence mass  is equal to $m_l = (m_u+m_d)/2$ instead of $m_d$.   
In this case, the EM errors in the heavy-light meson masses do affect the errors in the corresponding
decay constant difference because of the difference between the
EM effect in the charged $M_{D^+}$ and in the neutral $M_{D^0}$, which are averaged to obtain $M_D$.
We estimate this error when we quote $f_{D^+}-f_D$ below.

\begin{table}
\caption{ \label{tab:results_Phi} Results for $\Phi$ from the chiral analysis, for three choices of the light sea mass $m'_l$. $\Phi_{D}$ is the value 
of $\Phi$ when the light valence mass $m_\text{v}=m_l \equiv (m_u+m_d)/2$.  Valence masses here are always 
 taken to be the physical values $m_d$, $m_s$ or $m_l$, independent of the value of $m'_l$, and the strange sea mass is always physical 
 ($m'_s=m_s$).  In the EM errors on these quantities, we have not included the ``EM3'' error coming from the EM effects on the
 masses of the corresponding heavy-light mesons. Such errors largely cancel when we compute $f_{D^+}$ and $f_{D_s}$
 from $\Phi_{D^+}$  and $\Phi_{D_s}$ using the experimental meson masses.  For $\Phi_D$ and $f_D$, the situation is more complicated --- see text. 
 The negative central value of $\Phi_{D^+}-\Phi_{D}$ for $m'_l/m_s = 0.2$ is an effect of partial quenching, but note that the systematic
 errors are large in this case.
 }
 \begin{tabular}{|c|l|}
\hline
 $m'_l=m_l$  &
$\Phi_{D^+} = 9191\pm 16_{\rm stat}\;{}^{+38}_{-36}\vert_{a^2\,{\rm extrap}}\pm 13_{\rm FV}\pm 1_{\rm EM}\ {\rm MeV}^{3/2}$ \\ 
        &
$\Phi_{D_s} = 11046\pm 12_{\rm stat}\;{}^{+42}_{-38}\vert_{a^2\,{\rm extrap}}\pm 12_{\rm FV}\pm 4_{\rm EM}\ {\rm MeV}^{3/2}$ \\ 
        &
$\Phi_{D_s}/\Phi_{D^+} = 1.2018\pm 0.0010_{\rm stat}\;{}^{+0.0024}_{-0.0032}\vert_{a^2\,{\rm extrap}}\pm 0.0004_{\rm FV}\pm 0.0005_{\rm EM}$ \\ 
        &
$\Phi_{D} = 9168\pm 16_{\rm stat}\;{}^{+39}_{-40}\vert_{a^2\,{\rm extrap}}\pm 13_{\rm FV}\pm 1_{\rm EM}\ {\rm MeV}^{3/2}$ \\ 
        &
$\Phi_{D^+}-\Phi_{D} = 23.6\pm 0.3_{\rm stat}\;{}^{+4.7}_{-1.6}\vert_{a^2\,{\rm extrap}}\pm 0.1_{\rm FV}\pm 1.0_{\rm EM}\ {\rm MeV}^{3/2}$ \\ 
\hline
 $m'_l/m_s = 0.1$ &
$\Phi_{D^+} = 9412\pm 16_{\rm stat}\;{}^{+46}_{-86}\vert_{a^2\,{\rm extrap}}\pm 13_{\rm FV}\pm 1_{\rm EM}\ {\rm MeV}^{3/2}$ \\ 
        &
$\Phi_{D_s} = 11128\pm 13_{\rm stat}\;{}^{+36}_{-42}\vert_{a^2\,{\rm extrap}}\pm 12_{\rm FV}\pm 4_{\rm EM}\ {\rm MeV}^{3/2}$ \\ 
        &
$\Phi_{D_s}/\Phi_{D^+} = 1.1824\pm 0.0010_{\rm stat}\;{}^{+0.0078}_{-0.0036}\vert_{a^2\,{\rm extrap}}\pm 0.0004_{\rm FV}\pm 0.0003_{\rm EM}$ \\ 
        &
$\Phi_{D} = 9402\pm 16_{\rm stat}\;{}^{+48}_{-95}\vert_{a^2\,{\rm extrap}}\pm 13_{\rm FV}\pm 1_{\rm EM}\ {\rm MeV}^{3/2}$ \\ 
        &
$\Phi_{D^+}-\Phi_{D} = 10.4\pm 0.3_{\rm stat}\;{}^{+9.4}_{-2.4}\vert_{a^2\,{\rm extrap}}\pm 0.1_{\rm FV}\pm 0.5_{\rm EM}\ {\rm MeV}^{3/2}$ \\ 
\hline
 $m'_l/m_s = 0.2$ &
$\Phi_{D^+} = 9709\pm 19_{\rm stat}\;{}^{+53}_{-140}\vert_{a^2\,{\rm extrap}}\pm 13_{\rm FV}\pm 2_{\rm EM}\ {\rm MeV}^{3/2}$ \\ 
        &
$\Phi_{D_s} = 11250\pm 15_{\rm stat}\;{}^{+44}_{-47}\vert_{a^2\,{\rm extrap}}\pm 12_{\rm FV}\pm 4_{\rm EM}\ {\rm MeV}^{3/2}$ \\ 
        &
$\Phi_{D_s}/\Phi_{D^+} = 1.1588\pm 0.0011_{\rm stat}\;{}^{+0.0140}_{-0.0038}\vert_{a^2\,{\rm extrap}}\pm 0.0003_{\rm FV}\pm 0.0002_{\rm EM}$ \\ 
        &
$\Phi_{D} = 9714\pm 19_{\rm stat}\;{}^{+56}_{-154}\vert_{a^2\,{\rm extrap}}\pm 13_{\rm FV}\pm 2_{\rm EM}\ {\rm MeV}^{3/2}$ \\ 
        &
$\Phi_{D^+}-\Phi_{D} = -5.3\pm 0.3_{\rm stat}\;{}^{+15.0}_{-3.3}\vert_{a^2\,{\rm extrap}}\pm 0.1_{\rm FV}\pm 0.0_{\rm EM}\ {\rm MeV}^{3/2}$ \\
\hline
\end{tabular}
\end{table}

In \tabref{results_Phi_2}, we report additional results for the case when the light valence mass is kept equal to the light sea mass 
and $m'_l/m_s=0.1$ or $0.2$.  These unphysical results
may be useful for normalizing
other calculations, such as those of $B$-system decay constants, as described in \secref{conclusions}.

\begin{table}
\caption{ \label{tab:results_Phi_2} Results for $\Phi$ for two choices of light sea masses. Here the valence mass for $\Phi_D$ is taken equal to the light sea mass: $m_\text{v}=m'_l$.
The quantities denoted by ``phys" are those tabulated in Table~\ref{tab:results_Phi} for the case $m'_l=m_l$.
  }
\begin{tabular}{|c|l|}
\hline
 $m'_l/m_s = 0.1$ &
$\Phi_{D} = 9477\pm 15_{\rm stat}\;{}^{+39}_{-66}\vert_{a^2\,{\rm extrap}}\pm 13_{\rm FV}\pm 2_{\rm EM}\ {\rm MeV}^{3/2}$ \\
        &
$\Phi_{D_s} = 11128\pm 13_{\rm stat}\;{}^{+36}_{-42}\vert_{a^2\,{\rm extrap}}\pm 12_{\rm FV}\pm 4_{\rm EM}\ {\rm MeV}^{3/2}$ \\
        &
$\Phi_{D}/\Phi_{D}^{\rm ``phys"} = 1.0338\pm 0.0005_{\rm stat}\;{}^{+0.0009}_{-0.0031}\vert_{a^2\,{\rm extrap}}\pm 0.0000_{\rm FV}\pm 0.0001_{\rm EM}$ \\
        &
$\Phi_{D}/\Phi_{D^+}^{\rm ``phys"} = 1.0311\pm 0.0004_{\rm stat}\;{}^{+0.0010}_{-0.0036}\vert_{a^2\,{\rm extrap}}\pm 0.0000_{\rm FV}\pm 0.0002_{\rm EM}$ \\
        &
$\Phi_{D_s}/\Phi_{D_s}^{\rm ``phys"} = 1.0075\pm 0.0003_{\rm stat}\;{}^{+0.0005}_{-0.0006}\vert_{a^2\,{\rm extrap}}\pm 0.0000_{\rm FV}\pm 0.0000_{\rm EM}$ \\
\hline
 $m'_l/m_s = 0.2$ &
$\Phi_{D} = 9870\pm 17_{\rm stat}\;{}^{+39}_{-71}\vert_{a^2\,{\rm extrap}}\pm 13_{\rm FV}\pm 2_{\rm EM}\ {\rm MeV}^{3/2}$ \\
        &
$\Phi_{D_s} = 11250\pm 15_{\rm stat}\;{}^{+44}_{-47}\vert_{a^2\,{\rm extrap}}\pm 12_{\rm FV}\pm 4_{\rm EM}\ {\rm MeV}^{3/2}$ \\
        &
$\Phi_{D}/\Phi_{D}^{\rm ``phys"} = 1.0766\pm 0.0011_{\rm stat}\;{}^{+0.0017}_{-0.0038}\vert_{a^2\,{\rm extrap}}\pm 0.0001_{\rm FV}\pm 0.0002_{\rm EM}$ \\
        &
$\Phi_{D}/\Phi_{D^+}^{\rm ``phys"} = 1.0738\pm 0.0011_{\rm stat}\;{}^{+0.0017}_{-0.0043}\vert_{a^2\,{\rm extrap}}\pm 0.0001_{\rm FV}\pm 0.0002_{\rm EM}$ \\
        &
$\Phi_{D_s}/\Phi_{D_s}^{\rm ``phys"} = 1.0185\pm 0.0007_{\rm stat}\;{}^{+0.0014}_{-0.0010}\vert_{a^2\,{\rm extrap}}\pm 0.0000_{\rm FV}\pm 0.0000_{\rm EM}$ \\
\hline
\end{tabular}
\end{table}

At each $\beta$ value, we have reported, in  \tabref{results_a}, the
values for the lattice spacing  $a$ and the strange mass in lattice units $am_s$,  which come from our scale-setting 
procedure using  $M_{p4s}/F_{p4s}$ and $aF_{p4s}$. For the estimates of  the extrapolation errors in these quantities, we have used the six 
versions of the continuum extrapolation for the inputs, which are the quark-mass ratios,  $M_{p4s}/F_{p4s}$, and 
 $F_{p4s}$ in physical units.  Finite volume and electromagnetic errors come simply from propagating 
  the errors in  $f_{\pi^+}$ and the light quark masses through the analysis.

The self-contained chiral analysis of the current section gives:
\begin{eqnarray}
f_{D^+} &=& 212.6 \pm0.4_{\rm stat}\;{}^{+0.9}_{-0.8}\vert_{a^2\,{\rm extrap}}\pm 0.3_{\rm FV}\pm 0.0_{\rm EM}\pm 0.3_{f_\pi\, {\rm PDG}}    \ {\rm MeV} \,, \eqn{fD-chiral-result}\\
f_{D_s} &=& 249.0\pm 0.3_{\rm stat}\;{}^{+1.0}_{-0.9}\vert_{a^2\,{\rm extrap}}\pm 0.2_{\rm FV}\pm0.1_{\rm EM} \pm0.4_{f_\pi\, {\rm PDG}}  \  {\rm MeV} \,, \eqn{fDs-chiral-result}\\
f_{D_s}/f_{D^+} &=& 1.1712(10)_{\rm stat}({}^{+24}_{-31})_{a^2\,{\rm extrap}}(3)_{\rm FV}(5)_{\rm EM} \,, \eqn{ratio-chiral-result}\\
f_{D^+}- f_D  
 &=& 
 0.47 (1)_{\rm stat} ({}^{+11}_{-\phantom{0}4})_{a^2\,{\rm extrap}}(0)_{\rm FV}  (4)_{\rm EM}  \ {\rm MeV}\eqn{fDp-fD-chiral-result} \,,
\end{eqnarray}
where $f_D$ is the decay constant in the isospin limit, $m_u=m_d=m_l$.  In finding $f_{D^+}-f_D$ from $\Phi_{D^+}-\Phi_D$ in
\tabref{results_Phi}, we use the experimental value for $M_{D^+}$ and our result, $M_{D^+}-M_{D^0}=2.6$ MeV,  obtained from
the pure-QCD analysis in \secref{physical-mass-analysis}. Comparison
with the experimental mass difference $M_{D^+}-M_{D^0}=4.8$ MeV indicates that the EM effect on this difference is $\sim\!2.2$ MeV.
We take half of this difference, namely 1.1 MeV, as our estimate of the ``EM3'' effect on the heavy-light masses,
and propagate this error to  $f_{D^+}-f_D$,
adding it in quadrature with other EM errors to get the error quoted in \eq{fDp-fD-chiral-result}.

\section{Results and conclusions}\vspace{-2.0mm}
\label{sec:conclusions}

\begin{figure}[t]
\begin{center}
        \null\vspace{-08mm}
        \begin{tabular}{l l}
        \null\hspace{-2mm}\includegraphics[trim=0.5in 0 0.3in 0, clip, width=8.3cm]{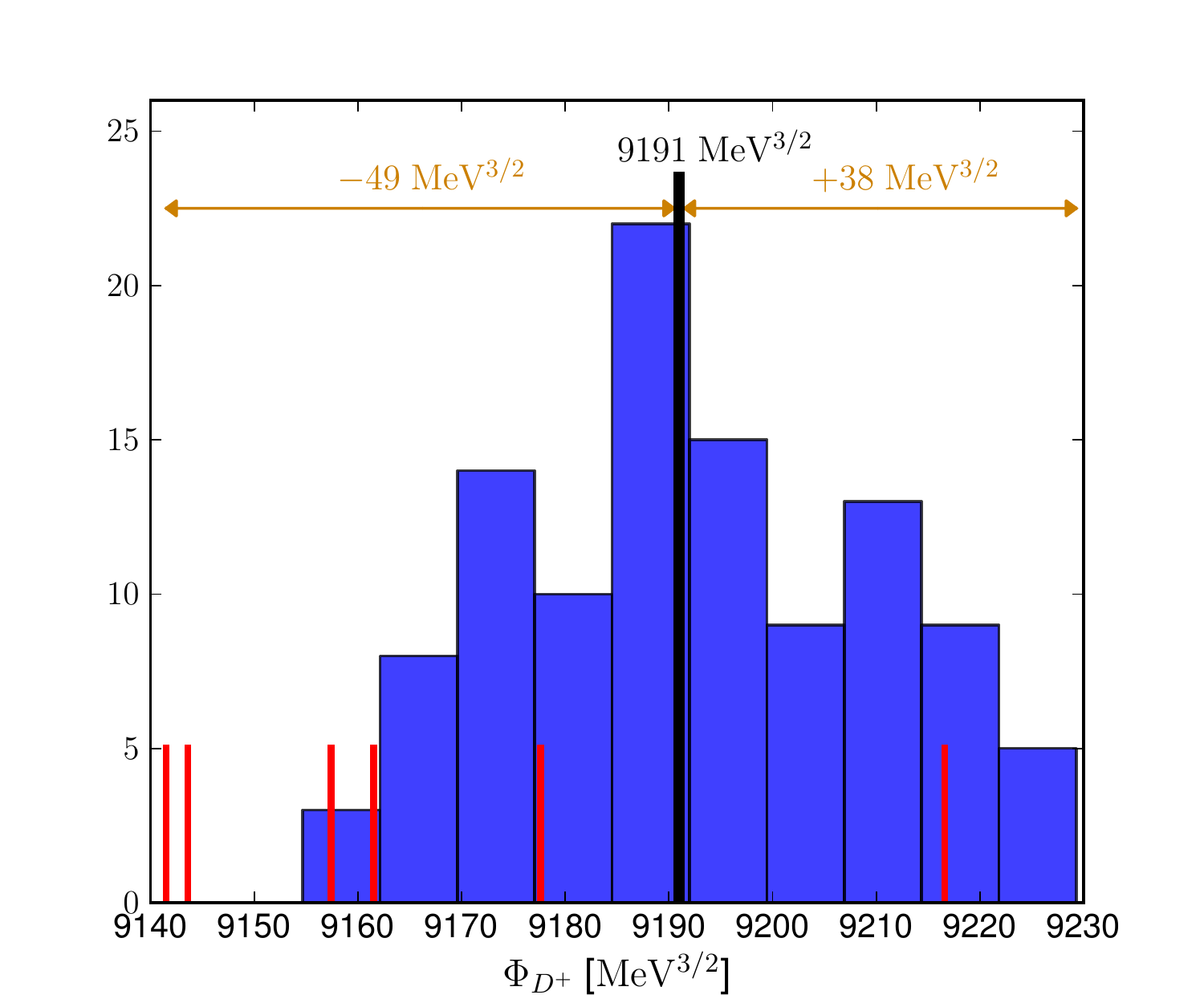}
        &\hspace{-2mm}
        \includegraphics[trim=0.5in 0 0.3in 0, clip, width=8.3cm]{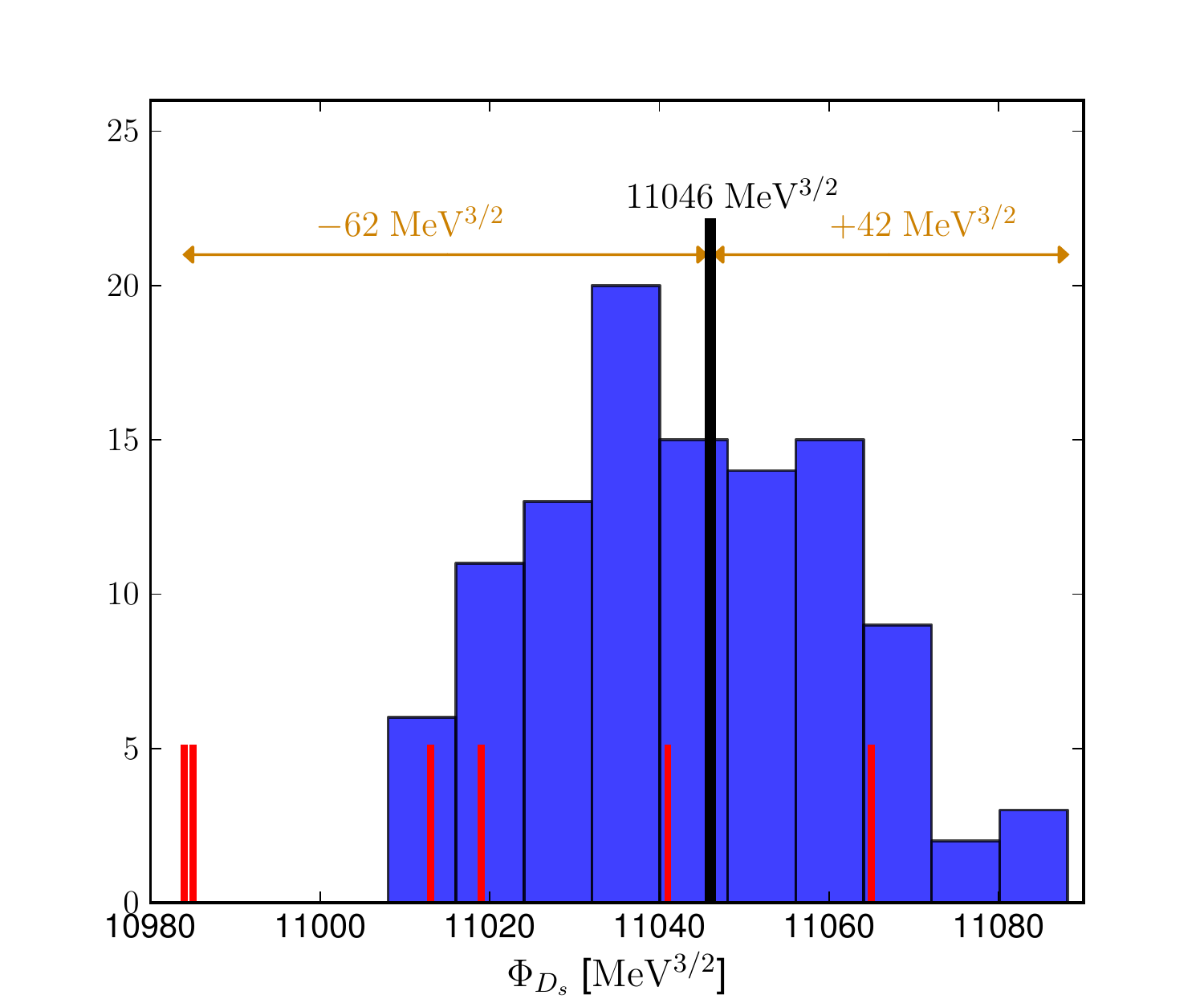}
        \end{tabular}
        \end{center}\vspace{-08mm}
\caption{
The same as \protect{\figref{hist_phi}}, but the histograms of $\Phi_{D^+}$ and $\Phi_{D_s}$ from the chiral analysis have been overlaid with results from various continuum extrapolations in the physical-mass
analysis, shown as vertical red lines. We take the full ranges shown at the top of each plot as the final estimates of the
systematic errors coming from the continuum extrapolation.
\label{fig:hist_phi_overlay}}
\end{figure}

Our main results are for the charm decay constants and their ratio.  We take the more precise determinations
from the self-contained chiral perturbation theory analysis using the full set of sea-quark ensembles,
Eqs.~(\ref{eq:fD-chiral-result})--(\ref{eq:ratio-chiral-result}), for our best estimate of the central values and
statistical errors.   We  then use the results of the simpler physical-mass analysis to help
estimate the systematic uncertainties.  For the continuum extrapolation error, we consider the differences
in the central values of $f_{D^+}$, $f_{D_s}$, and $f_{D_s}/f_{D^+}$, obtained with various
continuum-extrapolation Ans{\"a}tze in the physical-mass analysis,  and take those differences as the
uncertainty whenever they are larger than the error from the chiral analysis.  \Figref{hist_phi_overlay} shows the histograms
from \figref{hist_phi} overlaid with the results from the various continuum extrapolations  considered in \secref{physical-mass-analysis}  (vertical red lines), as well as our final estimates for the systematic errors of the continuum extrapolation.
The analysis on the
physical-mass ensembles also gives alternative, and comparably-sized,  estimates for the finite-volume and
EM errors to those in Eqs.~(\ref{eq:fD-chiral-result})--(\ref{eq:ratio-chiral-result}) (see
Table~\ref{tab:error_budget_1}), and we take the larger value as the uncertainty in each case. This
procedure yields our final results for $f_{D^+}$, $f_{D_s}$ and $f_{D_s}/f_{D^+}$:
\begin{eqnarray}
f_{D^+} &=& 212.6 \pm0.4_{\rm stat}\;{}^{+0.9}_{-1.1}\vert_{a^2\,{\rm extrap}}\pm 0.3_{\rm FV}
\pm 0.1_{\rm EM}\pm 0.3_{f_\pi\, {\rm PDG}}    \ {\rm MeV} \,, \eqn{fD-result}\\ 
f_{D_s} &=& 249.0\pm 0.3_{\rm stat}\;{}^{+1.0}_{-1.4}\vert_{a^2\,{\rm extrap}}\pm 0.2_{\rm FV}
\pm0.1_{\rm EM} \pm0.4_{f_\pi\, {\rm PDG}}
\  {\rm MeV} \,, \eqn{fDs-result}\\ 
f_{D_s}/f_{D^+} &=& 1.1712(10)_{\rm stat}({}^{+28}_{-31})_{a^2\,{\rm extrap}}(3)_{\rm FV}
(6)_{\rm EM} \,.  \eqn{ratio-result} \end{eqnarray}
For the effects of isospin violation we find
\begin{equation} 
f_{D^+}- f_D =
0.47 (1)_{\rm stat} ({}^{+25}_{-\phantom{0}4})_{a^2\,{\rm extrap}}(0)_{\rm FV}  (4)_{\rm EM}  \ {\rm MeV} ,
\eqn{fDp-fD-result}
\end{equation}
where the continuum-extrapolation error has been increased relative to that in \eq{fDp-fD-chiral-result} to take into
account the difference from the result of the physical-mass analysis.

We also update our determination of the decay-constant ratio $f_{K^+}/f_{\pi^+}$ in Ref.~\cite{FKPRL} from the
physical-mass analysis using additional configurations on the 0.06 fm physical quark mass
ensemble, and include results for quark-mass ratios coming from the tuning
procedure and continuum extrapolation described in Sec.~\ref{sec:physical-mass-analysis}:
\begin{eqnarray}
f_{K^+}/f_{\pi^+} &=& 1.1956 (10)_{\rm stat}\;{}^{+23}_{-14}\vert_{a^2\,{\rm extrap}} (10)_{\rm FV} (5)_{\rm
EM} \,, \eqn{fkpiratio-result} \\
m_s/m_l &=& 27.352 (51)_{\rm stat}\;{}^{+80}_{-20}\vert_{a^2\,{\rm extrap}} (39)_{\rm FV} (55)_{\rm
EM} \,, \eqn{slratio-result}\\ 
m_c/m_s &=& 11.747 (19)_{\rm stat}\;{}^{+52}_{-32}\vert_{a^2\,{\rm extrap}}
(6)_{\rm FV} (28)_{\rm EM}\eqn{csratio-result} \,.  \end{eqnarray}
Although our analysis also determines $m_u/m_d$, we do not quote a final result, because the errors in this
ratio are dominated by electromagnetic effects.  If we take the results from our preliminary study of EM
effects on pion and kaon masses reported in \rcite{CB-Lat14} at face value, we obtain a central value for
$m_u/m_d = 0.4482 (48)_{\rm stat}\;{}^{+21}_{-115}\vert_{a^2\,{\rm extrap}} (1)_{\rm FV}$, where we include
the uncertainties from all sources other than EM.  Once the full analysis of $m_u/m_d$ from our QCD+QED
simulations is complete, we expect the EM error to lie between 0.0150 and 0.0230.  Even the more conservative estimate for the EM
error on $m_u/m_d$,  however,  would not impact the uncertainties on our final results in
\eqsthru{fD-result}{csratio-result} significantly;  the electromagnetic error is subdominant
for most of these quantities, and one of several comparably sized errors in the case of $m_s/m_l$.
With the charm-quark mass tuned to match the $D_s$ mass, our analysis gives a
mass for the $\eta_c$ of $2982.33(0.35)({}^{+2.34}_{-2.07})$ MeV.  While this mass is in good
agreement with the experimental value, it should be remembered that our calculation does not include
the effects of disconnected contractions or decay channels to the $\eta_c$ mass.
 Finally, we note that we are computing the values of the decay constants as they are conventionally defined,
in a pure-QCD world. Comparison to experiment thus requires a matching of the decay rates between
 QCD and QCD+QED.  The errors in such a matching are not included in our error budgets for the decay constants, but are accounted for in our determinations of CKM matrix elements in Sec.~\ref{sec:CKM}.

\begin{figure}
\begin{tabular}{lll}
\includegraphics[width=0.47\textwidth]{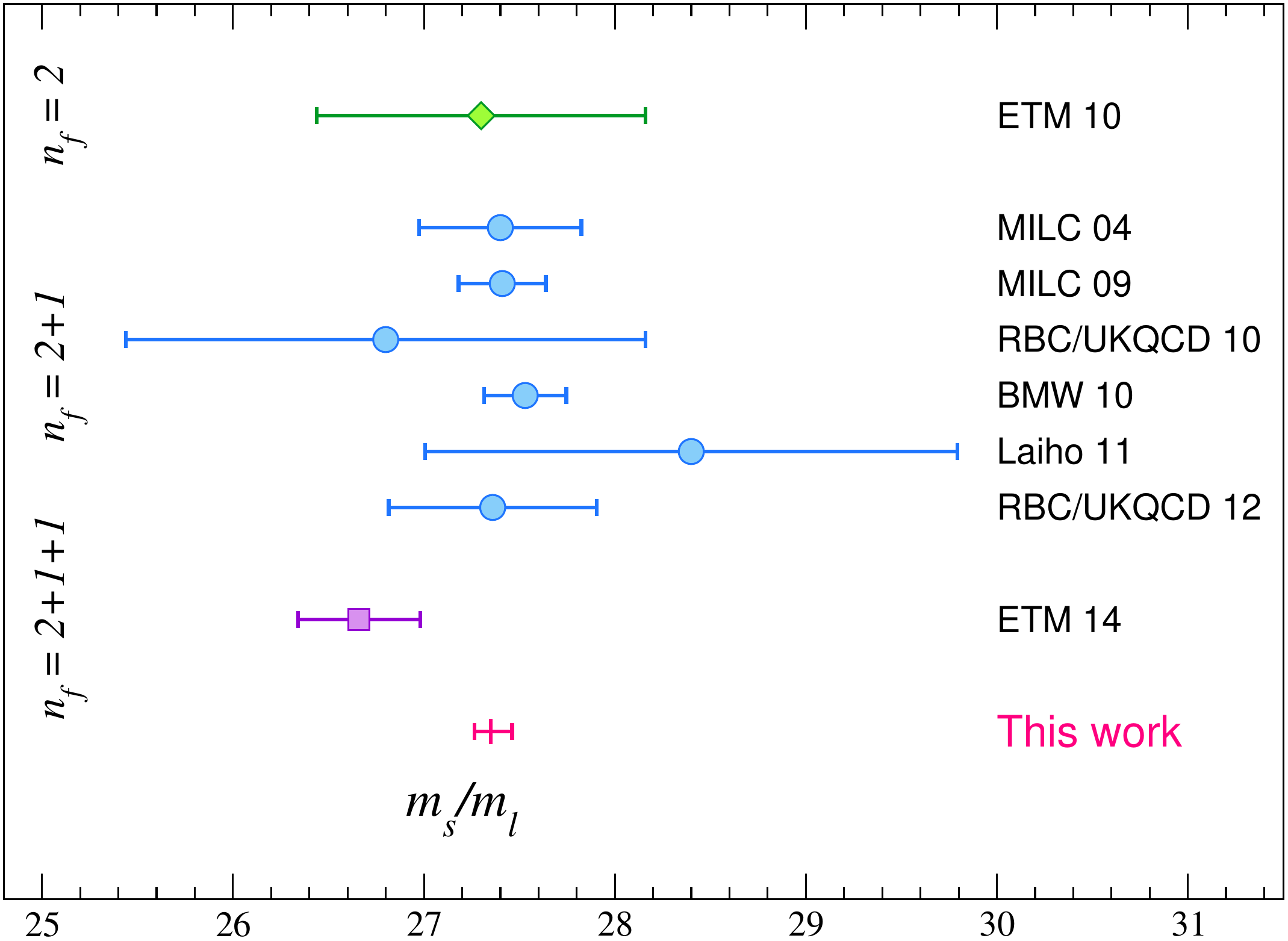} &
\phantom{x} &
\includegraphics[width=0.485\textwidth]{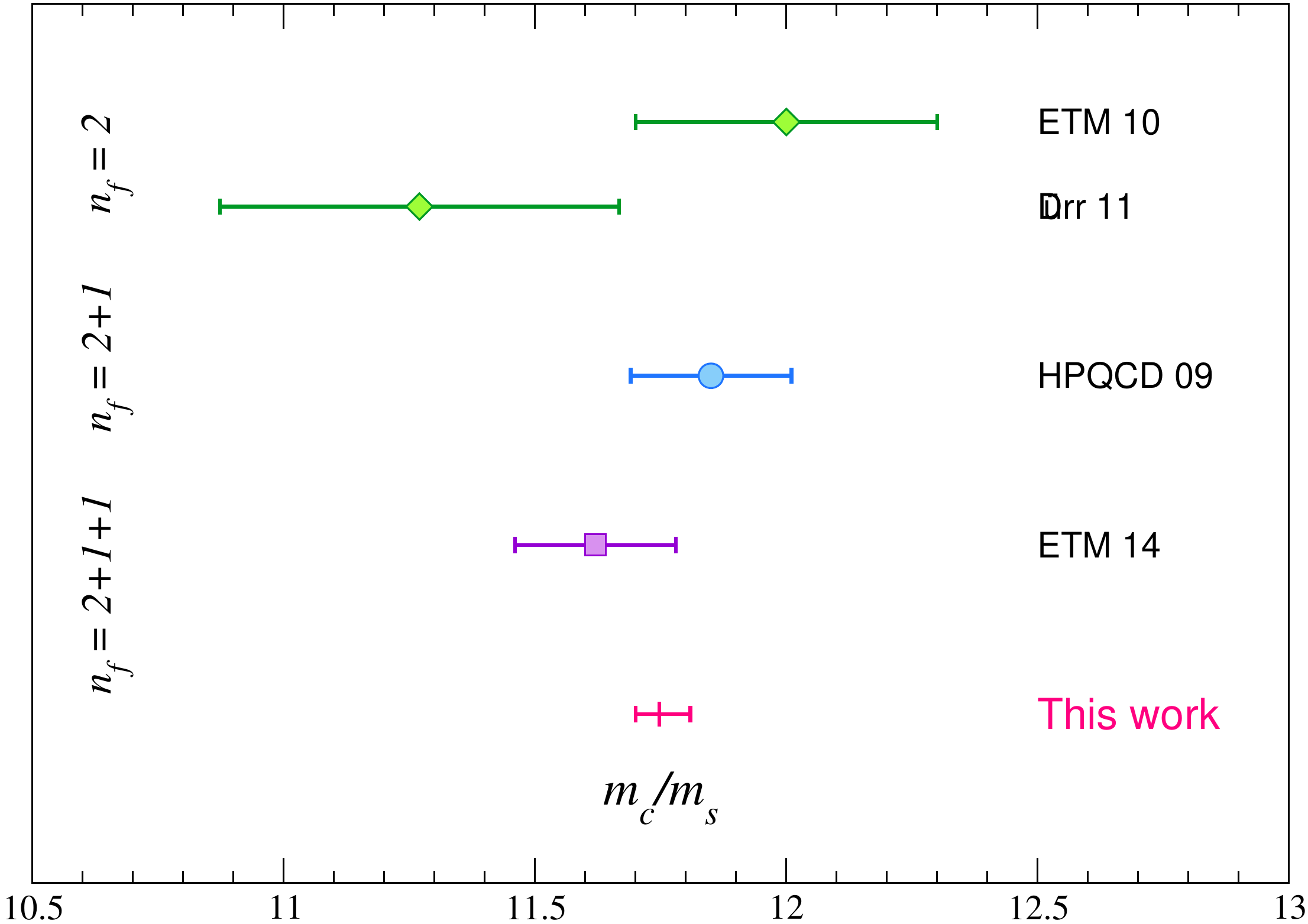} \\
\end{tabular}
\caption{\label{fig:msoml_mcoms_values} 
Unquenched lattice results for $m_s/m_l$
~\protect\cite{milc_rmp,Blossier:2010cr,FKOFPI_MILC04,FKOFPI_RBC-UKQCD10,
FKOFPI_BMW10,FKOFPI_LAIHO11,FKOFPI_RBCUKQCD12}
and $m_c/m_s$
~\protect\cite{Blossier:2010cr,Carrasco:2014cwa,Durr:2011ed,Davies:2009ih}.
Results are grouped by the
number of flavors from top to bottom: $n_f=2$ (green diamonds), $n_f = 2+1$ (blue
circles), and $n_f=2+1+1$ (purple squares).  Within each grouping, the results are in chronological order.
Our new results are denoted by magenta crosses and displayed at the bottom of each plot.  
}
\end{figure}

\begin{figure} \centerline{\includegraphics[width=0.65\textwidth]{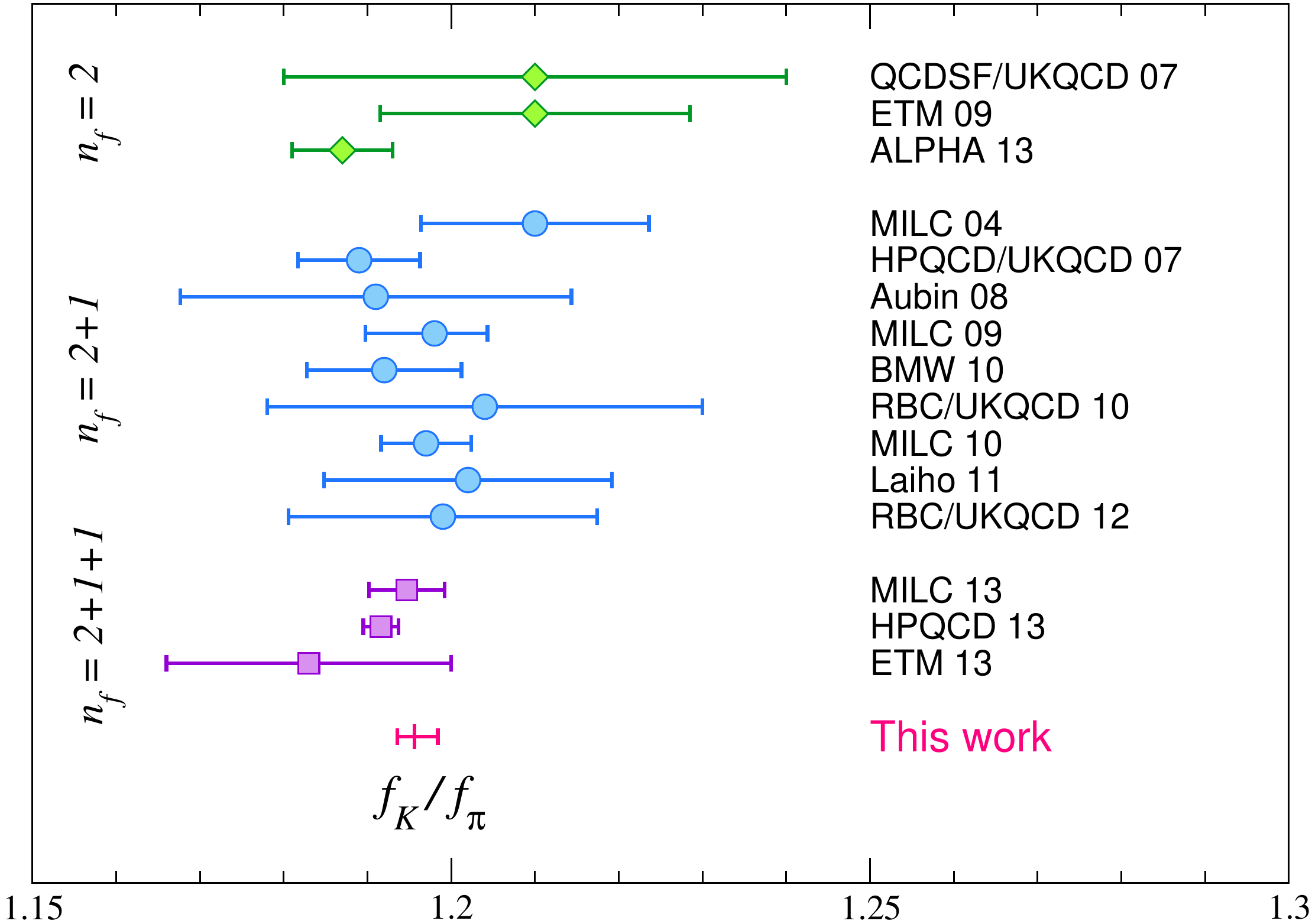}}
\caption{\label{fig:fkofpi_values} 
Unquenched lattice results for $f_K/f_\pi$
~\protect\cite{milc_rmp,FKOFPI_QCDSFUKQCD07,FKOFPI_ETM09,FKOFPI_ALPHA13,FKOFPI_MILC04,FKOFPI_HPQCD-UKQCD07,
FKOFPI_AUBIN08,FKOFPI_BMW10,FKOFPI_RBC-UKQCD10,FKOFPI_MILC10,FKOFPI_LAIHO11,FKOFPI_RBCUKQCD12,
FKOFPI_MILC13,FKOFPI_HPQCD13,ETMC13}.
The previous results are reviewed in \protect\cite{Aoki:2013ldr}.
Results are grouped by the
number of flavors from top to bottom: $n_f=2$ (green diamonds), $n_f = 2+1$ (blue
circles), and $n_f=2+1+1$ (purple squares).  Within each grouping, the results are in chronological order.
Our new result is denoted by a magenta cross and displayed at the bottom.  
In this plot we do not distinguish between results done in the isospin symmetric limit (degenerate
up and down quarks) and results including isospin violation.
The difference is small \protect\cite{Aoki:2013ldr} and does not affect the qualitative picture.
(Our result does include the up-down quark mass difference, and so is for $f_{K^+}/f_{\pi^+}$.)
}
\end{figure}

\begin{figure} \centerline{\includegraphics[width=0.65\textwidth]{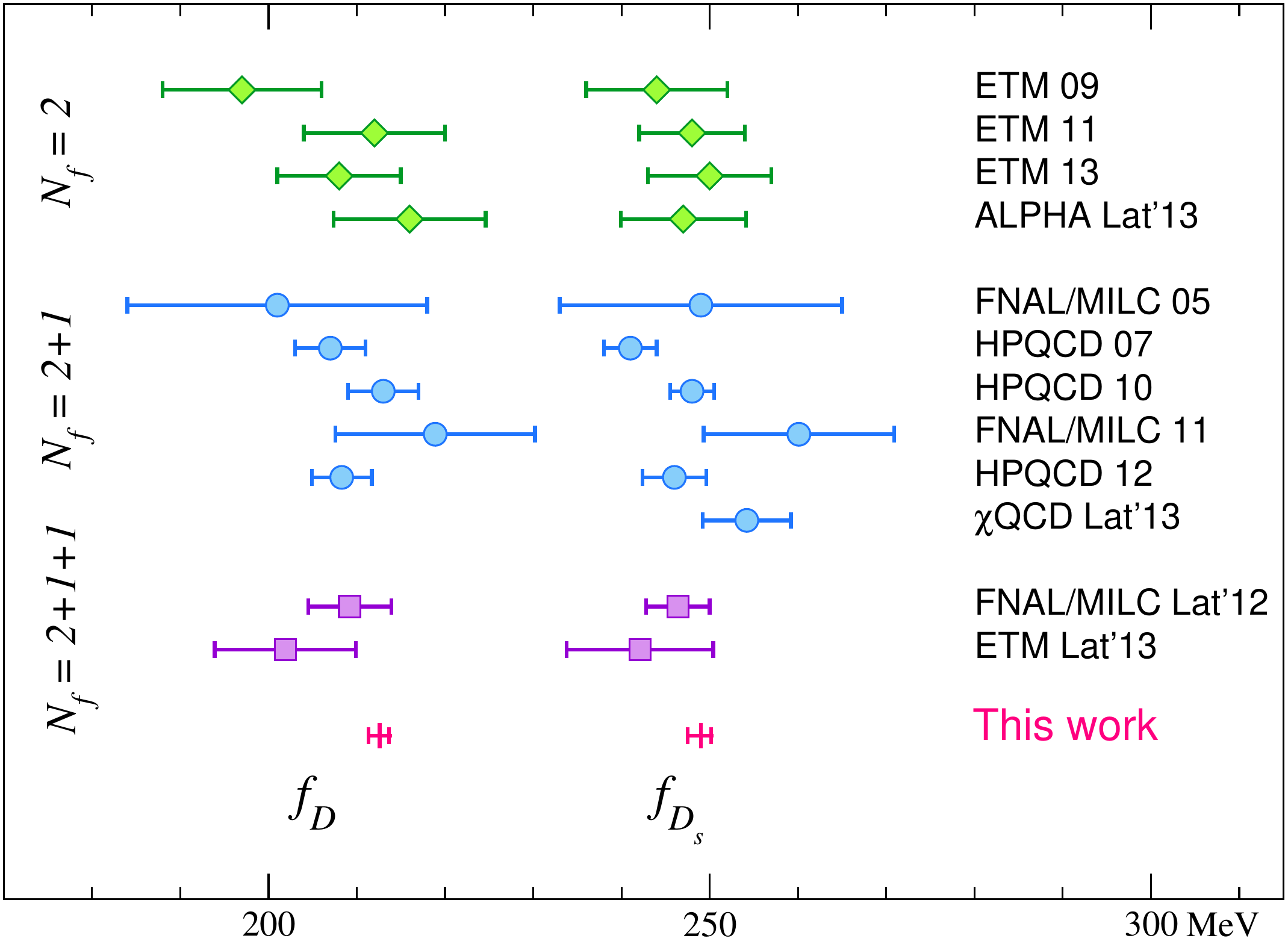}}
\caption{\label{fig:fdds_values} 
Unquenched lattice results for $f_D$ and
$f_{D_s}$~\protect\cite{FNAL95,HPQCD08,HPQCD10,FKOFPI_ETM09,
ETMC11,FNAL11,HPQCD12,LATTICE12_FD,ALPHA13,CHIQCD13,ETMC13}.
We do not include \rcite{TWQCD14} because
of the small volume used, and \rcite{PACS-CS11} because of the lack of a continuum extrapolation.
Results are grouped by the
number of flavors from top to bottom: $n_f=2$ (green diamonds), $n_f = 2+1$ (blue
circles), and $n_f=2+1+1$ (purple squares).  Within each grouping, the results are in chronological order.
Our new results are denoted by magenta pluses and displayed at the bottom.  
Again, we do not distinguish results in the isospin symmetric limit from those with
non-degenerate up and down quarks, where we have estimated the difference in
Eq.~\protect\ref{eq:fDp-fD-result}.
}
\end{figure}

\begin{figure} \includegraphics[width=0.65\textwidth]{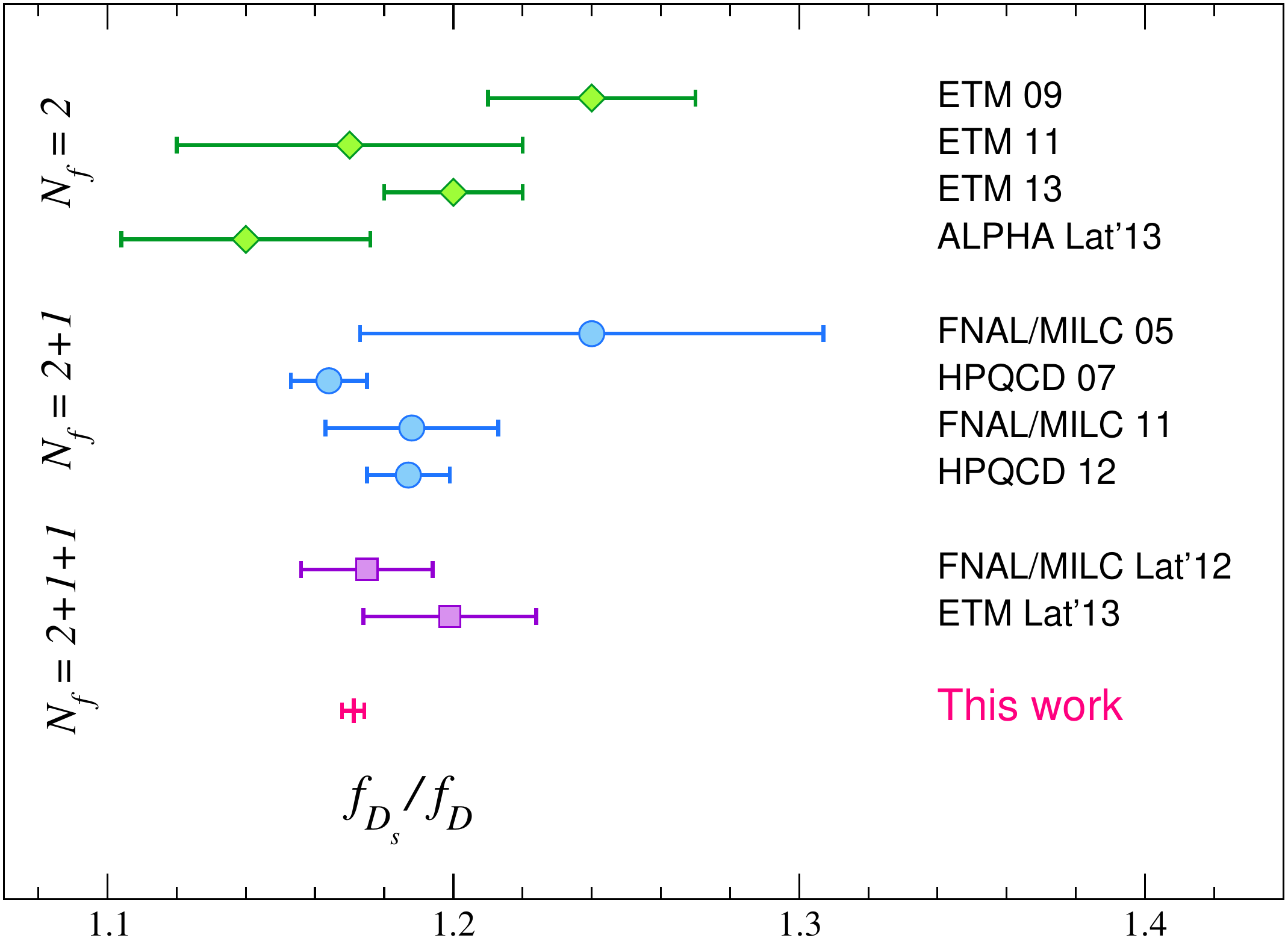}
\caption{\label{fig:fddsratio_values} Unquenched lattice results for
$f_{D_s}/f_D$~\protect\cite{FNAL95,HPQCD08,HPQCD10,FKOFPI_ETM09,
ETMC11,FNAL11,HPQCD12,LATTICE12_FD,ALPHA13,ETMC13}.  Results are grouped by the number of
flavors from top to bottom: $n_f=2$ (green diamonds), $n_f = 2+1$ (blue circles), and
$n_f=2+1+1$ (purple squares).  Within each grouping, the results are in chronological order.  Our new
result is  denoted by magenta crosses and displayed at the bottom.  }
\end{figure}

Figures~\ref{fig:msoml_mcoms_values}, \ref{fig:fkofpi_values}, \ref{fig:fdds_values}
and~\ref{fig:fddsratio_values}
compare our results for $m_s/m_l$, $m_c/m_s$, $f_{K^+}/f_{\pi^+}$ and the charm decay
constants with other unquenched calculations.  Our results agree with most determinations at the
1--2$\sigma$ level.  In particular, our value for $f_{D_s}$ agrees with the second-most-precise
determination from HPQCD obtained using HISQ valence quarks on the (2+1)-flavor MILC Asqtad
ensembles~\cite{HPQCD10}.  We disagree slightly with HPQCD's determination of the ratio
$f_{D_s}/f_{D^+}$~\cite{HPQCD12}, but only by 1.2$\sigma$.
Our result for $f_{D_s}$ is more precise than previous determinations primarily for
two reasons.  First, the statistical errors in our data points for the decay amplitudes are two or more
times smaller than those obtained by, for example, HPQCD \cite{HPQCD10}.  Second, our use of
ensembles with the physical light-quark mass eliminates the significant (although not dominant) uncertainty
from  the chiral extrapolation.  For $f_{D^+}$ and $f_{D_s}/f_{D^+}$, we also have significantly smaller
continuum-extrapolation errors due to the use of the HISQ sea-quark action and lattice spacings down
to $a \approx 0.06$~fm.

The dominant source of uncertainty in our results is from the continuum extrapolation, and will be reduced
once we include a still finer ensemble in our analysis with $a\approx 0.045$ fm 
and $m_l/m_s=0.2$, generation
of which is in progress.  In fact, we already have some preliminary data on this ensemble, albeit with small
statistics, and have tried including these data in the current chiral fits.  The fits have acceptable $p$
values and give results that are less than one statistical sigma away from those in
\eqsthru{fD-result}{fkpiratio-result}.  Once we have ensembles with lattice spacings as fine as
$a \approx 0.03$~fm, we expect to be able to use the same methods employed here to compute bottom decay
constants.  In the meantime, however, our results for $D$-meson decay constants using HISQ charm quarks can be
combined with calculations of the ratios $\Phi_{B_s}/\Phi_{D_s}$ using Fermilab heavy quarks to
improve the determinations of decay constants in the $B$
system, where the use of the HISQ action is more difficult.  The ratios of continuum-extrapolated decay constants at various unphysical
values of the light-quark mass may also be useful for this approach.  The analysis of $B$- and $D$-meson decay constants with Fermilab heavy quarks on the 2+1 flavor asqtad ensembles is presently being
finalized~\cite{ETHAN_INPREP}.

\section{Impact on CKM phenomenology} \label{sec:CKM}

We now use our decay constant results to obtain values for CKM matrix elements within the Standard Model,
and to test the unitarity of the first and second rows of the CKM matrix.

\begin{figure} \includegraphics[height=0.65\textwidth]{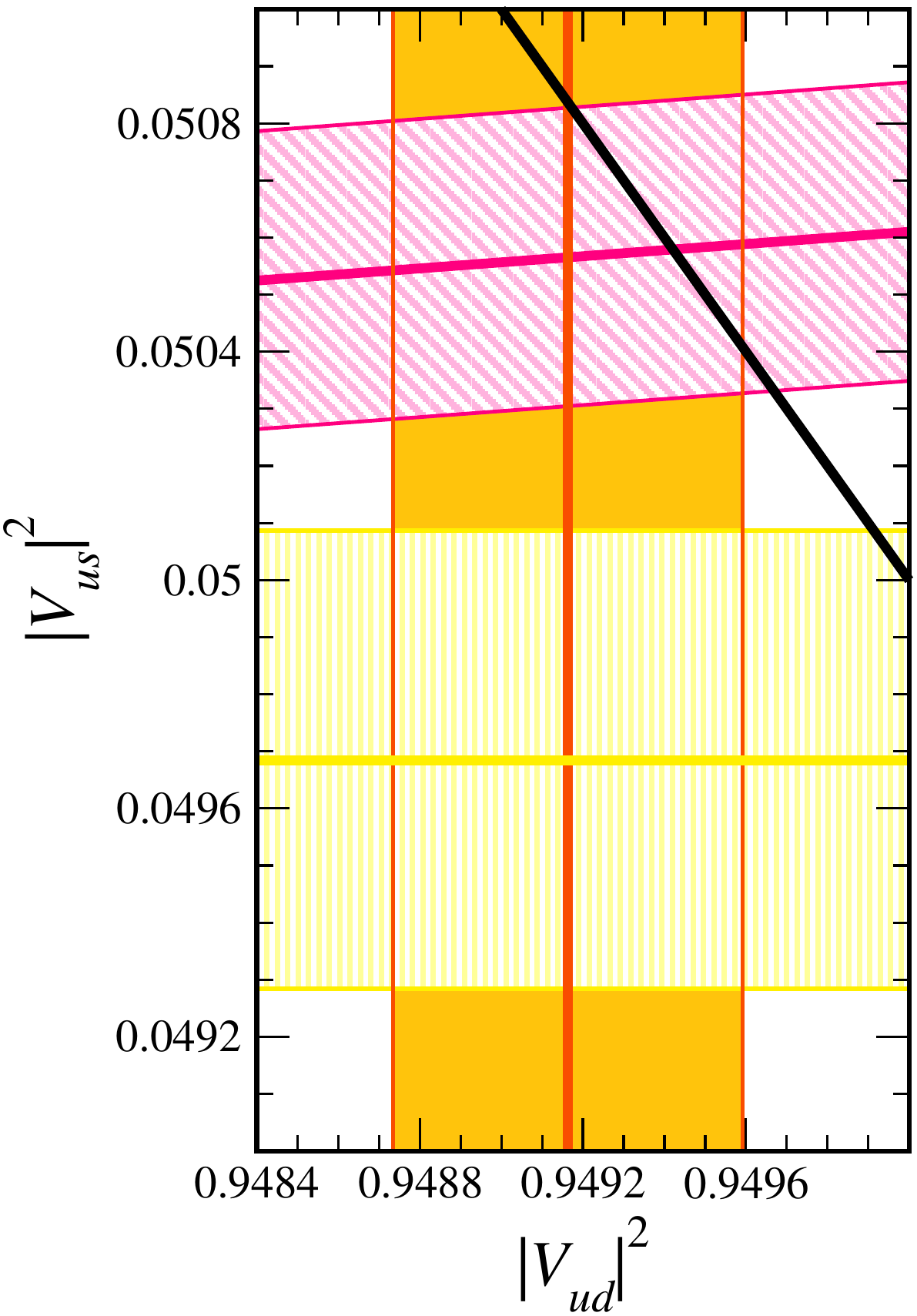} \hfill
\includegraphics[height=0.65\textwidth]{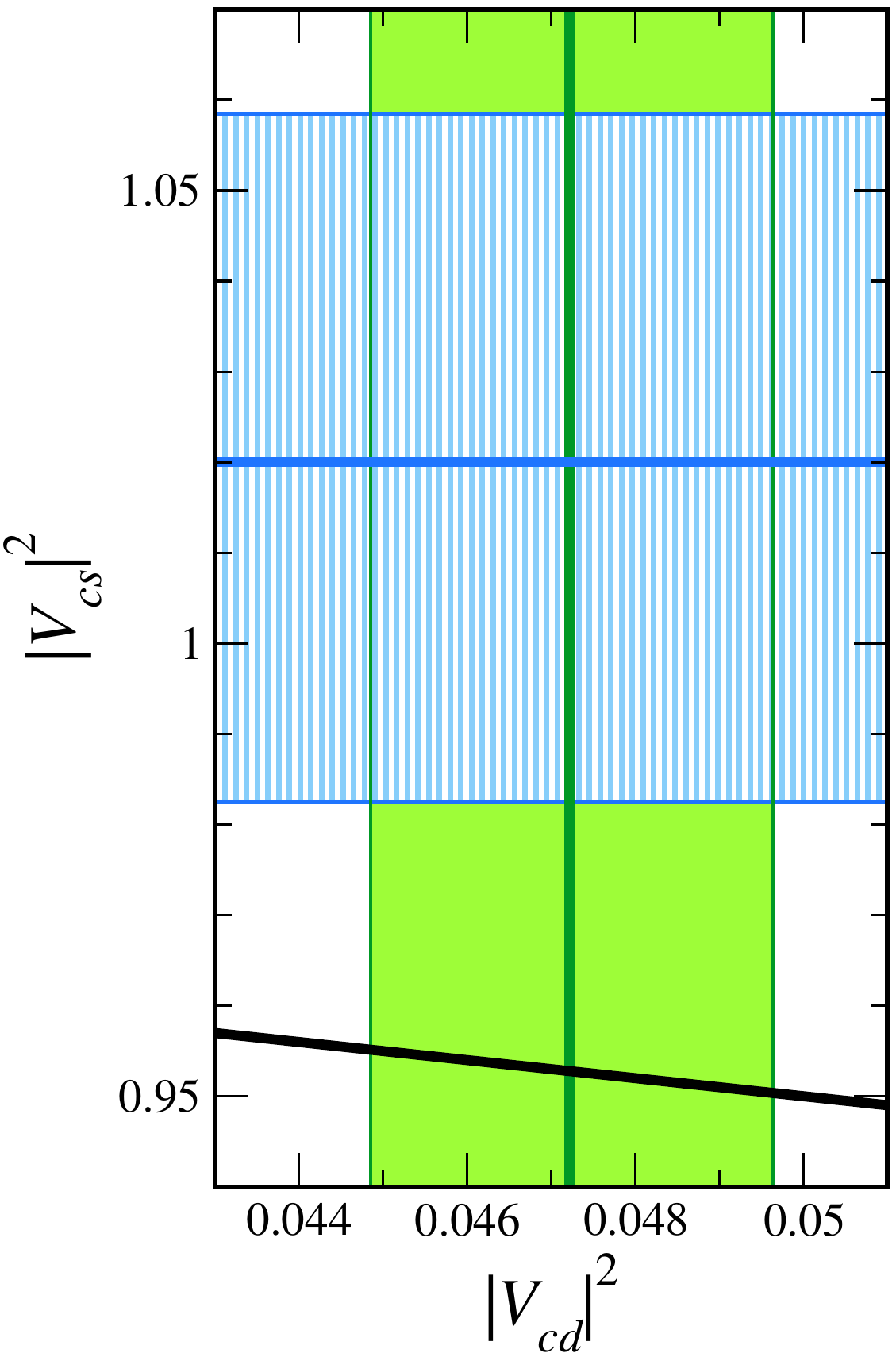} 
\caption{\label{fig:firstrow_unitarity} Unitarity
tests of the Cabibbo-Kobayashi-Maskawa matrix.  Left: squared magnitudes of elements of the first row of the
CKM matrix.  The magenta diagonal band shows $(|V_{us}|/|V_{ud}|)^2$ obtained using $f_{K^+}/f_{\pi^+}$ from
this work, the vertical orange band shows $|V_{ud}|^2$ from nuclear $\beta$ decay~\cite{Hardy:2008gy}, and the
horizontal yellow band shows $|V_{us}|^2$ obtained using our recent calculation of the kaon semileptonic form
factor at $q^2=0$~\cite{Bazavov:2013maa}.  The diagonal black line is the unitary prediction, and lies well
within the region of overlap of the magenta and orange bands.  Right: squared magnitudes of elements of the
second row of the CKM matrix.  The green vertical and blue horizontal bands show $|V_{cd}|^2$ and
$|V_{cs}|^2$ obtained using $f_{D^+}$ and $f_{D_s}$ from this work.  The black diagonal line does not
intersect with the region of overlap of the two colored bands, indicating a slight tension with CKM
unitarity.}
\end{figure}

The decay-constant ratio $f_{K^+}/f_{\pi^+}$ can be combined with experimental measurements of the
corresponding leptonic decay widths to obtain a precise value for the ratio
$|V_{us}|/|V_{ud}|$~\cite{Marciano:2004uf}. Combining our updated result for $f_{K^+}/f_{\pi^+}$ from
\eq{fkpiratio-result} with recent experimental results for the leptonic branching
fractions~\cite{PDG} and an estimate of the hadronic structure-dependent EM correction~\cite{Antonelli:2010yf},  we obtain 
\begin{equation} |V_{us}|/|V_{ud}| = 0.23081 (52) _{\rm LQCD} (29) _{{\rm BR}(K_{\ell 2})} (21)_{\rm EM} \,. 
\end{equation}
Taking $|V_{ud}|$ from nuclear $\beta$ decay~\cite{Hardy:2008gy}, we also obtain
\begin{equation} |V_{us}| = 0.22487 (51) _{\rm LQCD} (29) _{{\rm BR}(K_{\ell 2})} (20)_{\rm EM} (5)_{V_{ud}}\,. 
\end{equation}
This result for $|V_{us}|$ is more precise than our recent determination from a calculation of the kaon
semileptonic form factor on the physical-mass HISQ ensembles~\cite{Bazavov:2013maa}, and larger by
1.8$\sigma$.  Figure~\ref{fig:firstrow_unitarity} shows the unitarity test of the first row of the CKM
matrix using our result for $f_{K^+}/f_{\pi^+}$.   We find good agreement with CKM unitarity, and obtain a
value for the sum of squares of elements of the first row of the CKM matrix consistent with the
Standard-Model prediction zero at the level of $10^{-3}$:
\begin{equation} 1 - |V_{ud}|^2 - |V_{us}|^2 - |V_{ub}|^2 = 0.00026 (51) \,.  \end{equation}
Thus our result places stringent constraints on new-physics scenarios that would lead to deviations from
first-row CKM unitarity.  Finally, we note that, now that the uncertainty in $|V_{us}|^2$ is approximately the same as
that in $|V_{ud}|^2$, it is especially important
to scrutinize the current uncertainty estimate for $|V_{ud}|$.

The $D^+$- and $D_s$-meson decay constants can be combined with experimental measurements of the
corresponding leptonic decay widths to obtain $|V_{cd}|$ and $|V_{cs}|$.  The values $f_{D^+} |V_{cd}| =
46.06 (1.11)$ MeV and $f_{D_s} |V_{cs}| = 250.66 (4.48)$ MeV in the PDG~\cite{Rosner:2013ica} are obtained from
averaging the experimentally-measured decay rates into electron and muon final states including an estimate
of structure-dependent Bremsstrahlung effects that lowers the $D^+ \to \mu^+ \nu_{\mu}$ rate by $\sim
1\%$~\cite{Burdman:1994ip,Dobrescu:2008er}.  The PDG determinations of $f_{D^+} |V_{cd}|$ and $f_{D_s}
|V_{cs}|$ do not, however, take into account other  electroweak corrections ({\it c.f.}
Refs.~\cite{Marciano:2004uf} and \cite{Cirigliano:2011ny} and references therein).  Such contributions are estimated
for pion and kaon leptonic decay constants to be $\sim$ 1--2\%, and the uncertainties in these 
corrections, in particular from the contributions that depend on the hadronic structure, lead to $\sim 0.1\%$ uncertainties in
$|V_{us}|/|V_{ud}|$ and $|V_{us}|$ obtained from leptonic decays.  Now that the uncertainties in the charm
decay constants are at the half-a-percent level, it is timely to consider including electroweak
corrections when extracting $|V_{cd}|$ and $|V_{cs}|$ from leptonic $D$ decays, and we attempt to provide a
rough estimate of their possible size here.  We consider all of the contributions that have been estimated
for pion and kaon leptonic decays.  Not all of the necessary calculations have been performed for the charm
system, however, so, where necessary, we use results for the pion and kaon system as a guide and take a
generous uncertainty.

The universal long-distance EM contribution to leptonic decays of point-like charged particles was calculated
by Kinoshita~\cite{Kinoshita:1959ha}.   Evaluating this contribution for leptonic $D$ decays into
muons (because the experimental averages are dominated by measurements in the muon channel), the long-distance correction lowers both the $D^+$ and $D_s$ decay rates by about 2.5\%.  The universal
short-distance contribution to leptonic decays of charged pseudoscalar mesons, which accounts for
electroweak corrections not included in the definition of $G_F$, was computed by Sirlin~\cite{Sirlin:1981ie}.  Choosing $M_D$ for the factorization scale that enters $\ln(M_Z/\mu)$, the
``Sirlin factor" increases the $D^+$ and $D_s$ leptonic decay rates by about $1.8\%$.   Thus the net effect
of these two known corrections is a slight decrease in the $D^+$ and $D_s$ rates by less than a percent.
Finally, we consider EM effects that depend on the mesons' hadronic structure.  The expressions for the
structure-dependent contributions to charged pion and kaon decay rates have been computed at ${\mathcal O}(e^2
p^2)$ and  ${\mathcal O}(e^2 p^4)$ in chiral perturbation theory~\cite{Knecht:1999ag,Cirigliano:2007ga}.  The
dominant ${\mathcal O}(e^2 p^2)$ contribution takes the form $c_1^{(P)} \alpha/\pi$, and the coefficients
have been estimated numerically in the large-$N_c$ approximation to be $c_1^{(\pi)}=-2.4(5)$ and
$c_1^{(K)}=-1.9(5)$~\cite{DescotesGenon:2005pw}.  These calculations do not apply to the charm system,
however, because the $D_{(s)}$-meson masses are much heavier than the pion and kaon masses, and well outside
the range of validity of the light-meson chiral expansion.   We therefore consider the possibility that the analogous
coefficients for the $D$ system are 2--5 times larger than for the pion and kaon system.  With this
assumption, we find a range of the possible size for the hadronic correction to the $D^+$- and $D_s$-meson
leptonic decay rates from 1.1--2.8\%.  Corrections of this size would not be negligible compared to the
known short-distance and long-distance contributions;  thus it is important to obtain a more reliable
estimate of the contributions to charged $D$ decays due to hadronic structure in the future.

For the determinations of $|V_{cd}|$ and $|V_{cs}|$ given here, we first adjust the experimental decay rates
quoted in the PDG by the known long-distance and short-distance electroweak corrections.  We then add an estimate of the
uncertainty due to the unknown hadronic structure-dependent EM corrections, taking the lower estimate of 0.6\%.  With
these assumptions, and using our results for $f_{D^+}$ and $f_{D_s}$ from \eqs{fD-result}{fDs-result}, we obtain
\begin{eqnarray} |V_{cd}| &=& 0.217 (1) _{\rm LQCD} (5) _{\rm expt}  (1)_{\rm EM} \,,         \label{eq:Vcd} \\
|V_{cs}| &=& 1.010 (5) _{\rm LQCD} (18) _{\rm expt}  (6)_{\rm EM}                              \label{eq:Vcs} \,, \end{eqnarray}
where ``EM" denotes the error due to unknown structure-dependent EM corrections.  In both cases, the uncertainty is
dominated by the experimental error in the branching fractions.  Thus the significant improvement in $f_{D^+}$
and $f_{D_s}$ does not, at present, lead to direct improvement in $|V_{cd}|$ and $|V_{cs}|$.  Experimental
measurements of the $D^+$ decay rates have improved recently~\cite{Rosner:2013ica}, however, such that the
error on $|V_{cd}|$ from leptonic $D^+$ decays is now approximately half that of $|V_{cd}|$ obtained from
either neutrinos~\cite{PDG} or semileptonic $D \to \pi \ell \nu$ decay~\cite{Na:2011mc}.

Our result for $|V_{cd}|$ agrees with the determination from neutrinos.  Our $|V_{cd}|$ is 1.0$\sigma$ lower
than the determination from semileptonic $D$ decay  in Ref.~\cite{Na:2011mc}, while our $|V_{cs}|$ is 1.1$\sigma$ higher than that of  Ref.~\cite{Na:2010uf}.
Figure~\ref{fig:firstrow_unitarity} shows the unitarity test of the second row of the CKM matrix using our
results for $f_{D^+}$ and $f_{D_s}$.   We obtain a value for the sum of squares of elements of the second
row of the CKM matrix of
\begin{equation} 1 - |V_{cd}|^2 - |V_{cs}|^2 - |V_{cb}|^2 = -0.07(4) \,, \end{equation}
showing some tension with CKM unitarity.
This test will continue to become more stringent as
experimental measurements of the $D^+$ and $D_s$ decay rates become more precise.  At present, even if our
rough estimate of the uncertainty due to structure-dependent EM corrections in Eqs.~(\ref{eq:Vcd}) and~(\ref{eq:Vcs})
is too small by a factor of two,  the errors on $|V_{cd}|$ and
$|V_{cs}|$ would not change significantly.  It will be important, however, to obtain a more reliable estimate of the contributions to
charged $D$ decays due to hadronic structure in the future.   

\section*{Acknowledgements}
Computations for this work were carried out with resources provided by the 
USQCD Collaboration, the Argonne Leadership Computing Facility and
the National Energy Research Scientific Computing Center, which are funded 
by the Office of Science of the United States Department of Energy; 
and with resources provided by the National Center for Atmospheric Research,
the National Center for Supercomputing Applications, the National Institute 
for Computational Science, and the Texas Advanced Computing Center, which
are funded through the National Science Foundation's Teragrid/XSEDE Program;
and with resources provided by the Blue Waters Computing Project, which is
funded by NSF grants OCI-0725070 and ACI-1238993 and the state of Illinois.
This work is also part of the "Lattice QCD on Blue Waters" PRAC allocation supported by the
National Science Foundation grant OCI-0832315.
This work was supported in part by the
U.S.\ Department of Energy under grants 
No.~DE-FG02-91ER40628 (C.B., J.\ Komijani), 
No.~DE-FC02-12ER41879 (C.D., J.F., L.L.), 
No.~DE-FG02-91ER40661 (S.G., R.Z.), 
No.~DE-SC0010120 (S.G.),
No.~DE-FC02-06ER41443 (R.Z.),
No.~DE-FG02-13ER42001 (D.D., A.X.K.), 
No.~DE-FG02-04ER-41298 (D.T.); 
No.~DE-FG02-13ER-41976 (D.T.),
No.~DE-FC02-06ER-41439 (J.\ Kim),
by the National Science Foundation under Grants 
No.~PHY-1067881 (C.D., J.F., L.L.), 
No.~PHY-1212389 (R.Z.),
and No.~PHY-1316748 (R.S.); 
by the URA Visiting Scholars' program (A.X.K.);
by the MICINN (Spain) under grant FPA2010-16696 and Ram\'on y Cajal program (E.G.);
by the Junta de Andaluc\'ia (Spain) under Grants No.~FQM-101, No.~FQM-330, and No.~FQM-6552 (E.G.);
and by European Commission (EC) under Grant No.~PCIG10-GA-2011-303781 (E.G.).
A.S.K.\ thanks the DFG cluster of excellence ``Origin and Structure of the Universe'' at the Technische Universit\"at M\"unchen for hospitality while this work was being completed.
This manuscript has been co-authored by an employee of Brookhaven Science
Associates, LLC, under Contract No. DE-AC02-98CH10886 with the
U.S.\ Department of Energy. 
Fermilab is operated by Fermi Research Alliance, LLC, under Contract
No.\ DE-AC02-07CH11359 with the United States Department of Energy.

\appendix
\section{\label{sec:m_Q} Expansion of $\Phi_0$ in terms of $1/m_Q$}

\Equation{chpt-form} contains the effects of hyperfine splittings (\eg $M^*_D-M_D$) and flavor splittings (\eg $M_{D_s}-M_D$), but no other $1/m_Q$ effects. 
Boyd and Grinstein \cite{BoydGrinstein} find some other contributions at the same order as hyperfine and flavor splittings. However, 
 one can show that most of these terms only produce  $1/m_Q$ corrections to
 the  LECs relevant to the pseudoscalar-meson decay constants.  (Some of the terms violate heavy-quark spin symmetry,
and therefore give different contributions to the pseudoscalar and vector-meson decay constants at this order, but we are not concerned with vector-meson decay constants here.)
Following  Eq.~(20) of Ref.~\cite{BoydGrinstein}, at the order of $\cO(1/m_Q,m_q^0)$ where $m_q$ is a light quark mass,
the $1/m_Q$ terms can be included by replacing $\Phi_0$ by $\Phi_0(1+{\rm const}/m_Q)$.
This dependence can be simply absorbed in $\Phi_0$ for a fixed value of $m_Q$. However, in our analysis
the charm mass varies by about $10\%$, which leads to a correction comparable to that produced
by terms of $\cO(m_q)\sim \cO(m_\pi^2)$. 
Therefore, replacing $\Phi_0$ by $\Phi_0(1+{\rm const}/m_Q)$ in \eq{chpt-form} should be considered a
NLO correction.
At this order the rate for $D^* \rightarrow D\pi$ is governed by $g_\pi(1 + {\rm const}/m_Q)$ instead of $g_\pi$,
which is already taken into account by incorporating the range $g_\pi = 0.53(8)$ in the fits.
We do not allow any further dependence of $g_\pi$ on $m_Q$ in our analysis, because this dependence is 
formally  NNLO.

\begin{figure}[t]
\null\vspace{-5mm}
\hspace{-8mm}\includegraphics[width=16.2cm]{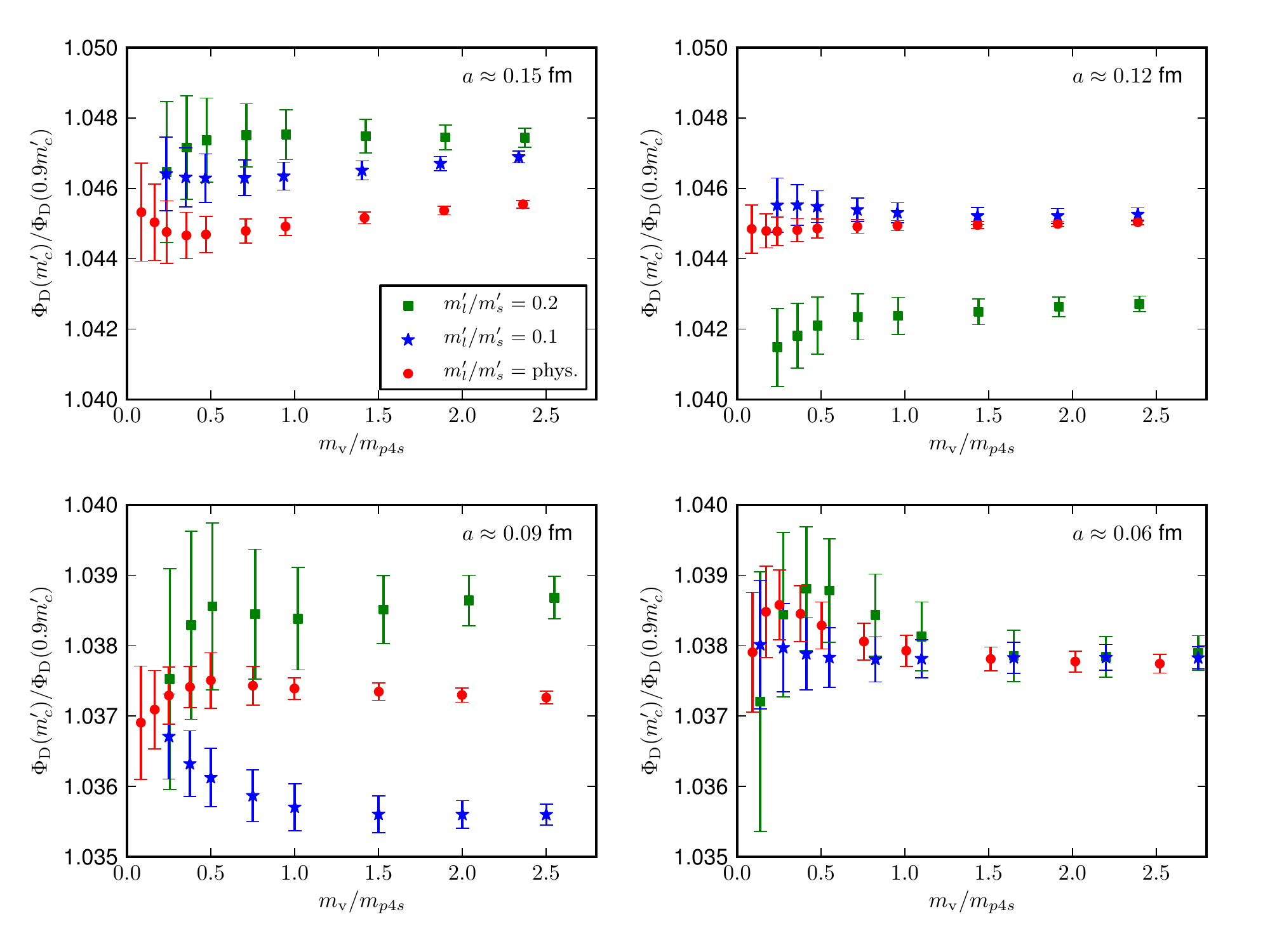}
\vspace{-5mm}
\caption{ The ratio $\Phi_D(m'_c)/\Phi_D(0.9m'_c)$ (where $m'_c$ is the charm sea mass of the ensembles) 
as a function of $m_{\rm v}$, the light valence-quark mass.
The upper left panel shows data at $a\approx 0.15$ fm.
The upper right panel shows the data at $a\approx 0.12$ fm from the ensembles with $m_s$ tuned close to its physical value.
In the second row, we show $a\approx 0.09$ fm (left)  and $a\approx 0.06$ fm (right) data.
\label{fig:PhiD_ratio_10mC_09mC}}
\end{figure}

On each ensemble, we have data with two different values of the valence charm mass: $m'_c$ and $0.9 m'_c$, where $m'_c$ is the charm sea mass of the ensemble.
In \figref{PhiD_ratio_10mC_09mC}, the ratio of $\Phi_D$ at $m'_c$ to $\Phi_D$ at $0.9 m'_c$ is shown in terms of $m_{\rm v}$ for our four lattice spacings. 
The fact that $\Phi_D(m'_c)/\Phi_D(0.9m'_c)$ does not vary much as a function of the light valence-quark
mass is  evidence 
that the $1/m_Q$ effects can be absorbed in the overall factor in front of the full one-loop result as discussed above. 
On the other hand, $\Phi_D$ computed at $m'_c$ and at $0.9 m'_c$ are highly correlated so that their ratio is known precisely. 
Since our fits take the correlations into account, the $p$~values will be low unless the chiral form is able to reproduce the ratio to high accuracy.
Therefore, the expansion of the overall factor, $\Phi_0$, in terms of $1/m_Q$  needs to be
taken beyond the first order; 
for acceptable fits we need to introduce a $1/m_Q^2$ term as well as the $1/m_Q$ term,
as indicated in \eq{Phi0-form}.
Furthermore, good fits require the LEC $k_1$ in \eq{Phi0-form} to have
generic dependence on $a$; such dependence for  $k_2$ is also strongly preferred by the fits.

Note finally that \figref{PhiD_ratio_10mC_09mC} shows a roughly $4\%$ difference  
 between $\Phi_D$ at $m'_c $ and at $0.9 m'_c$.  As claimed in the discussion above \eq{Phi0-form}, this is comparable to the
chiral NLO effects of a nonzero pion mass, which may be estimated 
from the fits shown in \figref{chiral-fit}.
Indeed,  those fits imply that the difference between the physical value of $\Phi_{D^+}$ 
and its value in the (two-flavor) chiral limit is roughly $3\%$.

\end{document}